\documentclass[PAPER,UKenglish,cernpreprint,atlasdraft=false, texmf, orcidlogo]{atlasdoc}
\usepackage{atlaspackage}
\usepackage{atlasbiblatex}

\usepackage{atlasphysics}
\graphicspath{{logos/}{figures/}}

\usepackage{ANA-TOPQ-2024-12-PAPER-defs}

\AtlasTitle{Precise measurement of the $t\bar{t}$ production cross-section and lepton differential distributions in $\boldmath e\mu$ dilepton events from $\boldmath\sqrt{s}=13\,\TeV$ $\boldmath pp$ collisions with the ATLAS detector}

\AtlasAbstract{%
The inclusive top quark pair ($t\bar{t}$) cross-section $\sigma_{t\bar{t}}$
has been measured in proton--proton collisions at $\sqrt{s}=13\,\TeV$, using
$\intlumi\,\ifb$ of
data collected  by the ATLAS experiment at the Large Hadron Collider.
Using events with an opposite-charge $e\mu$ pair and $b$-tagged
jets, the cross-section is measured to be:
\begin{equation}\nonumber
\sigma_{t\bar{t}} = \ttxval\pm\ttxstat\,\mathrm{(stat)}\ \pm\ttxsyst\,\mathrm{(syst)}\ \pm\ttxlumi\,\mathrm{(lumi)}\ \pm \ttxebeam\,\mathrm{(beam)}\,\mathrm{pb},
\end{equation}
where the uncertainties reflect the limited size of the data sample,
experimental and theoretical systematic
effects, the integrated luminosity, and the proton beam energy,
giving a total uncertainty of \ttxrel. The result is used to determine the
top quark pole mass via the dependence of the predicted cross-section on \mtpole,
giving $\mtpole=\mtpval^{+\mtpeup}_{-\mtpedn}$\,\GeV. The same event sample is used
to measure absolute and normalised differential cross-sections for the
$\ttbar\rightarrow e\mu\nu\bar{\nu}\bbbar$ process
as a function of single-lepton and dilepton kinematic variables.
Complementary measurements of $e\mu\bbbar$ production, treating both \ttbar\
and $Wt$ events as signal, are also provided. Both sets of differential
cross-sections are compared to the predictions of various Monte Carlo
event generators, demonstrating that the state-of-the-art generators
{\scr Powheg MiNNLO} and {\scr Powheg} $bb4l$ describe the data better than
{\scr Powheg} {\ttr hvq}.

The sensitivity of some of the measured differential distributions to
quasi-bound-state formation near the \ttbar\ threshold is investigated in an
addendum.
}

\AtlasRefCode{TOPQ-2024-12}

\PreprintIdNumber{CERN-EP-2025-211}

\AtlasJournal{EPJC}
\AtlasCoverEgroupAnalysisTeam{atlas-topq-2024-12--analysis-team@cern.ch}

\hypersetup{pdftitle={ATLAS document},pdfauthor={The ATLAS Collaboration}}

\begin{document}

\maketitle

\tableofcontents

\section{Introduction}\label{s:intro}

The study of top quark-antiquark (\ttbar) production at the CERN Large Hadron
Collider (LHC) allows quantum chromodynamics (QCD) to be probed at some
of the highest accessible energy scales, and forms a key part of the
experimental programme of the ATLAS experiment~\cite{TOPQ-2023-19}. The
large mass of the top quark, close to the scale of electroweak symmetry
breaking, gives it a unique role in the Standard Model of particle physics,
and \ttbar\ production is also a significant background in many searches
for physics beyond the Standard Model. Precise measurements of absolute
\ttbar\ production rates and differential distributions are therefore
vital to fully exploit the discovery potential of the LHC, and to refine
theoretical predictions and QCD calculational tools.

The inclusive \ttbar\ production cross-section \xtt\
in proton--proton ($pp$) collisions has
been calculated at next-to-next-to-leading-order (NNLO) accuracy in the
strong coupling constant \alphas, including the resummation of
next-to-next-to-leading-logarithmic (NNLL) soft gluon
terms~\cite{topxtheo1,topxtheo2,topxtheo3,topxtheo4,topxtheo5,topxtheo6}.
The predictions are in excellent agreement with measurements from
ATLAS and CMS at centre-of-mass energies $\sqrt{s}$ from 5.02\,\TeV\
to 13.6\,\TeV~\cite{TOPQ-2018-40,TOPQ-2013-04,TOPQ-2017-08,TOPQ-2016-08,TOPQ-2018-26,TOPQ-2020-02,TOPQ-2023-21,CMS-TOP-23-005,CMS-TOP-13-004,CMS-TOP-12-006,CMS-TOP-16-006,CMS-TOP-17-001,CMS-TOP-22-012}. At \sxyt, assuming a fixed top quark mass of
$\mtop=172.5$\,\GeV, the NNLO+NNLL prediction calculated by the {\ttr Top++ 2.0}
program~\cite{toppp} is $834\pm 21^{+21}_{-30}$\,pb, i.e.\ a total
uncertainty of $^{+3.5}_{-4.4}$\%. The first uncertainty component of the
prediction
corresponds to parton distribution function (PDF) uncertainties evaluated
using the PDF4LHC21~\cite{pdf4lhc3} combination of the
CT18~\cite{ct18}, MSHT20~\cite{msht20} and NNPDF3.1~\cite{nnpdf31} PDF sets.
The second represents QCD renormalisation (\mur) and factorisation (\muf) scale
uncertainties as a proxy for unknown higher-order corrections,
evaluated from the envelope of predictions with independent
variations of \mur\ and \muf\ up and down by factors of two from default values
of $\mur=\muf=\mtop$, whilst never letting them differ by more than a factor
of two~\cite{qcdscale1,qcdscale2}. The predicted cross-section also depends
strongly on \mtop, decreasing by 2.7\% for a 1\,\GeV\ increase in the top quark
mass. The mass parameter in the cross-section prediction is the top quark pole
mass \mtpole, corresponding to the definition of the mass of a free particle,
and allowing \xtt\ to be interpreted as a measurement of \mtpole, free of
the ambiguities linked to the direct reconstruction of the invariant mass of
its decay products~\cite{buckleypole,mochpole,hoangpole,nasonpole}.

In the Standard Model, 99.8\% of top quark decays are to a $W$ boson and
$b$-quark~\cite{pdg}, making the final-state topologies in \ttbar\ production
dependent on the $W$ boson decay modes.
The $e^+\mu^-$ dilepton channel\footnote{Charge-conjugate decay modes are implied
unless otherwise stated.}
($\ttbar\rightarrow W^{+}bW^{-}\bar{b}\rightarrow e^{+}\mu^{-}\nu\bar{\nu}\bbbar$)
with one or two jets tagged as likely to contain $b$-hadrons ($b$-tagged),
has been  exploited to make increasingly precise ATLAS measurements of \xtt\
at \sxyt\ using progressively larger
data samples~\cite{TOPQ-2015-09,TOPQ-2018-17}, culminating in a
measurement using the full \intlumi\,\ifb\ dataset that achieved
an uncertainty of 1.8\%~\cite{TOPQ-2018-26}. The same-flavour dilepton channels
($\ttbar\rightarrow e^+e^-\nu\bar{\nu}\bbbar$ and
$\ttbar\rightarrow\mu^+\mu^-\nu\bar{\nu}\bbbar$) in that dataset
were used to make a precise test of lepton flavour universality in
$W\rightarrow e\nu$ and $W\rightarrow\mu\nu$ decays~\cite{TOPQ-2023-28}.
This paper reports a further measurement of \xtt\ using the full \sxyt\ dataset,
profiting from the precise calibration of lepton efficiencies used in
Ref.~\cite{TOPQ-2023-28}, together with
recent improvements in the modelling of lepton
kinematics in \ttbar\ events using the \minnlops\ procedure~\cite{powhegminnlo}
for the computation of \ttbar\ production at NNLO matched to parton shower
simulation. The
resulting measurement is also used to extract an updated value for \mtpole.

The $e\mu$ + $b$-tagged jets sample also enables precise measurements of
the differential distributions of the leptons produced in \ttbar\ events
(\ttem). A set of eight one-dimensional and three two-dimensional distributions
was measured at \sxvt~\cite{TOPQ-2015-02} and with a partial \sxyt\
sample~\cite{TOPQ-2018-17}, and a slightly different set of distributions
was measured with the full \sxyt\ dataset~\cite{TOPQ-2018-26}. This paper
presents more precise measurements of the distributions studied in
Ref.~\cite{TOPQ-2018-17}, namely the absolute and normalised differential
cross-sections as functions
of the transverse momentum \ptl\ and absolute rapidity \etal\ of the
single leptons\footnote{ATLAS uses a right-handed coordinate system with its
origin at
the nominal interaction point in the centre of the detector, and the $z$ axis
along the beam line. Pseudorapidity is defined in terms of the polar angle
$\theta$ as $\eta=-\ln\tan{\theta/2}$, and transverse momentum and energy
are defined relative to the beam line as $\pt=p\sin\theta$ and
$\et=E\sin\theta$. The azimuthal angle around the beam line is denoted by
$\phi$, and distances in $(\eta,\phi)$ space by
$\Delta R=\sqrt{(\Delta\eta)^2+(\Delta\phi)^2}$.
The rapidity is defined as $y=\frac{1}{2}\ln\left(\frac{E+p_z}{E-p_z}\right)$,
where $p_z$ is the $z$-component of the momentum and  $E$ is the energy
of the relevant object or system.
} (combined for electrons and muons), the \pt, invariant mass and absolute
rapidity of the $e\mu$ system (\ptll, \mll\ and \rapll), the absolute
azimuthal angle \dphill\ between the two leptons in the transverse plane,
and the scalar sums of the transverse momenta (\ptsum) and energies (\esum)
of the two leptons. This set of distributions was shown to have sensitivity to
the gluon PDF and \mtpole~\cite{TOPQ-2015-02}.
Two-dimensional distributions of \etal, \rapll\ and \dphill\
as functions of \mll\ are also measured, and the single-lepton \ptl\ values
are used to form distributions \ptlmax\ and \ptlmin\ of the maximum and
minimum lepton \pt\ in each event. All these distributions are defined without
making any requirements on jets or $b$-jets. The distributions are compared with
the predictions from various next-to-leading-order (NLO) \ttbar\ matrix
element generators combined with parton showers,
as well as the \minnlops\ approach.

In kinematic regions with high lepton \pt, the differential cross-section
measurements are limited by uncertainties due to the interference that
occurs at NLO between \ttbar\ and $Wt$, the associated production of a
$W$ boson and a top quark, which can both lead to $WbW\bar{b}$ and hence
\embb\ final states~\cite{wtinter2,wtinter3,wtinter4,ATL-PHYS-PUB-2016-020}.
Such final states receive contributions from both \ttbar-like (doubly resonant)
and $Wt$ (singly resonant) processes. A second set of lepton differential
cross-sections are measured in this paper by considering the \embb\ final
state as the signal region, explicitly requiring the presence of
two $b$-tagged jets. The same set of leptonic differential
cross-sections as for \ttem\ production are measured using this signal region,
and compared
with the sum of predictions for \ttbar\ and $Wt$ production, and to predictions
from the $bb4l$ generator for the full NLO
$pp\rightarrow\ell^+\ell^-\nu\bar{\nu}\bbbar$ process matched to
parton showers~\cite{jezobb4l}. These measurements of leptonic distributions
are complementary to those also involving the jet kinematics described
in Ref.~\cite{TOPQ-2020-14}.

The event selection, analysis methodology and uncertainty evaluation
for the inclusive, \ttem\ differential and \embb\ differential cross-section
measurements are similar. The data and Monte Carlo simulation
samples are described in Section~\ref{s:datmc}, followed by the event
reconstruction and selection in Section~\ref{s:evtsel}. The measurement
techniques are discussed in Section~\ref{s:xsec} and the evaluation of
systematic uncertainties in Section~\ref{s:syst}. Inclusive cross-section results
are given in Section~\ref{s:ires}, together with the
determination of \mtpole. Section~\ref{s:dres} presents the \ttem\ and
\embb\ differential cross-section results and  comparisons
with the predictions of various event generators. Finally, conclusions are
drawn in Section~\ref{s:conc}.

\section{Data and simulated event samples}\label{s:datmc}

The ATLAS detector~\cite{PERF-2007-01,PIX-2018-001,ATLAS-TDR-19} at the LHC
covers nearly the entire solid angle around the collision point.
It consists of an inner tracking
detector surrounded by a thin superconducting solenoid producing
a 2T axial magnetic field, electromagnetic and hadronic calorimeters,
and an external muon spectrometer incorporating three large toroidal
magnet assemblies.
The analysis was performed on samples of $pp$ collision data
collected at \sxyt\ in 2015--18, corresponding to an integrated
luminosity of $\intlumia\pm\intlumie$\,\ifb\
after data quality requirements~\cite{DAPR-2021-01,DAPR-2018-01}.
Events were required to satisfy a single-electron or single-muon
trigger~\cite{TRIG-2018-05,TRIG-2018-01}, with transverse momentum
thresholds that were progressively raised during the data-taking period.
The electron trigger reached the efficiency plateau region for electrons with
reconstructed $\pt>25$\,\GeV\ in 2015 and for $\pt>27$\,\GeV in 2016--18.
The corresponding thresholds for the muon trigger are 21\,\GeV\ for
2015 and 27.3\,\GeV thereafter. Each recorded event also includes the
signals from on average~33 superimposed inelastic $pp$ collisions,
referred to as pileup.

Monte Carlo simulated event samples were used to develop the analysis
procedures, to evaluate signal and background contributions, and to compare
with data. Samples were processed using either the full ATLAS detector
simulation~\cite{SOFT-2010-01} based on {\scr Geant4}~\cite{geant4},
or with a faster simulation making use of parameterised showers in the
calorimeters~\cite{ATL-PHYS-PUB-2010-013}. The effects of pileup were
simulated by generating additional inelastic $pp$ collisions with
{\scr Pythia8} (v8.186)~\cite{pythia8} using the A3 set of parameter values
(tune)~\cite{ATL-PHYS-PUB-2016-017} and overlaying them on the primary
simulated events, so as to match the distribution of the number of inelastic
events per bunch crossing observed in the data. These combined
events were then processed using the same reconstruction and analysis
chain as the data~\cite{ATL-SOFT-PUB-2021-001}.
Small corrections were applied to lepton and jet energy
scales~\cite{EGAM-2018-01,MUON-2022-01,JETM-2018-05}, and to lepton and
$b$-tagging efficiencies~\cite{EGAM-2021-01,MUON-2018-03,FTAG-2018-01},
to improve agreement with the response
observed in data. Further topology-specific lepton isolation corrections
were applied as discussed in Section~\ref{ss:sylept}.

The baseline simulated \ttbar\ sample was generated using the \minnlops\
matched NNLO QCD plus parton shower procedure~\cite{powhegminnlo}, implemented
in {\scr Powheg}~\cite{powheg,powheg2,powheg3} with the settings $\pthard=0$
and $\ptdef=2$~\cite{pypthard,pyguide},
using the NNPDF3.0 PDF set~\cite{nnpdf3} in the matrix element
and interfaced to {\scr Pythia8.312} with the
A14 tune~\cite{ATL-PHYS-PUB-2014-021} and NNPDF2.3 PDF set~\cite{nnpdfffn}
for the parton shower, hadronisation and underlying event modelling.
The QCD renormalisation and factorisation scales were set to
$\mur=\muf=\hTtt/4$, where \hTtt\ is the sum of the transverse masses
($\sqrt{}(m^2_t+p^2_{\mathrm{T},t})$) of the top quark and antiquark.
This sample, referred to hereafter
as \powhegminnlo, predicts a softer top quark \pt\ spectrum in \ttbar\
events than the standard samples based on the NLO {\ttr hvq} process in
{\scr Powheg Box} v2~\cite{powheghvq}
interfaced to {\scr Pythia8} used for the previous \sxyt\
analyses~\cite{TOPQ-2018-17,TOPQ-2018-26}, and predicts a lepton \pt\ spectrum
closer to that seen in data. For comparisons with data and to study systematic
uncertainties, a {\scr Powheg\,+\,Pythia8} \ttbar\
simulation sample was generated using the {\ttr hvq} process,
with nominal settings of $\pthard=0$ and the \hdamp\ parameter,
which gives a cut-off scale for the first gluon emission, set to
$\hdamp=\frac{3}{2}\mtop$~\cite{ATL-PHYS-PUB-2016-020}. The renormalisation
and factorisation scales were set to the top quark transverse mass.
Variations of this
sample were generated with $\pthard=1$, with $\hdamp=3\mtop$,
and using {\scr Herwig} 7.1.3~\cite{herwigpp,herwig71} instead
of {\scr Pythia8} for the parton shower, hadronisation and underlying event.
Further variations were obtained from the \powhegminnlo\ and nominal
{\scr Powheg\,+\,Pythia8} samples by using event weights to change the
QCD renormalisation and factorisation scales, the amounts of initial and
final state radiation, and the PDF. In all samples, the top quark mass was set to
$\mtop=172.5$\,\GeV, the $W\rightarrow\ell\nu$ branching ratio to the
Standard Model prediction of 0.1082 for each lepton flavour
($e$, $\mu$ and $\tau$)~\cite{wpdg}, and {\scr EvtGen}~\cite{evtgen} was used
to handle the decays of $b$- and $c$-flavoured hadrons.

The measured \embb\ cross-sections were also compared with a sample generated
with the {\scr Powheg} $bb4l$ generator~\cite{jezobb4l}, which
includes the complete set of $\ell\ell'\nu\bar{\nu}\bbbar$
final states, taking into account interference between the \ttbar\
and $Wt$ final states as well as off-shell and non-resonant effects. This
sample uses the {\scr Powheg Box Res} framework~\cite{powhegboxres},
with matrix elements at NLO in QCD, and resonance-aware matching to the parton
shower~\cite{resaware1,resaware2}, with QCD scales and other configuration
parameters set as described in Ref.~\cite{TOPQ-2020-14}.

Backgrounds in the \xtt\ inclusive and \ttem\ differential cross-section
measurements are classified into two types: those with two real prompt leptons
(electrons or muons, including those produced via leptonic decays of
$\tau$-leptons), and those where at least one of the reconstructed leptons
is misidentified, i.e.\ a non-prompt lepton from the decay of a bottom or
charm hadron, an electron from a photon conversion, a hadronic jet misidentified
as an electron, or a muon produced from the decay in flight of a pion or kaon.
The background with two real prompt leptons is dominated by $Wt$ production,
modelled using {\scr Powheg\,+\,Pythia8} with the same settings
as the corresponding {\scr Powheg}\ {\ttr hvq} \ttbar\ sample.
The renormalisation and factorisation
scales were set to the dynamic scales $\mur=\muf=H_\mathrm{T}/2$ where
$H_\mathrm{T}$ is the scalar sum of the \pt\ of all outgoing partons, giving
a better description of data \cite{ATL-PHYS-PUB-2021-042}
than that obtained from the fixed scales $\mur=\muf=\mtop$ used in previous
analyses~\cite{TOPQ-2018-17,TOPQ-2018-26}.
The interference between \ttbar\ and $Wt$ was modelled using
the diagram removal scheme~\cite{wtinter1,wtinter2}. The $Wt$ cross-section
was taken to be $79.3\pm 2.2$ (PDF) $^{+1.9}_{-1.8}$ (QCD scale)\,pb,
based on an NLO calculation with the addition of third-order
corrections resumming NNLL soft gluon contributions~\cite{kidonakis21}.
The $Wt$ contribution is part of the signal for the \embb\ cross-section
measurement.

In all three measurements, smaller backgrounds with two
prompt leptons arise from $Z\rightarrow\tau\tau (\rightarrow e\mu)$+jets,
and from diboson production ($WW$, $WZ$ and $ZZ$) in association with jets.
The $Z$+jets background was modelled with
{\scr Sherpa 2.2.11}~\cite{sherpa22} for $Z\rightarrow ee/\mu\mu$ and
{\scr Sherpa 2.2.14} for \ztt, with NLO matrix elements for up to
two partons, and leading-order matrix elements for up to five partons,
calculated with the {\scr Comix}~\cite{comix} and
{\scr OpenLoops}~\cite{openloops} libraries and matched with the
{\scr Sherpa} parton shower~\cite{sherpaps} using the {\scr MEPS@NLO}
prescription~\cite{mepsnlo1,mepsnlo2,mepsnlo3,mepsnlo4}.
They were generated using the NNPDF3.0 PDF set and normalised to an
NNLO cross-section prediction~\cite{zjnnlo}. Diboson production was simulated
using {\scr Sherpa 2.2.2} using a similar approach.
Production of \ttbar\ in association
with a leptonically decaying $W$, $Z$ or Higgs boson, or
an additional \ttbar\ pair,  gives a negligible
contribution to the opposite-charge dilepton samples, but is significant
in the same-charge control samples used to assess the background from
misidentified leptons. These processes were
modelled at NLO using {\scr Powheg\,+\,Pythia8}, or {\scr MadGraph5\_aMC@NLO}
(referred to hereafter as {\scr aMC@NLO})~\cite{amcnlo}
interfaced to {\scr Pythia8}.

Backgrounds with one real and one misidentified lepton arise from \ttbar\
events with one leptonically decaying and one hadronically decaying $W$,
simulated with \powhegminnlo. Similar backgrounds arise from $W$+jets
production, modelled with {\scr Sherpa 2.2.11} as for $Z$+jets, and
$t$-channel single top production, modelled with {\scr Powheg\,+\,Pythia8}.
Other backgrounds, including processes with two misidentified leptons,
are negligible after the event selections used in the analysis.

\section{Event reconstruction and selection}\label{s:evtsel}

This analysis makes use of reconstructed electrons, muons and $b$-tagged jets.
Electron candidates were reconstructed from a localised cluster of energy
deposits in the
electromagnetic calorimeter matched to a track in the inner detector, passing
the `Medium' likelihood-based requirement of Ref.~\cite{EGAM-2018-01}.
They were required to have transverse momentum $\pt>20$\,\GeV\
and pseudorapidity
$|\eta|<2.47$, excluding the transition region between the barrel and endcap
electromagnetic calorimeters, $1.37<|\eta|<1.52$, and to be consistent with
originating from the signal primary vertex. The latter was defined as the
reconstructed vertex with the highest sum of $\pt^2$ of associated tracks.
To reduce background from
non-prompt electrons, electron candidates were further required
to pass the `Tight' isolation requirements of Ref.~\cite{EGAM-2018-01},
based on the amount of summed calorimeter energy and track transverse momentum
close to the electron. Muon candidates were reconstructed by combining
tracks from the inner detector with matching tracks reconstructed in the
muon spectrometer, and were required to
have $\pt>20$\,\GeV, $|\eta|<2.5$ and to satisfy the `Medium' requirements
of Ref.~\cite{MUON-2018-03}. Muons were also required to be consistent
with the signal primary vertex and to satisfy the `Tight' isolation requirements
of Ref.~\cite{MUON-2018-03}.

Jets were reconstructed using the anti-$k_t$ algorithm~\cite{antikt,fastjet}
with radius parameter $R=0.4$, starting from particle-flow objects that combine
information from topological clusters of calorimeter energy deposits and
inner-detector tracks~\cite{PERF-2015-09}.
After calibration using information from
both simulation and data~\cite{JETM-2018-05}, jets were required to have
$\pt>25$\,\GeV\ and $|\eta|<2.5$, and jets with $\pt<60$\,\GeV\
and $|\eta|<2.4$ were subject to additional pileup rejection criteria
using the multivariate jet-vertex tagger (JVT)~\cite{PERF-2014-03}.
To prevent double counting of electron energy deposits as jets, the closest
jet to an electron candidate was removed if it was within $\Delta R=0.2$
of the electron. Furthermore, to reduce the contribution of leptons
from heavy-flavour hadron decays inside jets, leptons within $\Delta R=0.4$
of selected jets were discarded, unless the lepton was a muon and the jet
had fewer than three associated tracks, in which case the jet was discarded.
Jets likely to contain $b$-hadrons were tagged using the DL1r
algorithm~\cite{FTAG-2018-01,FTAG-2019-07}, a multivariate discriminant based
on deep-learning techniques making use of track impact parameters and
reconstructed secondary vertices.
For the inclusive \xtt\ and \ttem\ differential cross-section measurements,
a tagger working point with an efficiency of 70\% for tagging $b$-quark jets
from top-quark decays in simulated
\ttbar\ events was used, corresponding to rejection factors of about~380
against light quark and gluon jets, and~10 against jets originating
from charm quarks. A looser working point with 77\% $b$-tagging efficiency
and a factor of two less background rejection was used for the
\embb\ measurement.

Selected events were required to have exactly one electron and exactly one
muon passing the requirements given above, with at least one of the leptons
matched to a corresponding electron or muon trigger. Although the lepton
\pt\ threshold is 20\,\GeV, the trigger matching requirements imply that at least
one lepton must have offline reconstructed $\pt>27$\,\GeV (if an electron) or
$\pt>27.3$\,\GeV (if a muon) for most of the dataset. However, the
lower thresholds in 2015 data give access to the phase space where the
two leptons satisfy $(p_\mathrm{T,1}, p_\mathrm{T,2})>(21,20)$\,\GeV.
Events where the electron and muon were separated in angle by
$|\Delta\theta|<0.15$ and $|\Delta\phi|<0.15$
were rejected. Events with an opposite-charge $e\mu$ pair comprise the main
analysis sample, whilst events with a same-charge $e\mu$ pair were used to
estimate the background from misidentified leptons.

\section{Cross-section measurements}\label{s:xsec}

The inclusive \xtt\ and \ttem\ differential cross-sections were measured
using the double-tagging technique described in Ref.~\cite{TOPQ-2018-17},
employing subsets of the opposite-charge $e\mu$ sample with exactly one
and exactly two $b$-tagged jets, as described in Sections~\ref{ss:incmeas}
and~\ref{ss:ttdiff}. The \embb\ differential cross-sections were derived
mainly from the two $b$-tagged jet sample using the 77\% efficiency
$b$-tagging working point, and are described in Section~\ref{ss:embbdiff}.
Background estimates are discussed in Section~\ref{ss:bkgd}, and
the validation of the differential measurements using simulation in
Section~\ref{ss:diffval}.

\subsection{Inclusive cross-sections}\label{ss:incmeas}

The inclusive
\ttbar\ cross-section \xtt\ was determined by counting the numbers of
opposite-charge $e\mu$ events with exactly one ($N_1$) and exactly two ($N_2$)
$b$-tagged jets, using the 70\% $b$-tagging efficiency working point.
The two event counts satisfy the tagging equations:
\begin{equation}
\begin{array}{lll}
N_1 & = & L \xtt\ \epsem 2\epsb (1-\cb\epsb) + \nib , \\*[2mm]
N_2 & = & L \xtt\ \epsem \cb\epsb^2 + \niib,
\end{array}\label{e:tags}
\end{equation}
where $L$ is the integrated luminosity of the sample, \epsem\ the efficiency
for a \ttbar\ event to pass the opposite-charge $e\mu$
selection, and \cb\ is a tagging correlation coefficient close to unity.
The combined probability for a jet from the quark $q$ in the $t\rightarrow Wq$
decay to fall within the acceptance of the detector, be reconstructed as a jet
with transverse momentum above the selection threshold, and be tagged as a
$b$-jet, is denoted by \epsb. The parameter $\cb=\epsbb/\epsb^2$,
where \epsbb\ is the probability to reconstruct and $b$-tag both $b$-quark jets
from the two top quark decays, accounts for the fact that the tagging
probabilities of the two $b$-quark jets are not completely independent.
It was evaluated
from simulation as $\cb=4\nttem N^{\ttbar}_2/(N^{\ttbar}_1+2N^{\ttbar}_2)^2$,
where \nttem\ is the number of selected $e\mu$ \ttbar\ events and $N^{\ttbar}_1$
and $N^{\ttbar}_2$ are the numbers of such events with one and two $b$-tagged
jets.\footnote{
The formula for \cb\ follows from the definitions of \epsb\ as the number
of $b$-tagged jets divided by the total number of $b$-jets from \ttbar\ events,
i.e.\ $\epsb=(N^{\ttbar}_1+2N^{\ttbar}_2)/(2\nttem)$ and \epsbb\ as the number
of \ttbar\ events with two $b$-tagged jets divided by the total number of \ttbar\
events, i.e. $\epsbb=N^{\ttbar}_2/\nttem$.}
In the baseline \powhegminnlo\ simulation, $\epsem\approx 0.89\%$,
$\epsb\approx 54\%$ and $\cb\approx 1.009$, indicating a small positive
correlation.
The backgrounds from sources other than \ttbar\ production are given
by the terms \nib\ and \niib, evaluated using a combination of simulation
and data control samples as discussed in Section~\ref{ss:bkgd}.
The tagging equations Eqs.~(\ref{e:tags}) were then solved
to determine \xtt\ and \epsb.

The selection efficiency \epsem\ can be written as the product of two terms:
$\epsem=\aem\gem$. The acceptance $\aem\approx 1.7$\% represents the fraction
of simulated \ttbar\ events that have a true opposite-charge $e\mu$ pair
from $W\rightarrow e/\mu$ decays, with each lepton satisfying
$\pt>20$\,\GeV\ and $|\eta|<2.5$. These requirements define the fiducial region.
The contributions via leptonic $\tau$ decays
($t\rightarrow W\rightarrow\tau\rightarrow e/\mu$) are included.
The lepton four-momenta were evaluated at particle level
after final-state radiation, and `dressed' by including the four-momenta of
any photons within a cone of size
$\Delta R=0.1$ around the lepton direction, excluding photons produced
from hadron decays or interactions with the detector material.
The reconstruction efficiency \gem\ represents the probability that the
two leptons are reconstructed and pass all the trigger, identification and
isolation requirements. It also corrects for the 0.5\% of selected
\ttbar\ events that are outside the fiducial region and only selected due to
the limited lepton resolution. The fiducial
cross-section \xfid\ corresponding to the production of \ttbar\ events with a
particle-level electron and muon satisfying the requirements on \pt\ and
$\eta$ is $\xfid=\aem\xtt$, and can be measured by replacing
$\xtt\epsem$ with $\xfid\gem$ in Eqs.~(\ref{e:tags}).
As in Refs.~\cite{TOPQ-2015-02,TOPQ-2018-17}, this fiducial cross-section
definition makes no requirements on the presence of jets, as the
tagging formalism allows the number of \ttbar\ events with no reconstructed
and $b$-tagged jets to be inferred from the event counts $N_1$ and $N_2$.
Measurement of the fiducial cross-section avoids the systematic uncertainties
associated with the evaluation of the acceptance.

A total of 1\,483\,775 events passed the opposite-charge $e\mu$ selection in
data. The numbers of observed events with one and two $b$-tagged jets are shown
in Table~\ref{t:evtcount}, together with the expected non-\ttbar\
contributions from $Wt$ and dibosons evaluated from simulation,
and $Z$+jets and misidentified leptons evaluated using both data and simulation.
The one $b$-tag sample is expected to be about 88\% pure and the two $b$-tag
sample 96\% pure in \ttbar\ events, with the largest backgrounds coming
from $Wt$ production in both cases. The distribution of the $b$-tagged jet
multiplicity \nbtag\ is shown in Figure~\ref{f:dmcjlept}(a) and compared with the
expectation from simulation, broken down into contributions from
\ttbar\ events (modelled using the \powhegminnlo\ sample) and various
background processes. The predictions using
{\scr Powheg\,+\,Pythia8} and {\scr Powheg\,+\,Herwig7} \ttbar\
events are also shown.\footnote{The names {\scr Pythia} and {\scr Herwig}
are abbreviated as PY and HW in figure legends and some tables for compactness.}
All predictions are normalised to the same integrated
luminosity as the data, using the cross-sections discussed in
Sections~\ref{s:intro} and~\ref{s:datmc}. The three \ttbar\ samples have been
reweighted to increase
the number of events with at least three particle-level $b$-jets by 50\%.
As discussed in Section~\ref{ss:sytt}, this reweighting brings the number of
events with $\nbtag\geq 3$  into good agreement with data.
The purple dot-dashed line in Figure~\ref{f:dmcjlept}(a) shows the prediction
from \powhegminnlo\ without this reweighting, which gives a 25\% deficit in the
number of events with $\nbtag\geq 3$. Since the
predicted value of \cb\ is sensitive to the production of \ttbar\ with extra
heavy quarks, this reweighting was applied to the \powhegminnlo\ sample
throughout the analysis.

\begin{table}
\caption{\label{t:evtcount}Observed numbers of opposite-charge $e\mu$ events
with one ($N_1$) and two ($N_2$) $b$-tagged jets, together with the estimates
of backgrounds and associated uncertainties described in
Sections~\ref{ss:bkgd} and~\ref{s:syst}.}
\centering
\begin{tabular}{l
r@{$\,\pm\,$}l
r@{$\,\pm\,$}l
}
\toprule
Event counts & \multicolumn{2}{c}{$N_1$} & \multicolumn{2}{c}{$N_2$} \\
\midrule
Data & \multicolumn{2}{l}{576\,725} & \multicolumn{2}{l}{314\,961} \\
\midrule
$Wt$ single top & \phantom{0}59\,100 & 3\,200 & \phantom{0}10\,200 & 1\,800 \\
$Z(\rightarrow\tau\tau\rightarrow e\mu)$+jets & \phantom{00}2\,650 & 110 & \phantom{000}204 & 8 \\
Diboson & \phantom{00}1\,320 & 270 & \phantom{000\,0}54 & 13 \\
Misidentified leptons & \phantom{00}3\,410 & 910 & \phantom{00}1\,250 & 730 \\
\midrule
Total background & \phantom{0}66\,500 & 3\,300 & \phantom{0}11\,700 & 2\,000 \\
\bottomrule
\end{tabular}
\end{table}

\begin{figure}[tp]
\vspace{-5mm}
\parbox{83mm}{\includegraphics[width=76mm]{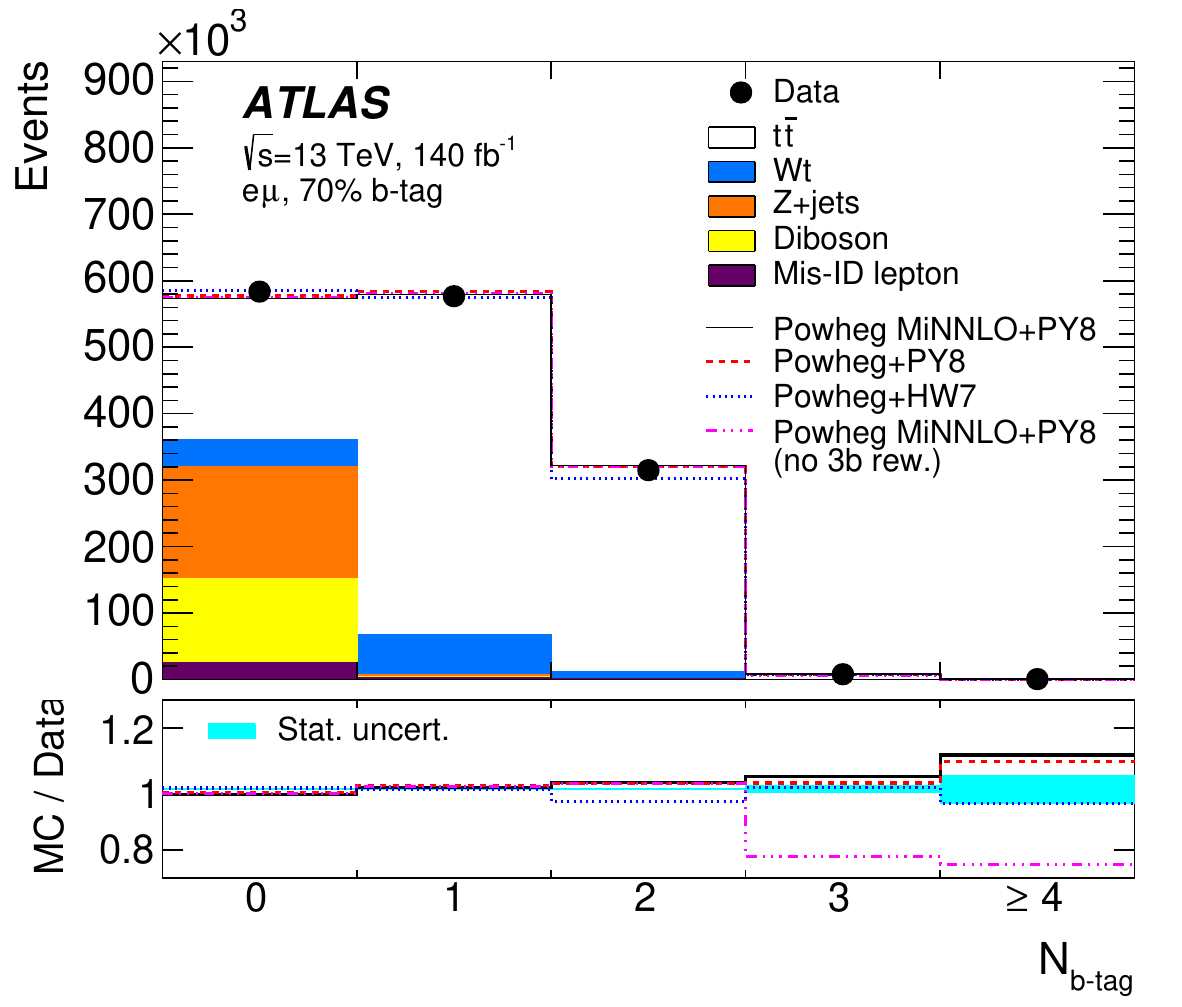}\vspace{-7mm}\center{(a)}}
\parbox{83mm}{\includegraphics[width=76mm]{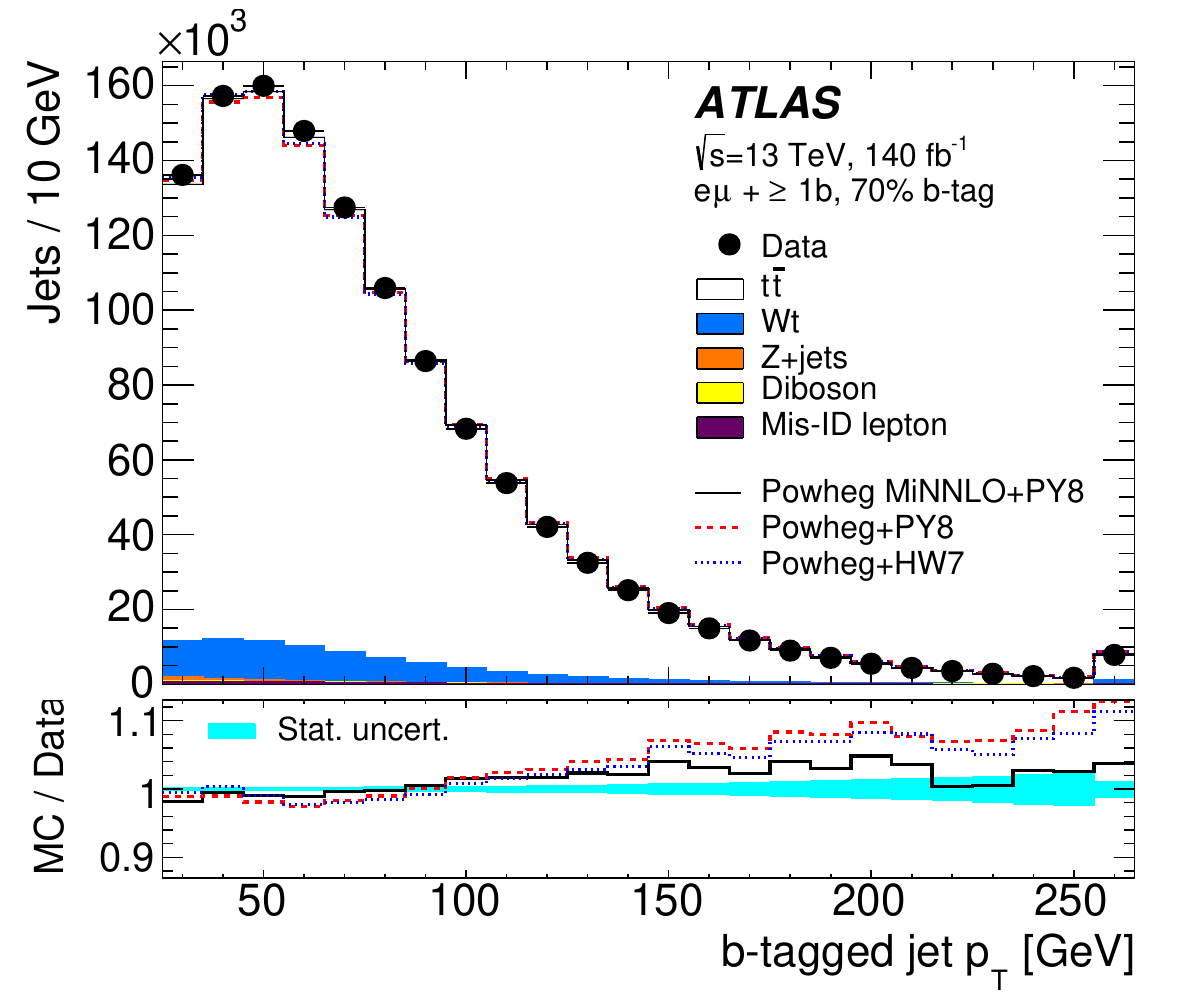}\vspace{-7mm}\center{(b)}}
\parbox{83mm}{\includegraphics[width=76mm]{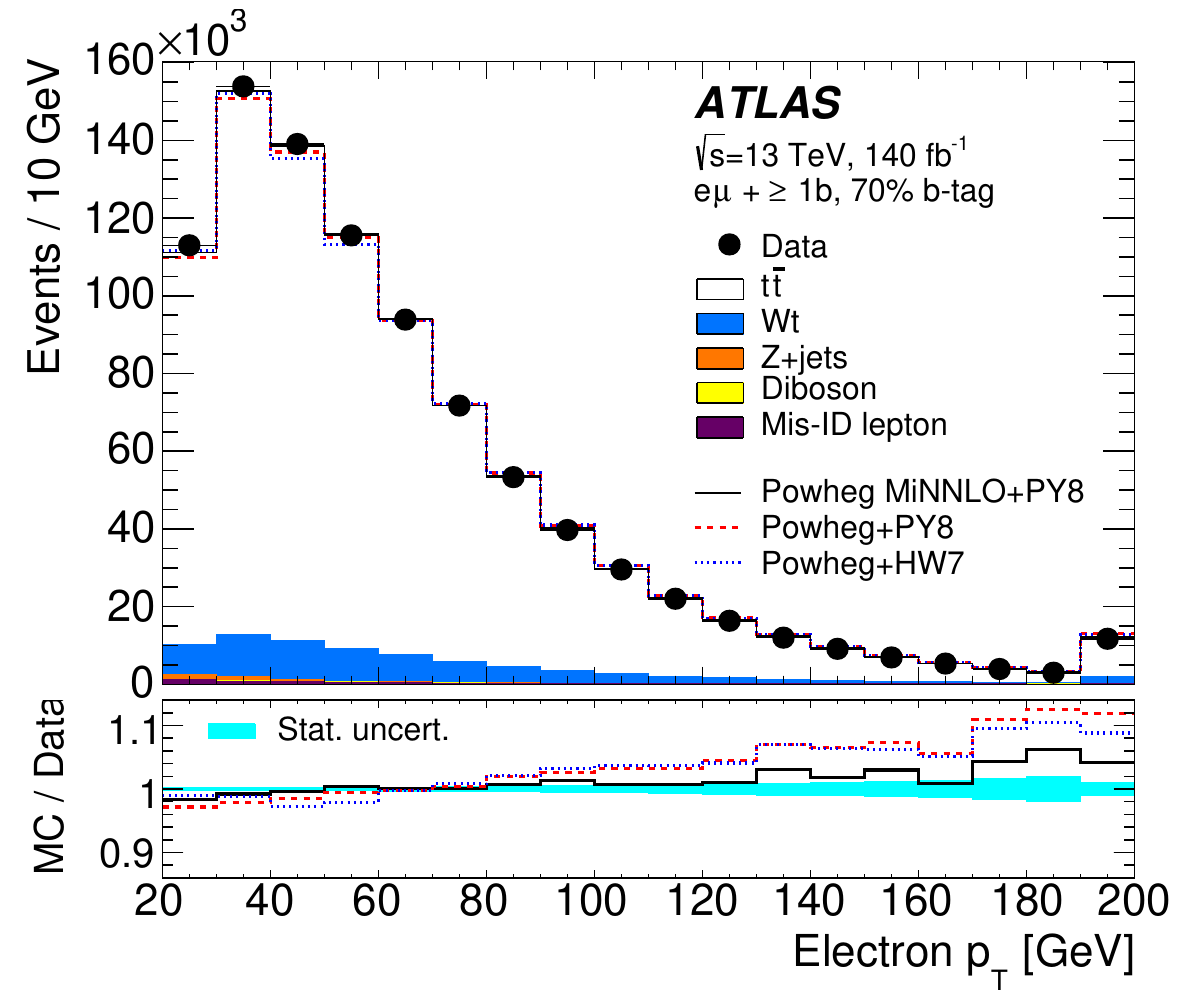}\vspace{-7mm}\center{(c)}}
\parbox{83mm}{\includegraphics[width=76mm]{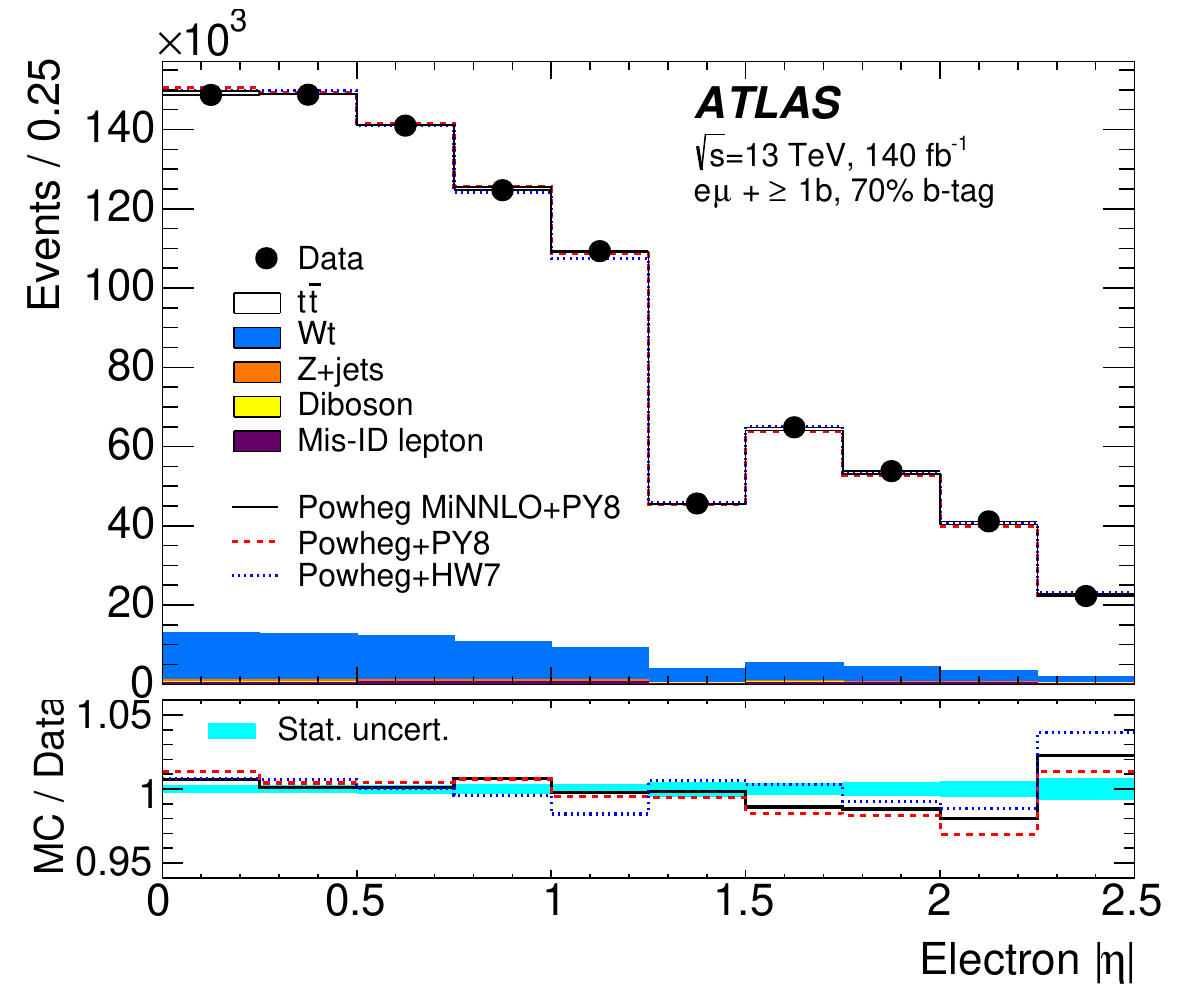}\vspace{-7mm}\center{(d)}}
\parbox{83mm}{\includegraphics[width=76mm]{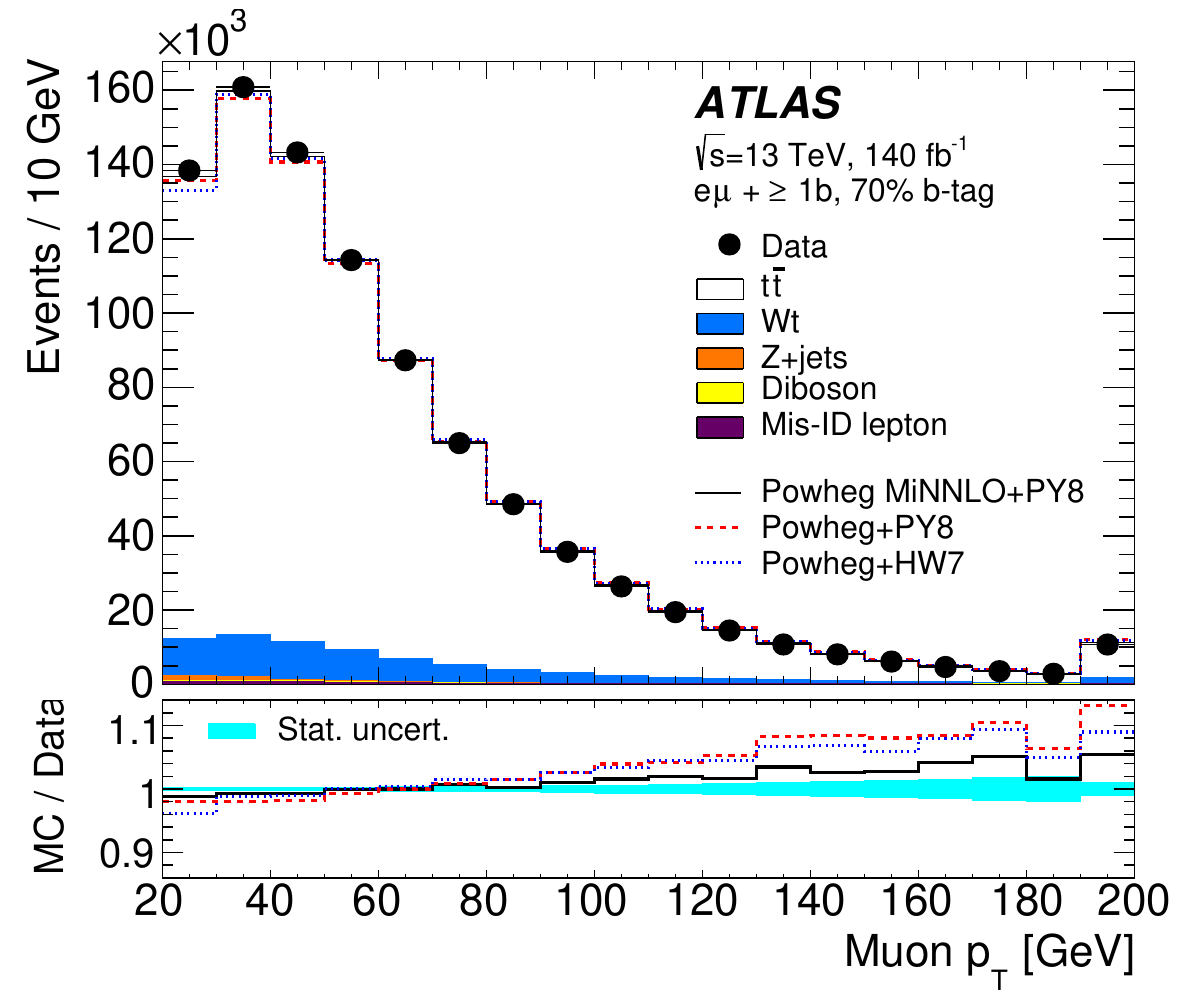}\vspace{-7mm}\center{(e)}}
\parbox{83mm}{\includegraphics[width=76mm]{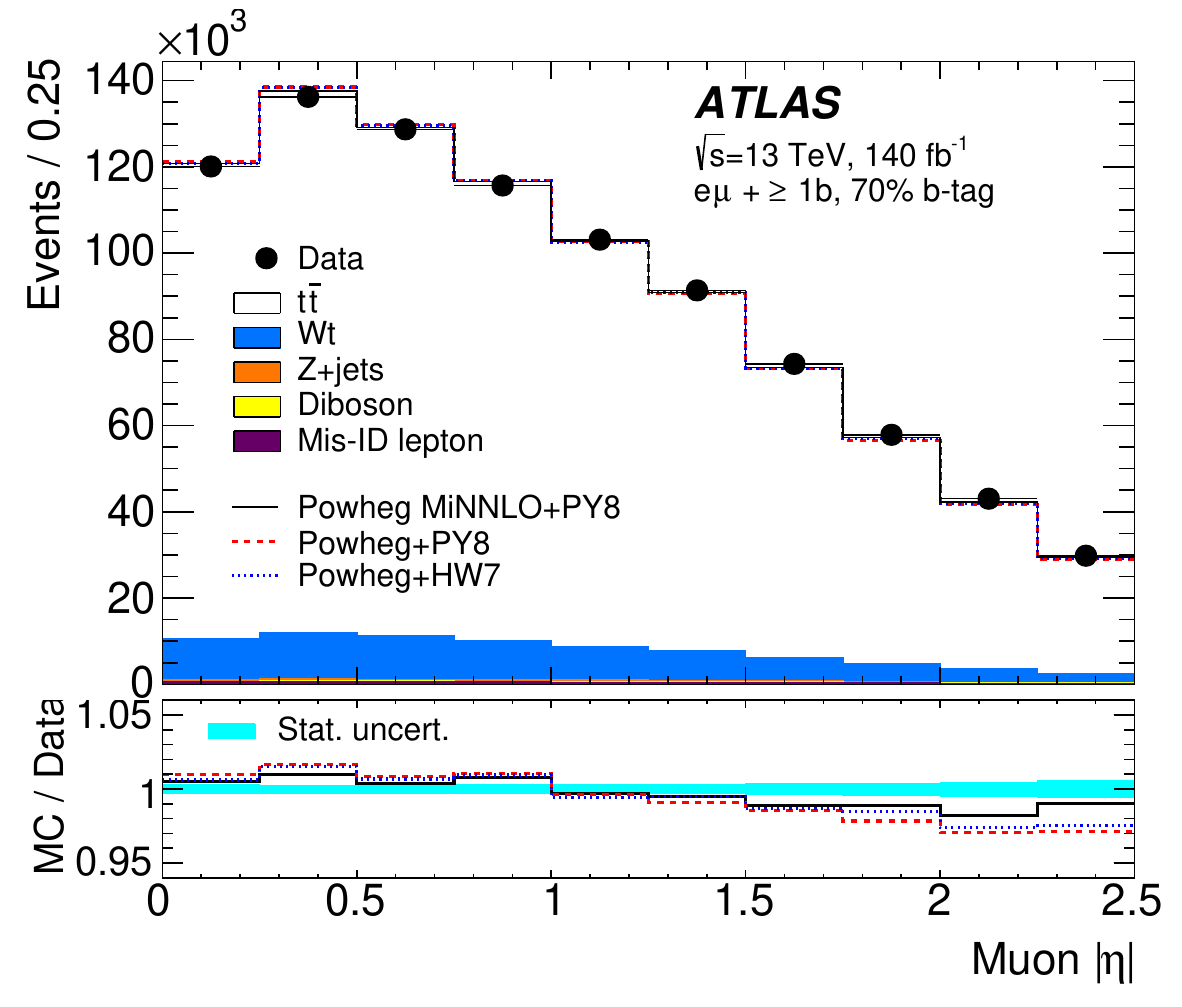}\vspace{-7mm}\center{(f)}}
\caption{\label{f:dmcjlept}Distributions of (a) the number of $b$-tagged jets
in selected opposite-charge $e\mu$ events; and (b) the \pt\ of $b$-tagged jets,
(c) the \pt\ of the electron, (d) the $|\eta|$ of the electron,
(e) the \pt\ of the muon and (f) the $|\eta|$ of the muon, in events with
an opposite-charge $e\mu$ pair and at least one $b$-tagged jet.
The reconstruction-level data are compared with the expectation from simulation,
broken down into contributions from \ttbar\ (modelled with \powhegminnlo),
$Wt$, $Z$+jets, dibosons, and events with misidentified electrons or muons.
The simulation prediction is normalised to the same integrated luminosity as the
data in (a) and to the same number of entries as the data in (b--f).
The lower panels show the ratios of simulation to data, using various \ttbar\
simulation samples and with the cyan shaded band indicating the statistical
uncertainty. The last bin includes the overflows in panels (b), (c) and (e).
}
\end{figure}

Figures~\ref{f:dmcjlept}(b) to~\ref{f:dmcjlept}(f) show distributions of the
\pt\ of the $b$-tagged jets (one entry per jet),
and the \pt\ and $|\eta|$ of the electron and muon,
in opposite-charge $e\mu$ events with at least one $b$-tagged jet. The total
simulation prediction is normalised to the same number of selected events
as the data, to facilitate shape comparisons. In general, the simulation
predictions describe the data well, though the data show softer lepton and
$b$-tagged jet
\pt\ distributions than all the simulation predictions, with \powhegminnlo\
being significantly closer to the data than {\scr Powheg\,+\,Pythia8} and
{\scr Powheg\,+\,Herwig7}. The differences in $|\eta|$ distributions
between electrons and muons reflect their differing reconstruction efficiencies
as a function of lepton $\eta$.

\subsection{$\boldmath\ttbar\rightarrow e\mu$ differential cross-sections}\label{ss:ttdiff}

The differential cross-sections as functions of the lepton and dilepton
variables defined in Section~\ref{s:intro} were measured
using an extension of Eqs.~(\ref{e:tags}), by counting the number of leptons or
events with one (\nxi) or two (\nyi) $b$-tagged jets where the lepton(s)
fall in bin $i$ of a differential distribution at reconstruction level.
There are two counts per event in the single-lepton distributions \ptl\ and
\etal, in the two bins corresponding to the electron and muon. In the
dilepton distributions, each event contributes a single count corresponding
to the bin in which the appropriate dilepton variable falls. For each bin
of each differential distribution, these counts satisfy the tagging equations:
\begin{equation}
\begin{array}{lll}
\nxi & = &  L \xtti\ \gemi 2\epsbi (1-\cbi\epsbi) + \nibi , \\*[2mm]
\nyi & = &  L \xtti\ \gemi \cbi(\epsbi)^2 + \niibi  ,
\end{array}\label{e:fidtags}
\end{equation}
where \xtti\ is the absolute fiducial differential cross-section in bin $i$.
The reconstruction efficiency \gemi\ represents the ratio of the number
of reconstructed $e\mu$ events (or leptons for \ptl\ and \etal) in bin $i$
defined using the reconstructed lepton(s), to the number of true $e\mu$ events
(or leptons) in the same bin $i$ at particle level,
evaluated using \ttbar\ simulation. In the definition of \gemi,
the particle-level electron and muon were required to have $\pt>20$\,\GeV\
and $|\eta|<2.5$, but no requirements were made on reconstructed or
particle-level jets, nor on the removal of overlaps between leptons and jets
in contrast to Ref.~\cite{TOPQ-2018-26} where such overlap requirements were
applied. The efficiency \gemi\ corrects for both the lepton
reconstruction efficiency and the effects of event migration, where events
in bin $j$ at particle level (or outside the fiducial region) appear in a
bin $i\neq j$ at reconstruction
level. The integral of any dilepton differential cross-section is equal to the
fiducial cross-section \xfid\ defined in Section~\ref{ss:incmeas}, and the
integrals of the single-lepton \ptl\ and \etal\ distributions are equal to
$2\xfid$. The  values of \gemi\ were taken from the \powhegminnlo\ \ttbar\
simulation, and are generally
around 0.4--0.6. The corresponding values of \cbi\ are always within 3\%
of unity, even at the edges of the differential distributions. The background
terms \nibi\ and \niibi\ were determined from simulation and data control
samples, allowing the tagging equations Eqs.~(\ref{e:fidtags}) to be solved to
give the absolute fiducial differential cross-sections \xtti\ and
associated \epsbi\ values for each bin $i$ of each differential distribution.

The bin ranges for each differential distribution were based on those used
in Ref.~\cite{TOPQ-2018-17}, splitting or redefining a few bins to profit from
the larger data sample. The \dphill\ distribution was measured in 20~rather
than 10~bins. For other distributions, the bin widths were chosen according
to the experimental lepton
resolution, in order to keep the fractions of events reconstructed in bin $i$
that do not originate from bin $i$ at particle level at the level of 10\% or
lower. The chosen bin ranges can be seen in
Figures~\ref{f:distresa}--\ref{f:distrese} in Section~\ref{ss:diffres} below,
and in Ref.~\cite{ttx8aux}. The last bins
of the \ptl, \ptll, \mll, \ptsum, \esum, \ptlmax\ and \ptlmin\ distributions
include overflow events falling above the last bin boundary.
The normalised fiducial differential cross-sections were calculated
from the absolute cross-sections \xtti\ as discussed in Ref.~\cite{TOPQ-2018-17},
and are statistically correlated between bins because of the normalisation.
The absolute dilepton differential cross-sections are not statistically
correlated between bins, but kinematic correlations between the electron
and muon within an event introduce small correlations within the absolute
single-lepton \ptl\ and \etal\ distributions. Figure~\ref{f:dmcdilept}
shows the reconstructed \ptll, \mll, \rapll, \dphill, \ptsum\ and \esum\
distributions for events with at least
one $b$-tagged jet, comparing data with predictions using the same set of
\ttbar\ simulation samples as Figure~\ref{f:dmcjlept}. The predictions
are in reasonable qualitative agreement with the data. However there are
some significant differences, in particular for \mll, \dphill\ and \ptsum, which
are discussed quantitatively in terms of the particle-level differential
cross-sections in Section~\ref{ss:diffcomp} below.

\begin{figure}[tp]
\vspace{-5mm}
\parbox{83mm}{\includegraphics[width=76mm]{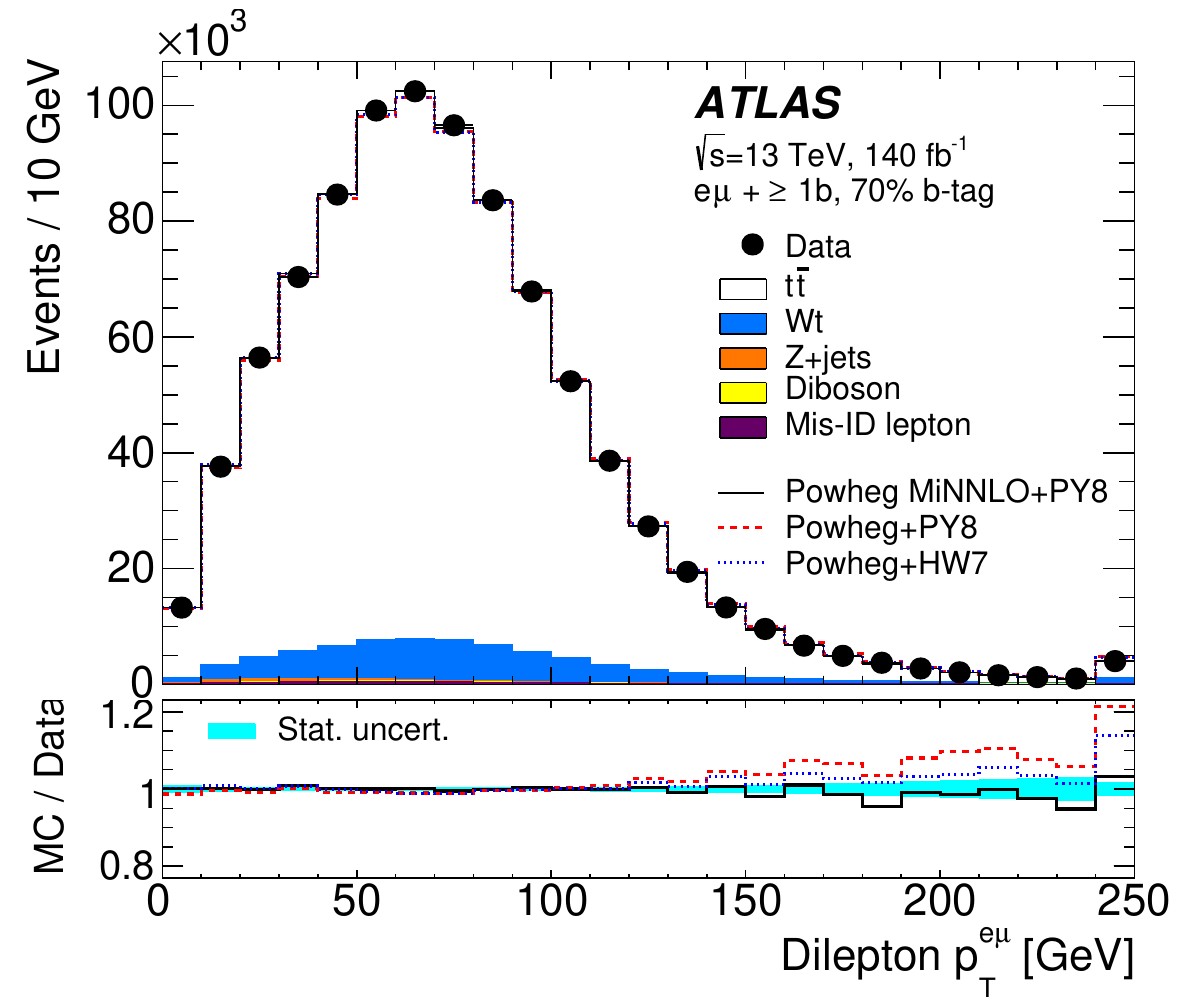}\vspace{-7mm}\center{(a)}}
\parbox{83mm}{\includegraphics[width=76mm]{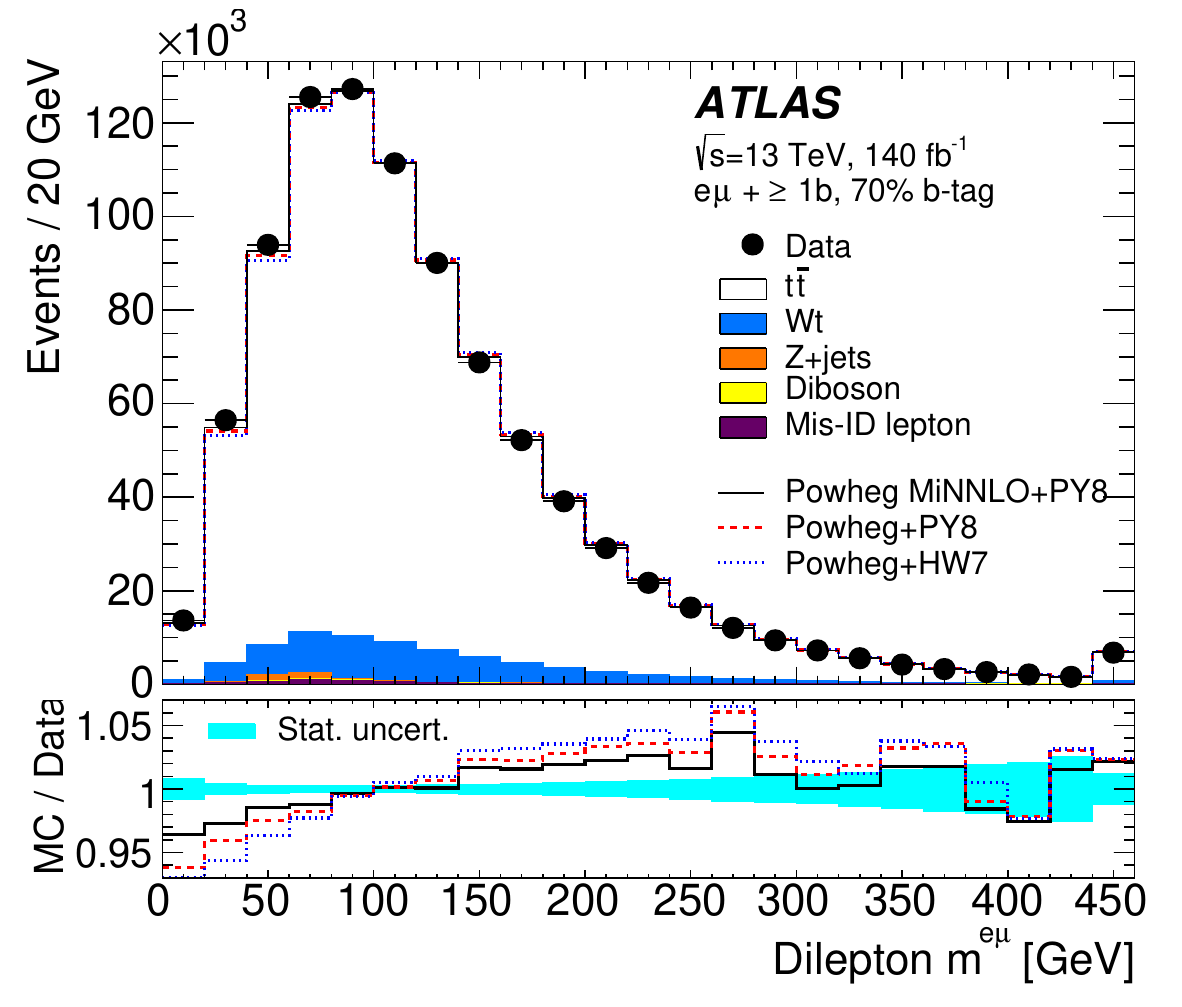}\vspace{-7mm}\center{(b)}}
\parbox{83mm}{\includegraphics[width=76mm]{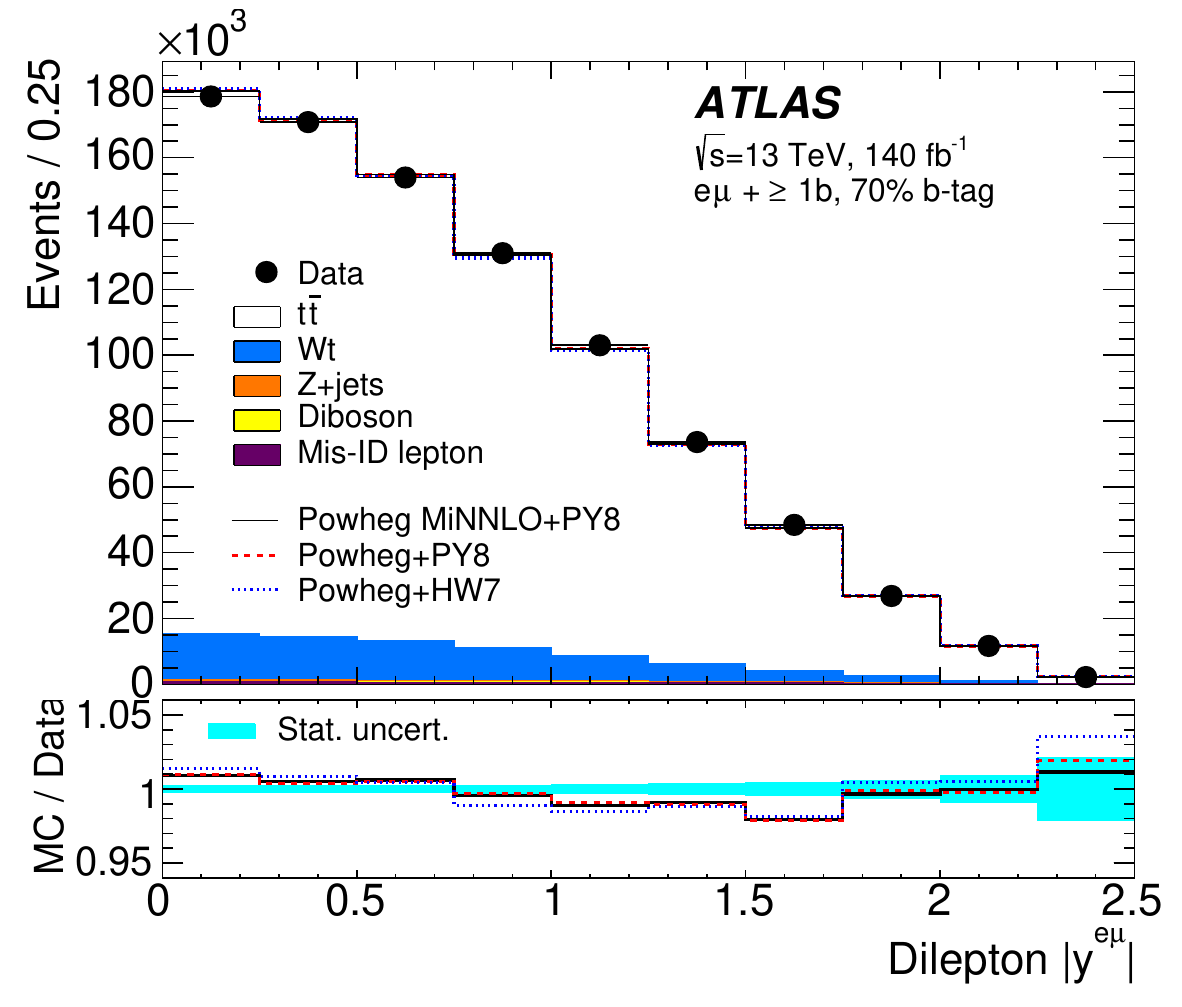}\vspace{-7mm}\center{(c)}}
\parbox{83mm}{\includegraphics[width=76mm]{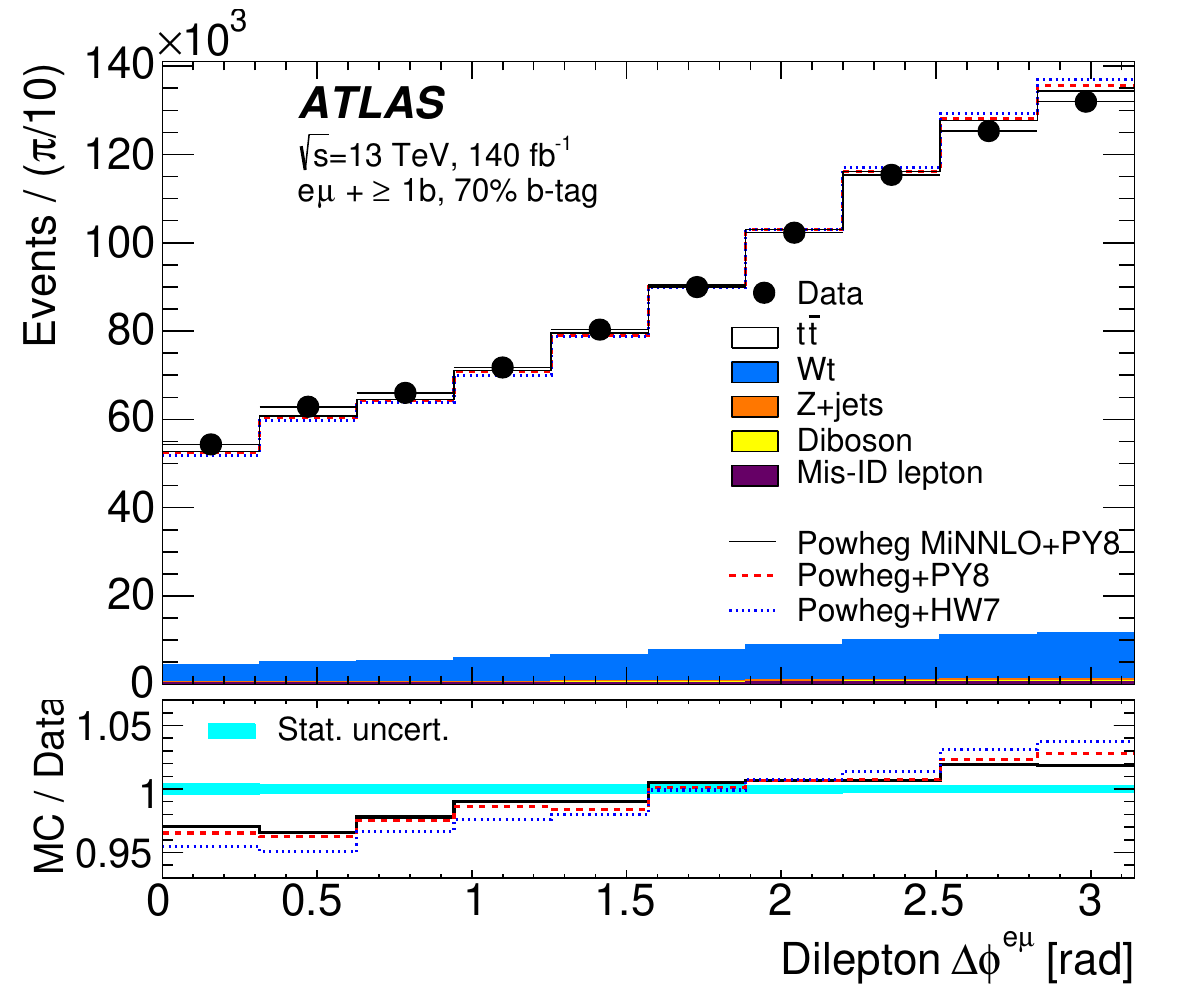}\vspace{-7mm}\center{(d)}}
\parbox{83mm}{\includegraphics[width=76mm]{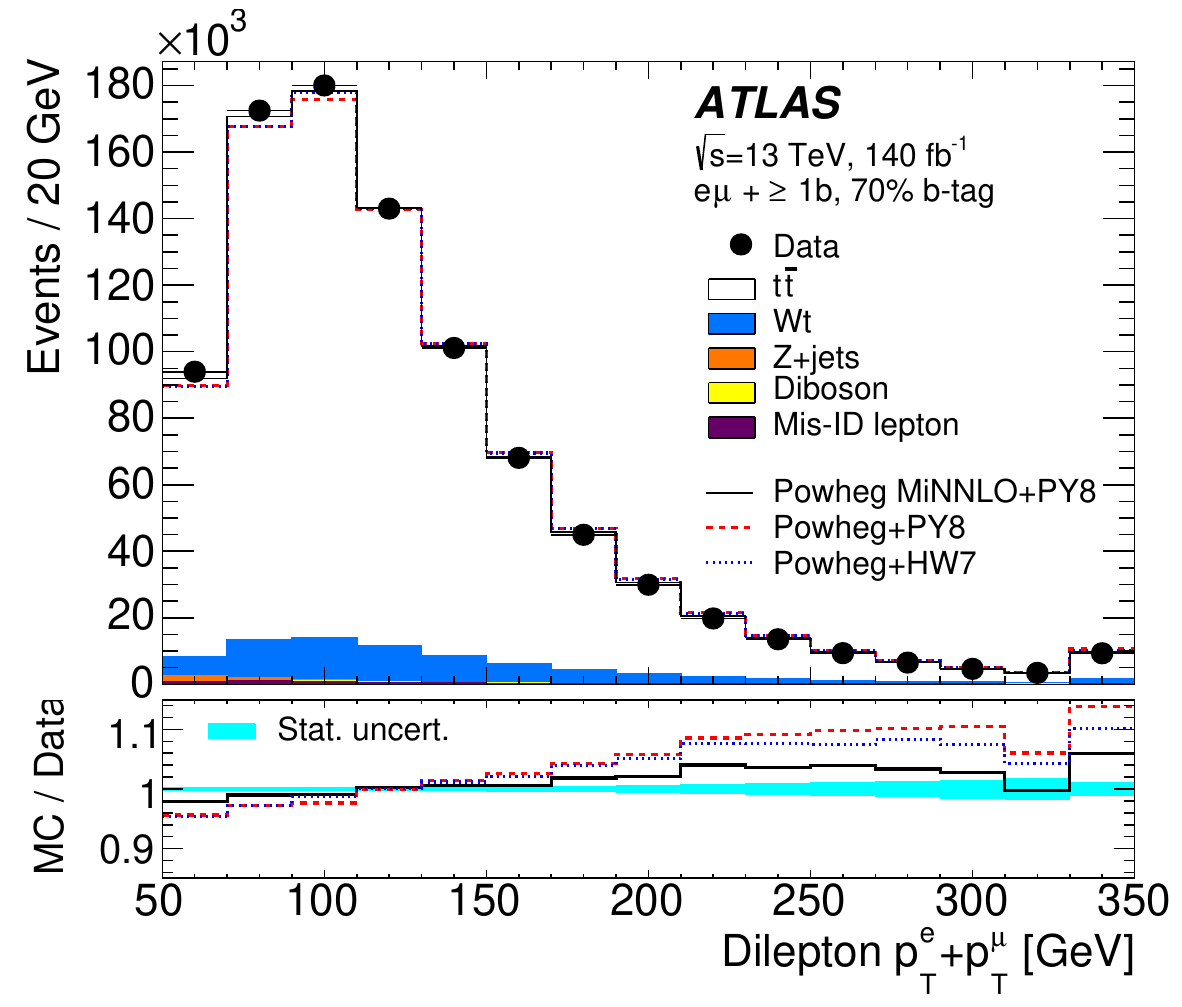}\vspace{-7mm}\center{(e)}}
\parbox{83mm}{\includegraphics[width=76mm]{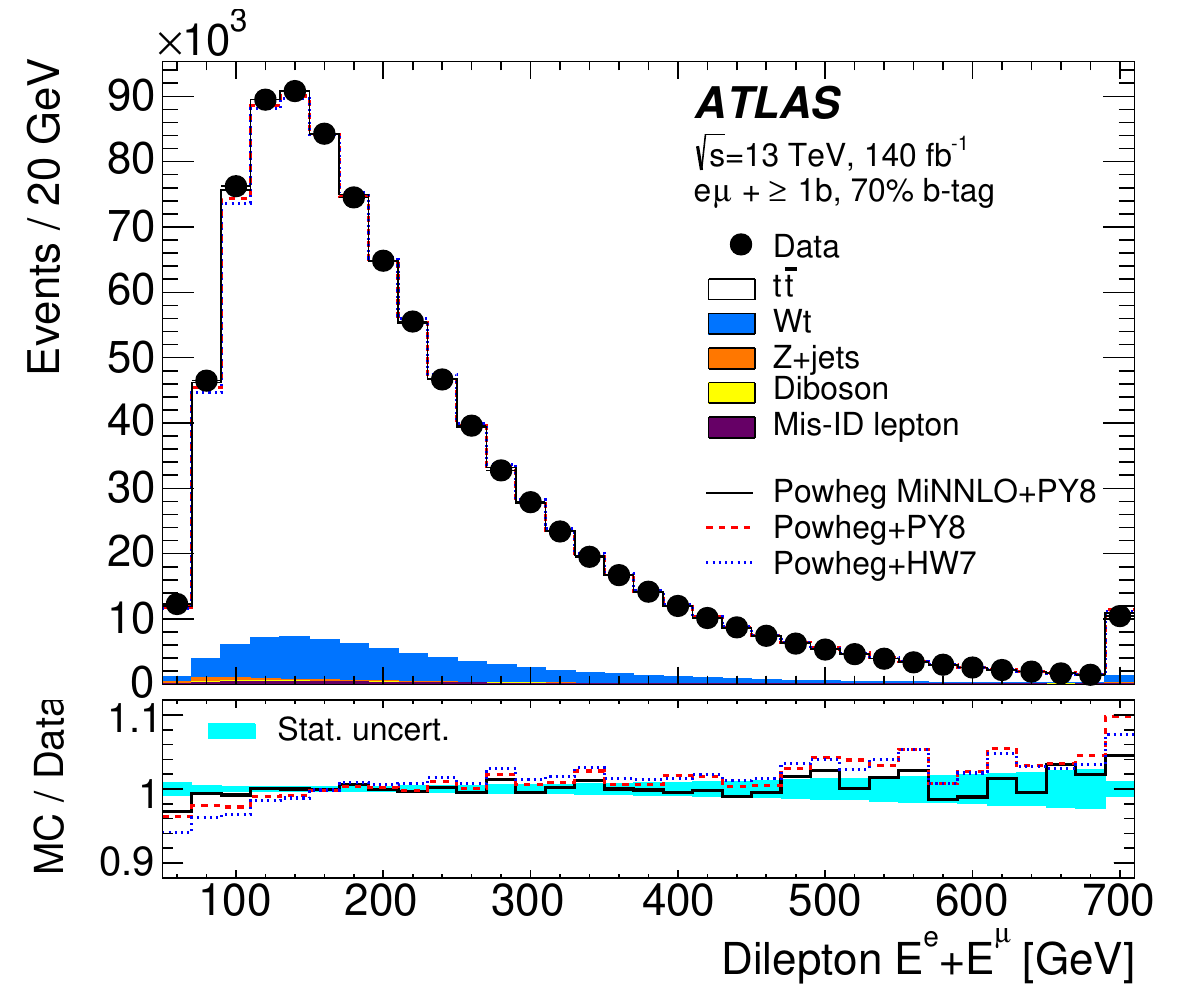}\vspace{-7mm}\center{(f)}}
\caption{\label{f:dmcdilept}Distributions of
(a) the dilepton \ptll, (b) invariant mass \mll, (c) rapidity \rapll,
(d) azimuthal angle difference \dphill, (e) lepton \pt\ sum \ptsum\ and
(f) lepton energy sum \esum, in events with an opposite-charge $e\mu$ pair and
at least one $b$-tagged jet. The reconstruction-level data are
compared with the expectation from simulation, broken down into contributions
from \ttbar\ \,(modelled with \powhegminnlo), $Wt$, $Z$+jets,
dibosons, and events with misidentified electrons or muons, normalised to the
same number of entries as the data.
The lower panels show the ratios of simulation to data,
using various \ttbar\ signal samples and with the cyan shaded band indicating
the statistical uncertainty. The last bin includes the overflow in panels
(a), (b), (e) and (f).}
\end{figure}

Two-dimensional distributions with \etal, \rapll\ or \dphill\ as the first
variable, and \mll\ as the second variable, were also measured, using
the same binning as Ref.~\cite{TOPQ-2018-17}. The excellent resolution
in \etal, \rapll\ and \dphill\ results in migration effects only being
significant between the four \mll\ bins, defined as $\mll<80$\,\GeV,
$80<\mll<120$\,\GeV, $120<\mll<200$\,\GeV\ and $\mll>200$\,\GeV.
The formalism of Eqs.~(\ref{e:fidtags})
was used, with the index $i$ running over the two-dimensional grid of bins
in both variables.

The measured differential cross-sections include contributions where one
or both leptons are produced via leptonic decays of $\tau$-leptons
($t\rightarrow W\rightarrow\tau\rightarrow e/\mu$). To enable comparisons
with theoretical predictions that only include direct
$t\rightarrow W\rightarrow e/\mu$ decays, a second set of cross-section results
was derived with a bin-by-bin multiplicative correction \fntaui\ to remove
the $\tau$ contributions:
\begin{eqnarray}\label{e:notau}
\xtti\,(\mbox{no-$\tau$}) = \fntaui\xtti\ ,
\end{eqnarray}
and similarly for the normalised cross-sections.
The corrections \fntaui\ were calculated from \ttbar\ simulation as the
fractions of leptons (for the \ptl\ and \etal\ distributions) or events (for the other distributions) in each particle-level bin, that originate
from events where neither lepton was produced from a $\tau$-lepton decay.
These fractions are typically in the range 0.8--0.9.

\subsection{$\boldmath e\mu b\bar{b}$ differential cross-sections}\label{ss:embbdiff}

The \ttem\ differential cross-sections include leptons from \ttbar\ events
only, treating leptons from $Wt$ events as background that is estimated and
subtracted using simulation. In some kinematic regions, e.g. those with high
lepton or dilepton \pt, the interference between
\ttbar\ and $Wt$ contributions becomes large and this subtraction has large
uncertainties that limit the measurement precision. Recent theoretical work
has focused on a combined description of the \ttbar\ and $Wt(b)$ processes,
through e.g.\ the {\scr Powheg} $bb4l$ generator~\cite{jezobb4l} that
includes the complete set of contributions to $e\mu\nu\bar{\nu}\bbbar$
production. To probe this final state, the fiducial cross-section definition
was modified to explicitly
include a requirement for particle-level $b$-jets, as well as leptons.
An \embb\ fiducial region was therefore defined, requiring at least two
particle-level $b$-jets with $\pt>25$\,\GeV\ and $|\eta|<2.5$, in addition
to the lepton requirements of the \ttem\ measurement. The particle-level
jets were reconstructed with the anti-$k_t$ algorithm with $R=0.4$ from
all truth particles with lifetime $c\tau>10$\,nm, excluding leptons and
neutrinos from $W$ and $Z$ boson decays. A particle-level jet was
assumed to be a $b$-jet if ghost-associated~\cite{ghostasoc} to at least one
weakly decaying $b$-hadron with $\pt>5$\,\GeV. The fiducial region
was then divided into $i$ bins of lepton or dilepton kinematics, defining
the same distributions as in the \ttem\ measurement but always requiring
there to be at least two particle-level $b$-jets.

The cross-section \xembbi\ in bin $i$ of a kinematic distribution was then
determined from:
\begin{equation}\label{e:xembbi}
\xembbi=\left(\frac{\nyi-\niibi}{L\ \gemembbi}\right)\cdot \stagsq
\end{equation}
where \nyi\ is the number of data $e\mu$ events with at least two $b$-tagged
jets in lepton kinematic bin $i$, \niibi\ is the estimated number of
background events from $Z$+jets, dibosons and events with misidentified
electrons or muons, and \stag\ is a $b$-tagging efficiency scale factor
defined below. As the background from events without two true $b$-jets
is very small, the looser 77\% $b$-tagging efficiency working point was used
for this measurement. The efficiency
\gemembbi\ is defined as $\gemembbi=\nembbirec/\nembbitrue$, where
\nembbirec\ is the number of selected opposite-charge $e\mu$ events from
\ttbar\ or $Wt$
with two $b$-tagged jets in lepton kinematic bin $i$, and \nembbitrue\
is the number of particle-level \ttbar\ and $Wt$ events that pass the
fiducial selection on both leptons and $b$-jets, and are in lepton kinematic
bin $i$ at particle level. The efficiency \gemembbi\ accounts for the efficiency
to $b$-tag the two $b$-jets in the \embb\ final state, as well as the
lepton efficiencies and migration in and out of kinematic bin $i$ due to
both lepton and jet resolution effects. The values of \gemembbi\ are around
0.2--0.35, smaller than \gemi\ in the \ttem\ measurement as \gemembbi\
includes the $b$-tagging efficiency for the two $b$-jets, evaluated from
simulation. The bin-to-bin migration is very similar to that for the
\ttem\ differential cross-sections.

To reduce the sensitivity to the modelling of the $b$-tagging efficiency
in simulation, the second term in Eq.~(\ref{e:xembbi}) corrects the
measured cross-section by the square of the ratio of the values of
\epsb\ in simulation (\epsbmc) and data (\epsbdata),
i.e.\ $\stag=\epsbmc/\epsbdata$. These efficiencies were evaluated
from the inclusive double-tagging formalism, Eqs.~(\ref{e:tags}), applied
to the total numbers of events with one and two $b$-tagged jets using the
77\% working point, treating \ttbar\ as signal and $Wt$ as background.
Since \stag\ is calculated inclusively (and not in bins $i$),
it is relatively insensitive to the $\ttbar/Wt$ interference uncertainties
and reduces the uncertainties in the absolute differential cross-sections
\xembbi. As the \stag\ correction is the same for all bins of all
differential cross-sections, it has no effect on the normalised \embb\
differential cross-sections, which were calculated from the \xembbi\ values
in the same way as for the \ttem\ measurement. These cross-sections include
the contributions from leptonic decays of $\tau$-leptons, and a second
set of results with the $\tau$ contributions removed analogously to
Eq.~(\ref{e:notau}) was also derived, using the \powhegminnlo\ \ttbar\ and
{\scr Powheg\,+\,Pythia8} $Wt$ (with the diagram removal scheme) samples
to calculate the corrections.

A total of 378\,266 opposite-charge $e\mu$ events with two $b$-tagged jets
were selected in data, with an estimated background from processes other
than \ttbar\ and $Wt$ of 0.4\%. The contribution from non-resonant $WW\bbbar$
events to the fiducial \embb\ cross-section is negligible.
A further 564\,501 one $b$-tag events were used
in the determination of \stag. Figure~\ref{f:dmcwb}(a) shows the $b$-tagged
jet multiplicity using the 77\% working point, and Figures~\ref{f:dmcwb}(b)
to~\ref{f:dmcwb}(f) show kinematic distributions for events with two $b$-tagged
jets, with data compared with the baseline simulation with \powhegminnlo\
\ttbar\ events and {\scr Powheg\,+\,Pythia8} $Wt$ events with the diagram
removal scheme, and alternative predictions using diagram subtraction $Wt$
events~\cite{wtinter1,wtinter2}, and the \powhegbbfl\ sample that covers both
\ttbar\ and $Wt$ final states.  All samples were reweighted to
enhance the number of events with at least three particle-level $b$-jets by 50\%.
The modelling of the data lepton \pt\ and $|\eta|$
distributions by all three predictions in Figure~\ref{f:dmcwb} is similar,
although the sample using diagram subtraction $Wt$ events has a slightly
softer lepton \pt\ distribution than that using diagram removal $Wt$ events,
and the \powhegbbfl\ sample has a slightly more central lepton
$|\eta|$ distribution.

\begin{figure}[tp]
\vspace{-5mm}
\parbox{83mm}{\includegraphics[width=76mm]{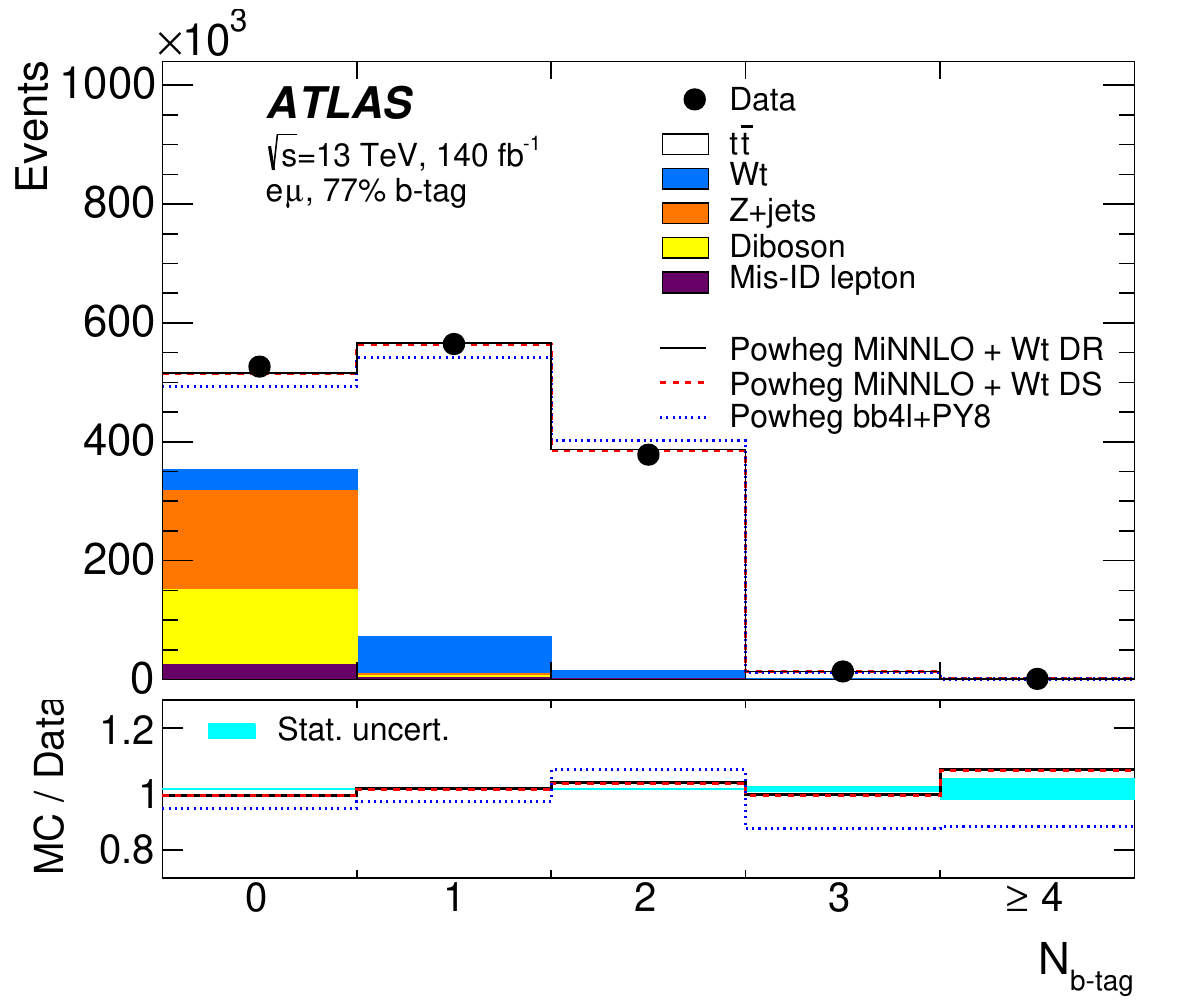}\vspace{-7mm}\center{(a)}}
\parbox{83mm}{\includegraphics[width=76mm]{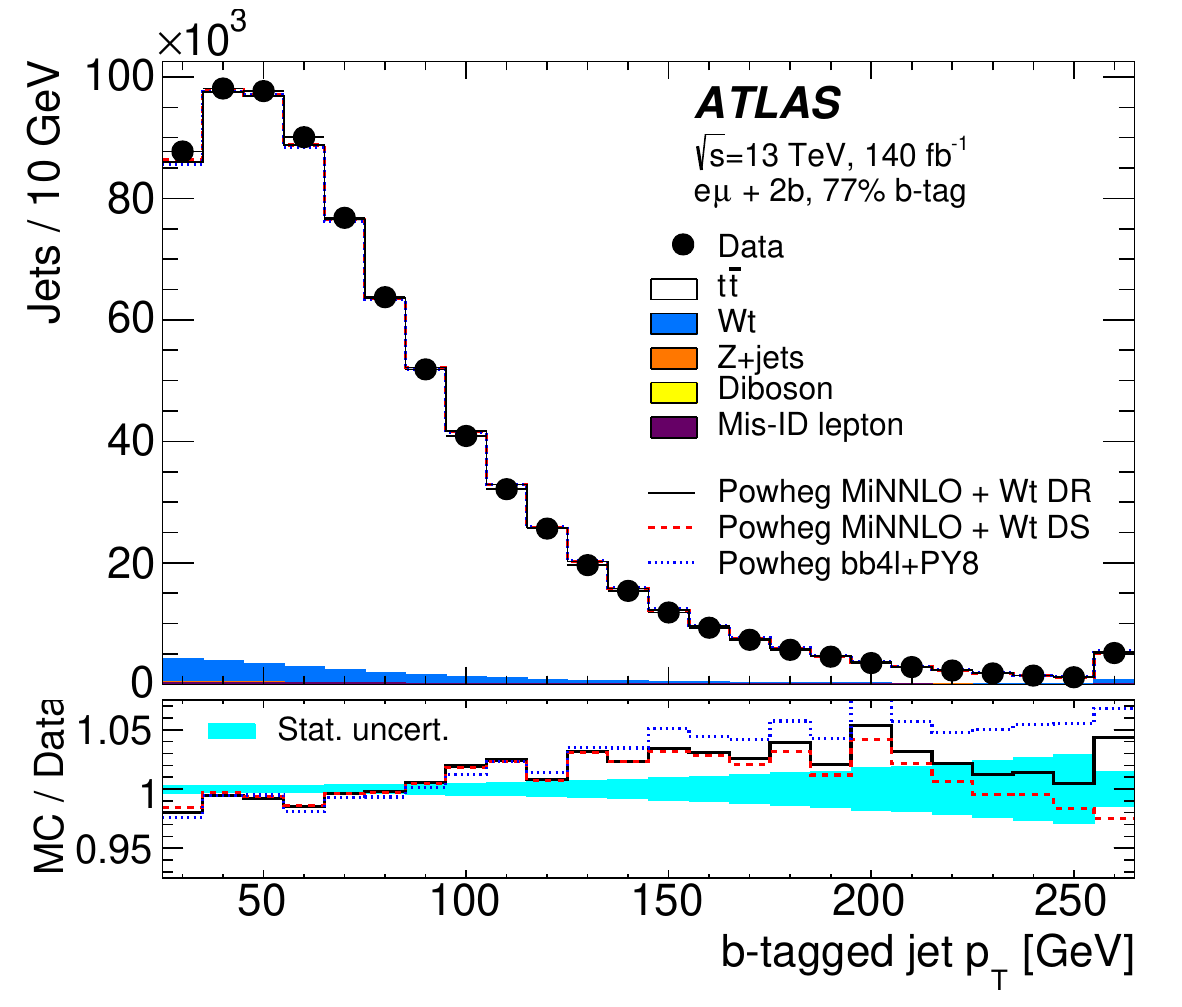}\vspace{-7mm}\center{(b)}}
\parbox{83mm}{\includegraphics[width=76mm]{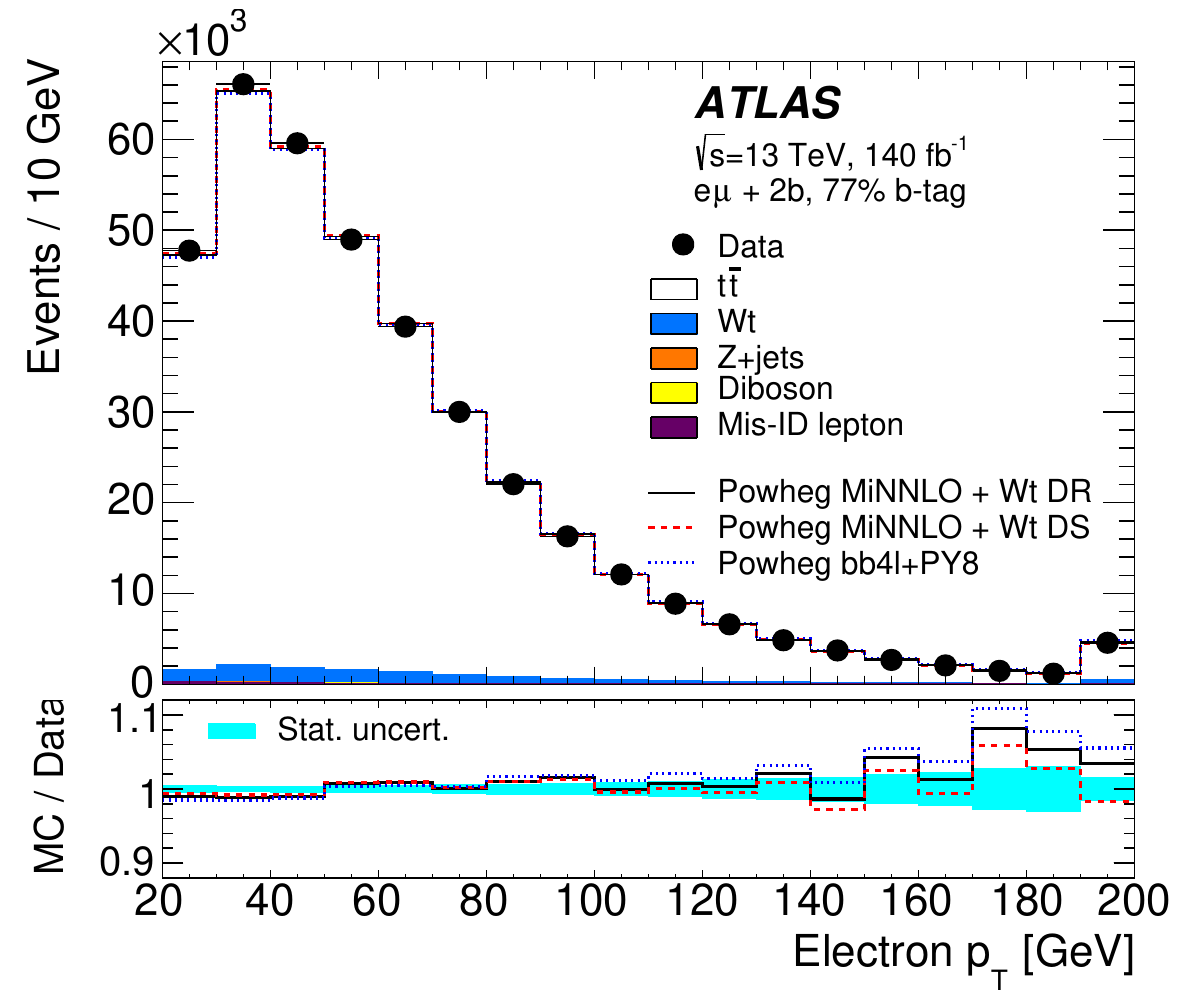}\vspace{-7mm}\center{(c)}}
\parbox{83mm}{\includegraphics[width=76mm]{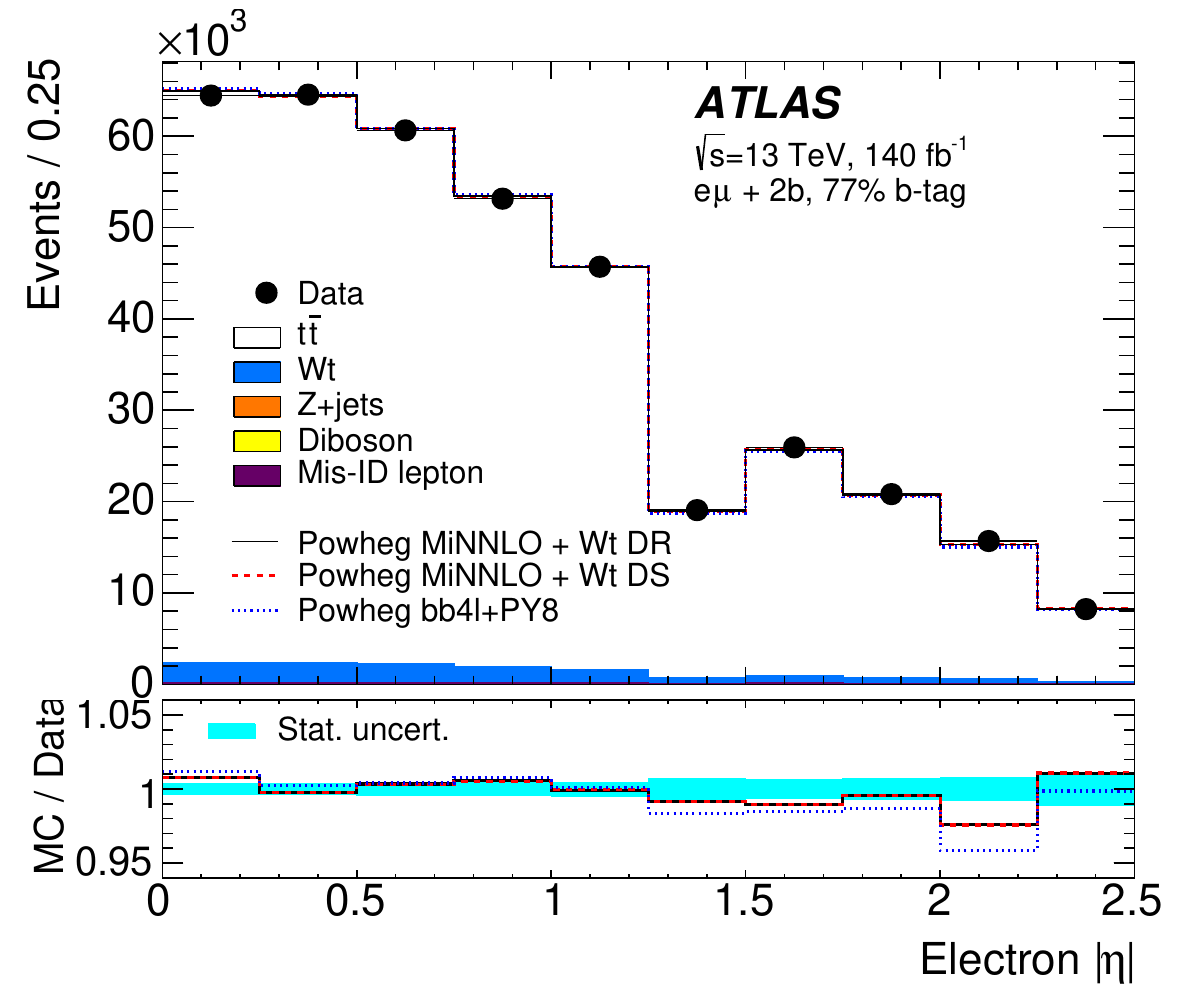}\vspace{-7mm}\center{(d)}}
\parbox{83mm}{\includegraphics[width=76mm]{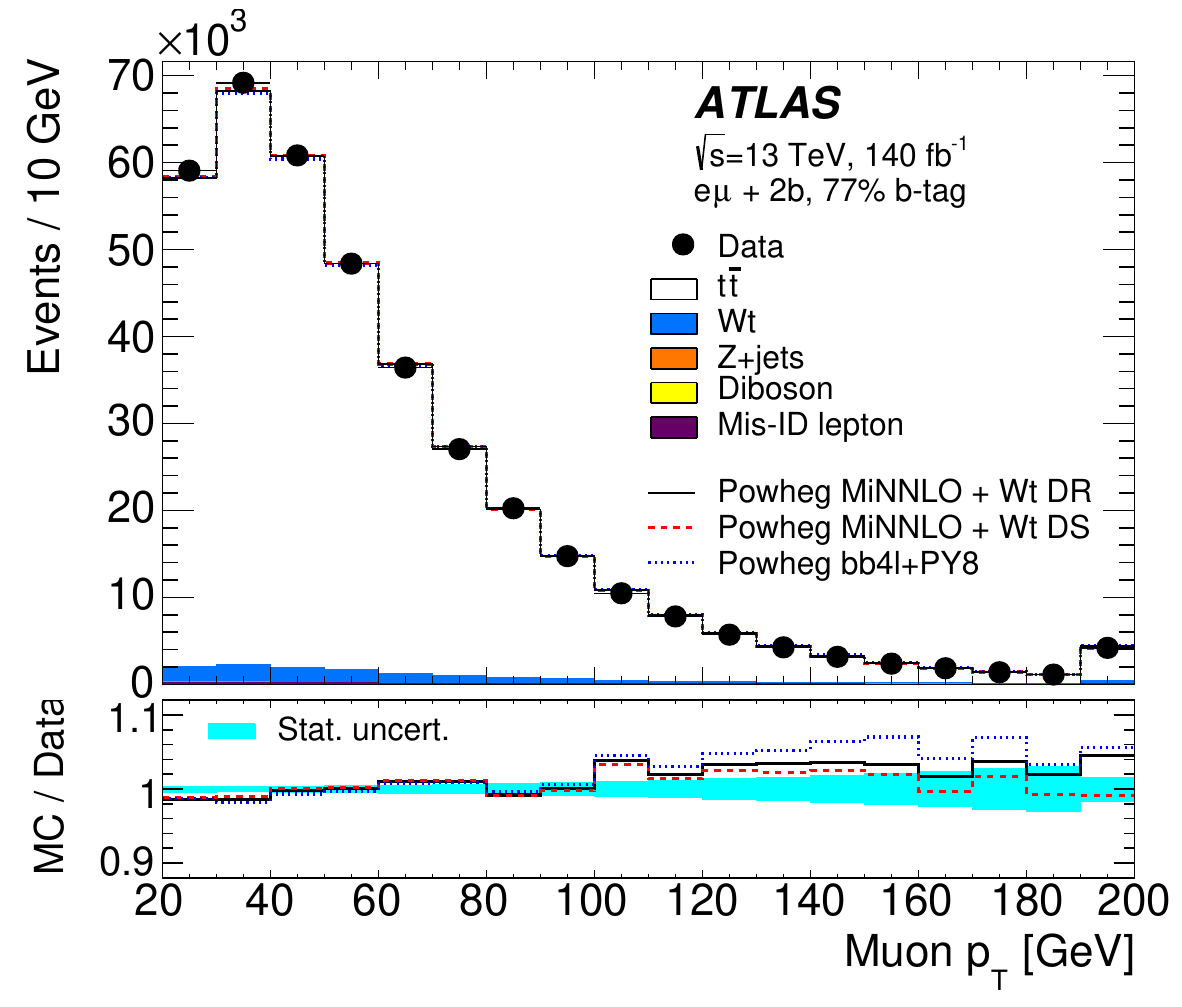}\vspace{-7mm}\center{(e)}}
\parbox{83mm}{\includegraphics[width=76mm]{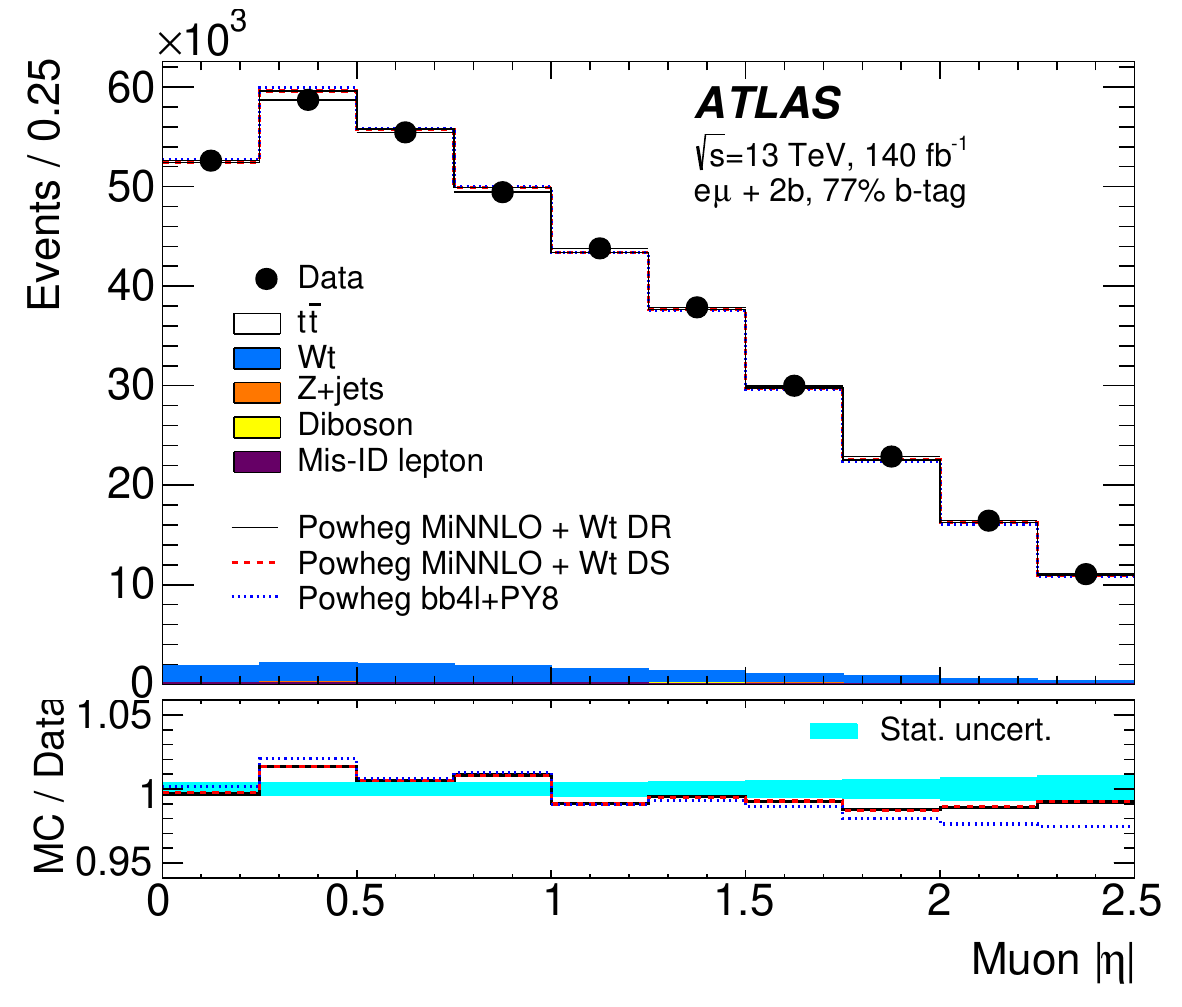}\vspace{-7mm}\center{(f)}}
\caption{\label{f:dmcwb}Distributions of (a) the number of $b$-tagged jets
in selected opposite-sign $e\mu$ events; and (b) the \pt\ of $b$-tagged jets,
(c) the \pt\ of the electron, (d) the $|\eta|$ of the electron,
(e) the \pt\ of the muon and (f) the $|\eta|$ of the muon, in events with
an opposite-sign $e\mu$ pair and exactly two $b$-tagged jets with the
77\% efficiency working point.
The reconstruction-level data are compared with the expectation from simulation,
broken down into contributions from \ttbar\ (modelled with \powhegminnlo),
$Wt$, $Z$+jets, dibosons, and events with misidentified electrons or muons.
The simulation prediction is normalised to the same integrated luminosity as the
data in (a) and to the same number of entries as the data in (b--f).
The lower panels show the ratios of simulation to data, using various
simulation samples and with the cyan shaded band indicating the statistical
uncertainty. The last bin includes the overflows in panels (b), (c) and (e).
}
\end{figure}

\subsection{Background estimates}\label{ss:bkgd}

The dominant background in the inclusive \xtt\ and \ttem\ differential
cross-section analyses comes from $Wt$ production, and was evaluated from
simulation as discussed in Section~\ref{s:datmc}. In the \embb\ analysis,
this process is part of the signal. The small diboson contribution
(dominated by $WW$ production with additional $b$-tagged jets) is treated as
background in all analyses and was evaluated using simulation. The production
of a $Z$ boson accompanied by heavy-flavour jets is subject to large
theoretical uncertainties, so the
$Z(\rightarrow\tau\tau\rightarrow e\mu)$+$b$-tagged jets
background predictions from {\scr Sherpa 2.2} were scaled using
the corresponding yield ratios of
$Z(\rightarrow ee/\mu\mu)$+$b$-tagged jets events in data and simulation.
These yields were extracted from template fits to the $ee$ and $\mu\mu$
invariant mass distributions, giving yield ratios of
$0.91\pm 0.04$ (one $b$-tagged jet) and $1.10\pm 0.04$ (two $b$-tagged jets).
The uncertainties include the effect of the harder $Z$-boson \pt\ spectrum
in selected $Z\rightarrow\tau\tau$ vs.\ $Z\rightarrow ee/\mu\mu$ events
and residual differences between the ratios measured in $ee$ and $\mu\mu$ events.

The background from events with misidentified leptons was estimated using the
same charge-sign (SS) $e\mu$ sample. Following the method used in
Ref.~\cite{TOPQ-2023-28}, the SS sample was selected as described in
Section~\ref{s:evtsel}, but additionally requiring the electron
to be accepted by a charge misidentification boosted decision tree
(BDT)~\cite{EGAM-2018-01}. This requirement reduces the rate of electron charge
misreconstruction by up to an order of magnitude and suppresses the
contribution of dilepton \ttbar\ events with a misreconstructed
electron charge to the SS sample. The misidentified-lepton background
\nijfake\ in lepton kinematic bin $i$ with $j$ $b$-tagged jets was estimated
from the number of SS events in data, \nijdss, after subtracting the number
of SS events with two prompt leptons \nijphss\ estimated using simulation, and then scaling by the
ratio $R^i_j$ of misidentified-lepton events in the opposite charge-sign (OS)
and SS samples in simulation:
\begin{eqnarray}
\nijfake & = & R^i_j (\nijdss-\nijphss) \nonumber \ ,\\
R^i_j & = & \frac{\nijfakeos}{\nijfakess} \label{e:fakettr} \ .
\end{eqnarray}
This method relies on simulation to predict the ratio of OS to SS
misidentified-lepton events, and the prompt SS contribution, but not the
absolute number of misidentified-lepton events \nijfake, which is calculated
using the SS event counts in data. The same formalism was used with a single
bin $i$ for the inclusive cross-section measurement.

The electron and muon \pt\ and $|\eta|$ distributions in the SS samples
with at least one $b$-tagged jet are shown in Figure~\ref{f:sslept}, and
compared with the simulation prediction broken down into various sources
of prompt and misidentified leptons. The simulation models the overall
numbers of events in the one $b$-tagged sample to better than 5\% and
the two $b$-tagged sample to 10\%, and also reproduces the shapes of the
kinematic distributions. The values of $R_j$ for the inclusive cross-section and
their total uncertainties are $R_1=1.9\pm 0.4$ and $R_2=3.9\pm 1.6$,
with variations of
$R^i_1$ in the range 1.5--3 and $R^i_2$ in the range 3--5 for both the \ttem\ and
\embb\ differential analyses. Deviations of $R^i_j$ from unity are caused
by the charge misidentification BDT, which has an inefficiency of up to 50\%
for misidentified electrons
and is applied to the SS sample only, and differences in the composition
of misidentified leptons between the OS and SS samples. The uncertainties
in $R^i_j$ were assessed by removing the photon conversion, misidentified hadron
and muon decay-in-flight contributions in turn, and recalculating $R^i_j$.
A 25\% uncertainty in the prompt SS contribution
with correctly reconstructed electron charge sign
was assumed, covering the uncertainties in modelling the major contributing
processes ($\ttbar+W$, $\ttbar+Z$ and $\ttbar+H$, and $WZ$).
The modelling of the charge misidentification BDT was validated using
$Z\rightarrow ee$ events, leading to a 10\% uncertainty in the prompt SS
component where the electron charge is misreconstructed.
Overall, the misidentified lepton background
contributes about $0.6\pm 0.2$\% of the OS sample with one $b$-tagged jet and
$0.4\pm 0.2$\% of the
two $b$-tagged jet sample (see Table~\ref{t:evtcount}), concentrated at low
lepton \pt. Further studies were performed with SS control samples where the
lepton isolation cuts were inverted to increase the misidentified lepton
contributions. The simulation also gives a good description
of these samples, with absolute rates within 15\% of those observed in data.

\begin{figure}[tp]
\vspace{-5mm}
\parbox{83mm}{\includegraphics[width=76mm]{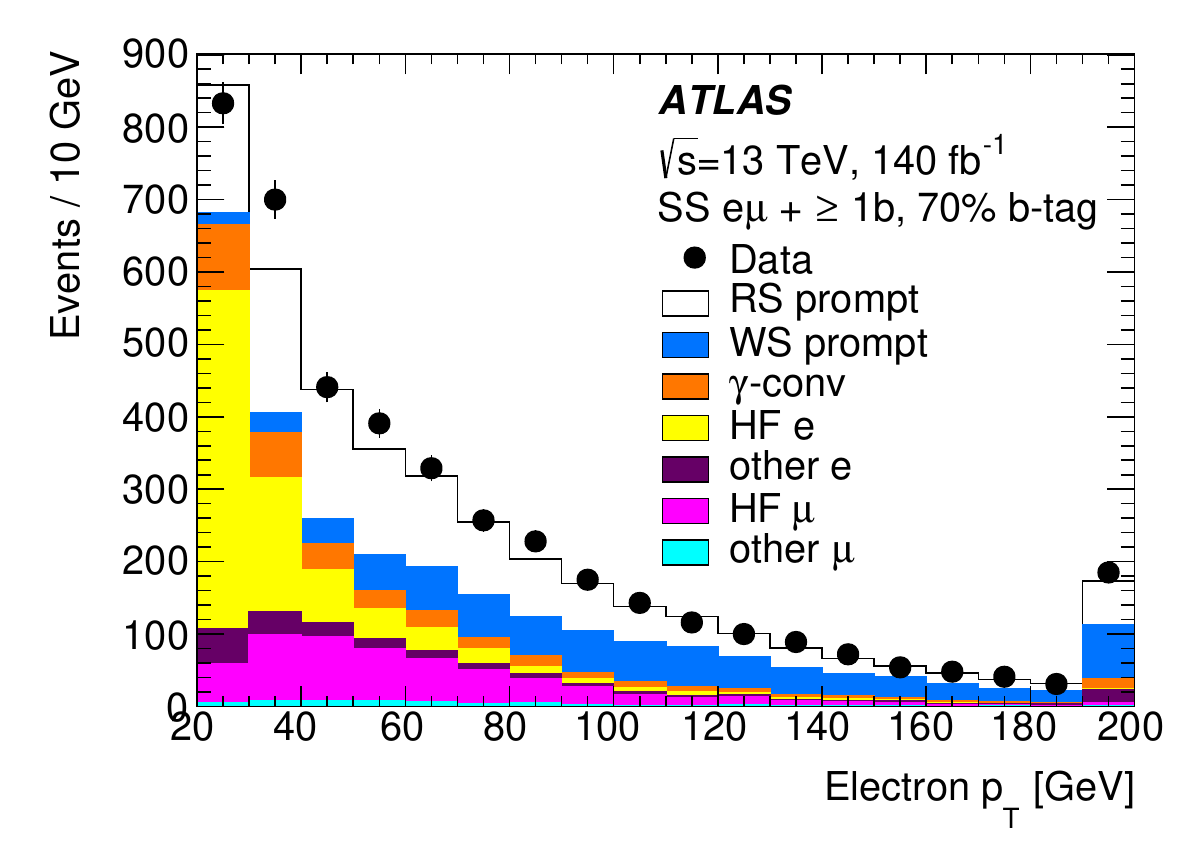}\vspace{-7mm}\center{(a)}}
\parbox{83mm}{\includegraphics[width=76mm]{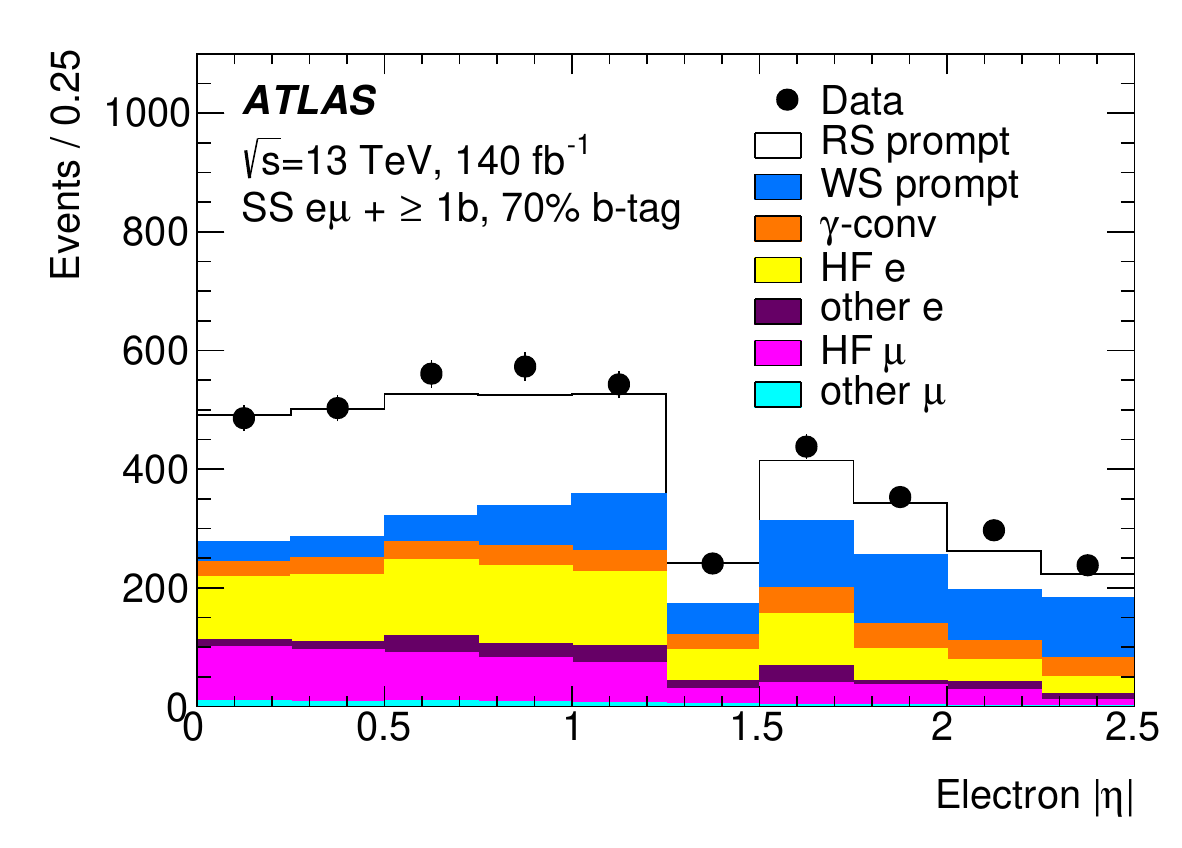}\vspace{-7mm}\center{(b)}}
\parbox{83mm}{\includegraphics[width=76mm]{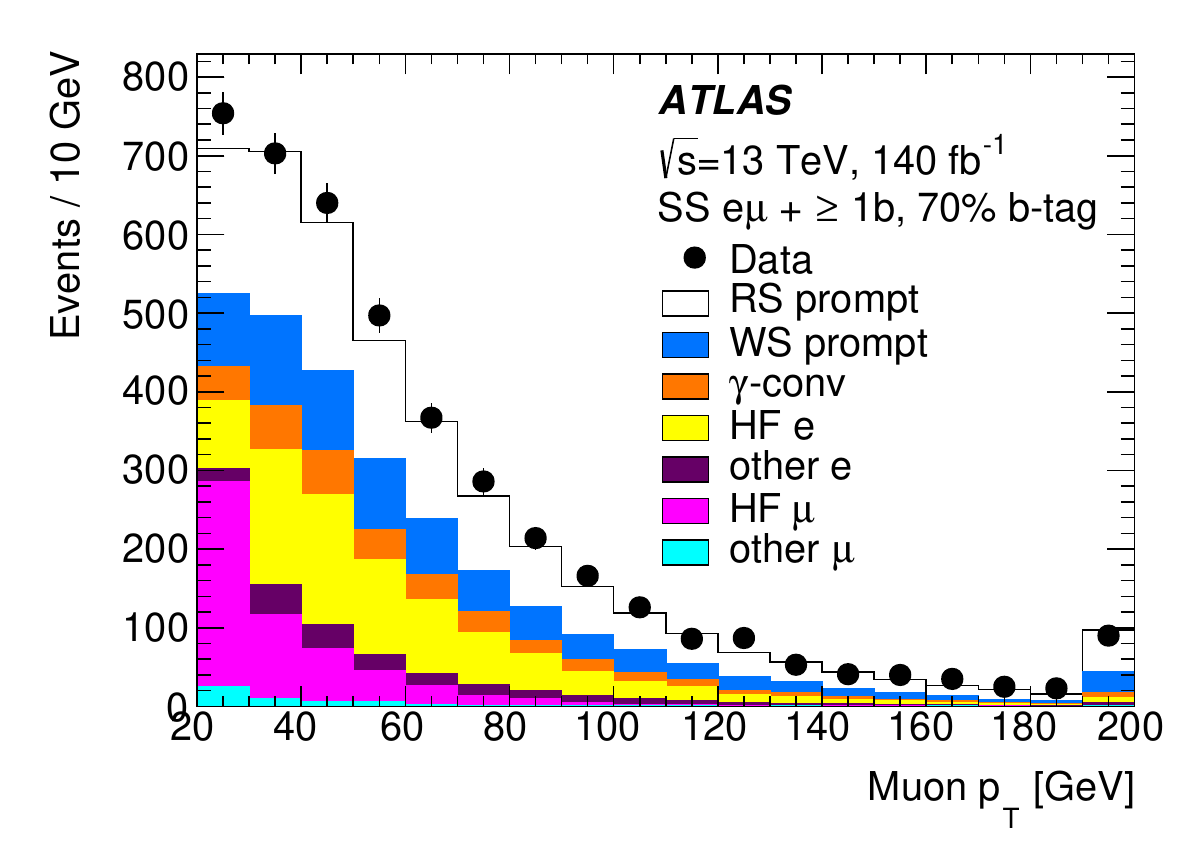}\vspace{-7mm}\center{(c)}}
\parbox{83mm}{\includegraphics[width=76mm]{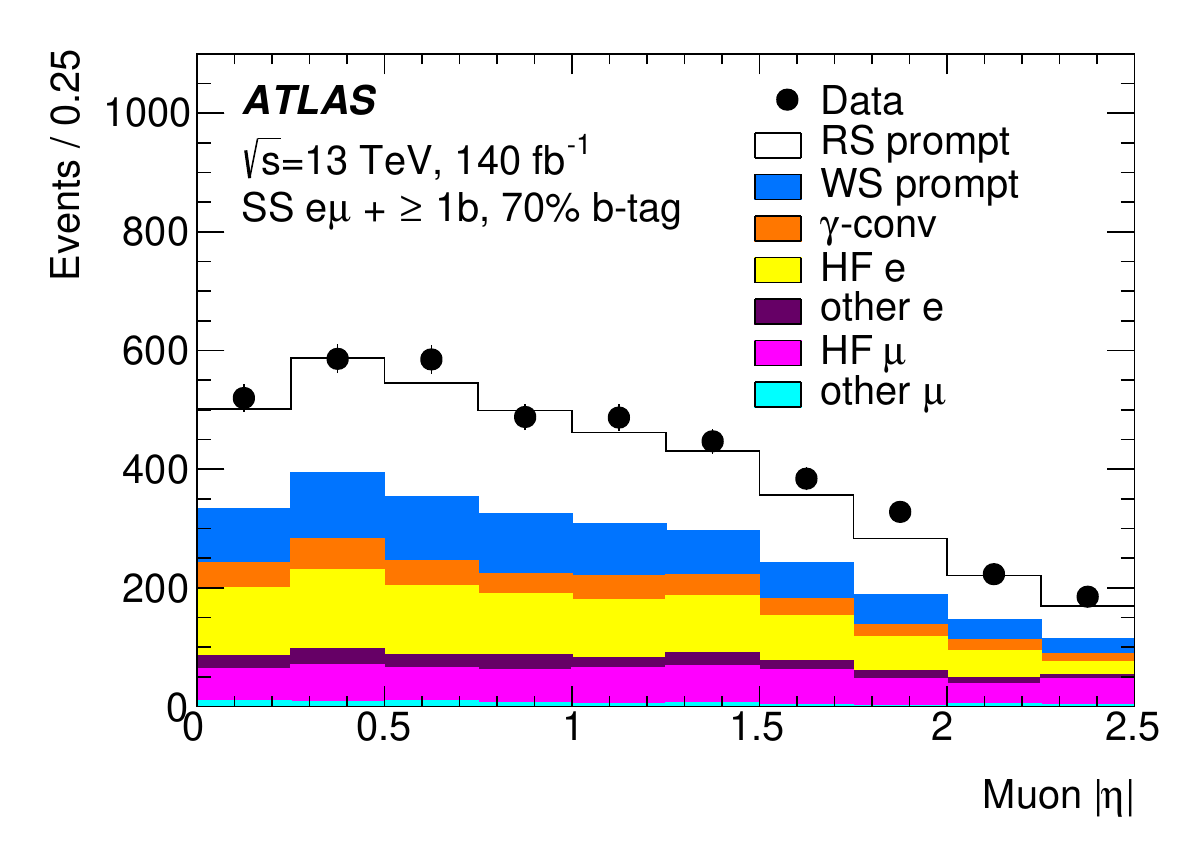}\vspace{-7mm}\center{(d)}}
\caption{\label{f:sslept}Distributions of (a) the electron \pt, (b) the
electron $|\eta|$, (c) the muon \pt\ and (d) the muon $|\eta|$, in events with
a same-charge $e\mu$ pair and at least one $b$-tagged jet, also requiring
the electron to pass the charge misidentification BDT.
The simulation prediction is normalised to the same integrated luminosity
as the data, and broken down into contributions where both leptons are prompt
and reconstructed with correct charge signs (RS), both leptons are prompt but
one has a misreconstructed charge sign (WS), or one is a misidentified lepton
from a photon conversion, a heavy-flavour (HF) hadron decay to an electron,
a heavy-flavour hadron decay to a muon, or other sources of misidentified
electrons (such as misidentified hadrons) and muons (such as decays in flight
of pions and kaons).
In the \pt\ distributions, the last bin includes the overflows.}
\end{figure}

\subsection{Validation of the differential measurements}\label{ss:diffval}

Tests using pseudo-experiment datasets generated from simulation were used to
validate the analysis procedures for the differential measurements, as
discussed in Refs.~\cite{TOPQ-2015-02,TOPQ-2018-17}. These tests demonstrated
that the method is unbiased and correctly estimates the statistical uncertainties
in each bin of each distribution. Figure~\ref{f:biastesta} shows examples
for the \ptl, \ptll, \etal\ and \rapll\ distributions in the \ttem\ analysis.
The filled black points
show the relative differences between the mean of the normalised differential
cross-sections obtained from 1000 pseudo-experiments and the true cross-sections
in each bin, divided by the true cross-sections to give fractional differences.
The pseudo-experiments were generated from a reference sample with
{\scr Powheg\,+\,Pythia8} \ttbar\ events plus backgrounds,
which was also used to determine the values of \gemi\ and \cbi.
The compatibility of the filled black points with zero
confirms that the method is unbiased for these samples.
For the alternative samples with different underlying distributions,
the open red points show the mean pseudo-experiment results and the dotted red
lines the true values, again
expressed as fractional deviations from the true cross-sections in the
reference sample, and obtained using \gemi\ and  \cbi\ values calculated
from the reference sample. An independent {\scr Powheg\,+\,Pythia8} \ttbar\
sample with $\mtop=176$\,\GeV\ was used as the alternative sample
in Figures~\ref{f:biastesta}(a) and~\ref{f:biastesta}(b),
and the {\scr Powheg\,+\,Pythia8} sample generated with NNPDF3.0 was reweighted
to the predictions of the CT14 PDF set~\cite{ct14} for
Figures~\ref{f:biastesta}(c) and~\ref{f:biastesta}(d).
In general, the results obtained from the
pseudo-experiments with alternative samples are consistent with the
true values within statistical uncertainties for all distributions and
a variety of alternative samples. These tests demonstrate that the
simple bin-by-bin correction procedure using \gemi\
correctly recovers the alternative distributions, without the need for
iteration or a matrix-based unfolding procedure. The two-dimensional
differential cross-sections were validated using similar procedures
and no significant biases were observed.

\begin{figure}[tp]
\vspace{-5mm}
\parbox{83mm}{\includegraphics[width=76mm]{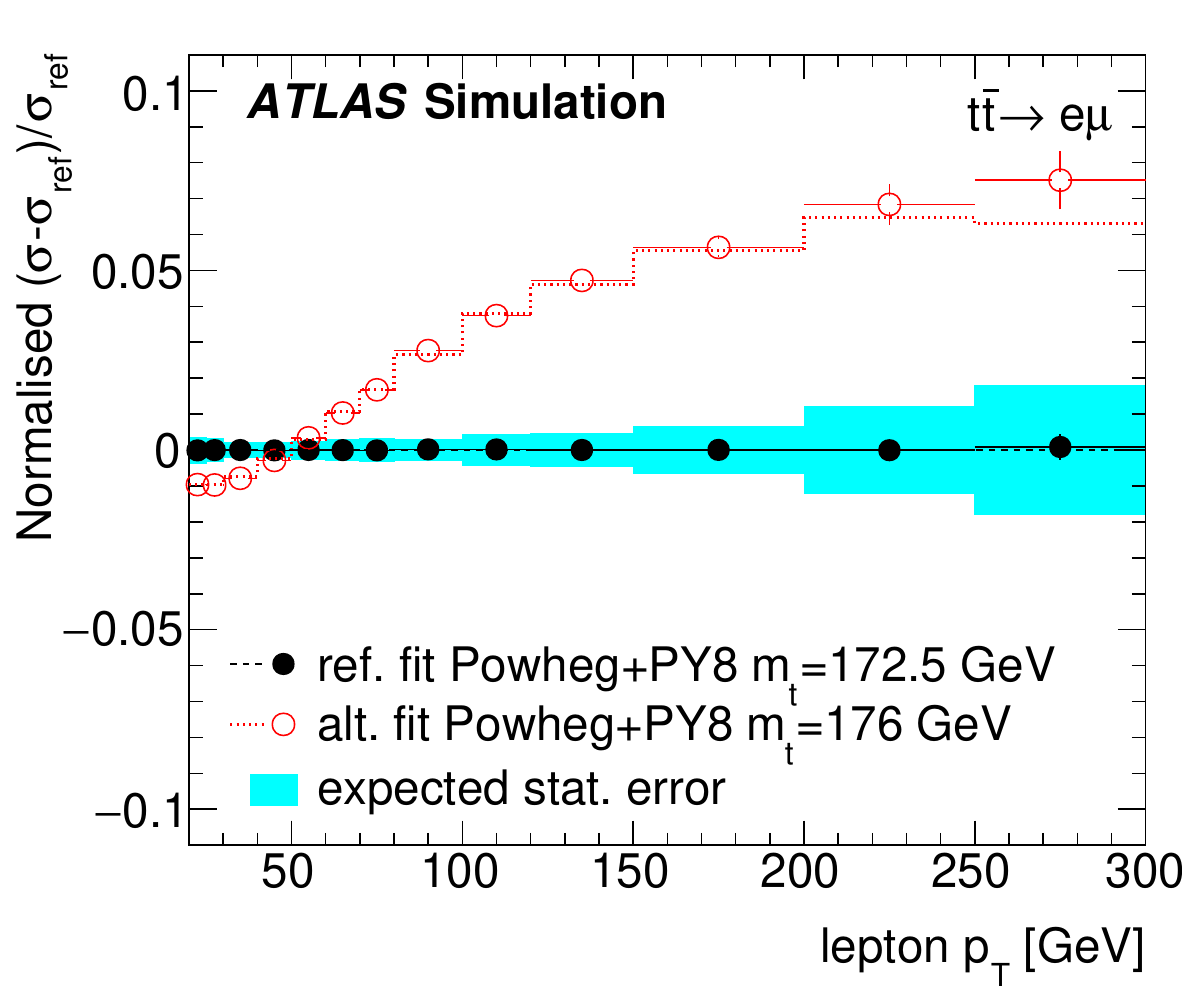}\vspace{-7mm}\center{(a)}}
\parbox{83mm}{\includegraphics[width=76mm]{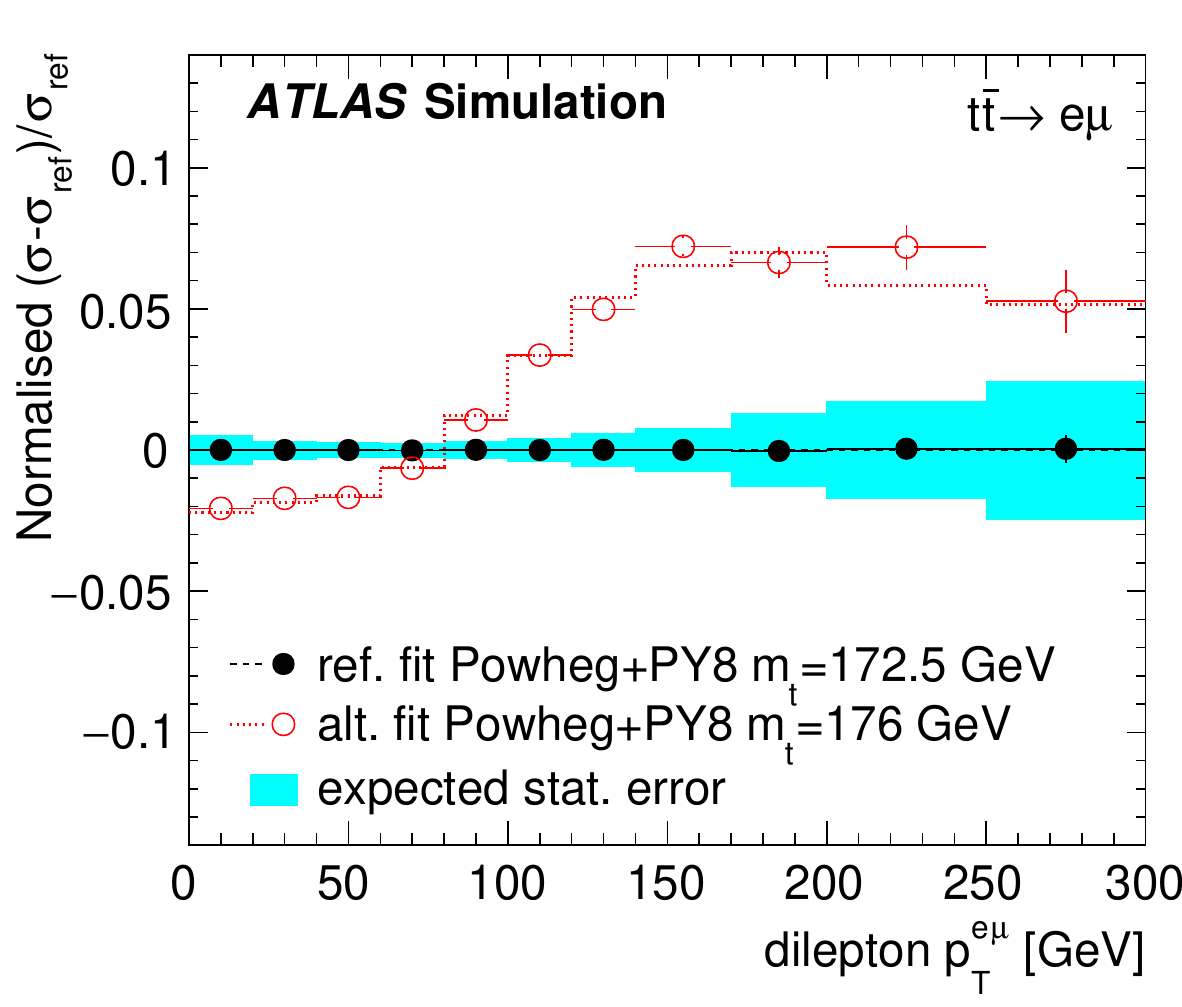}\vspace{-7mm}\center{(b)}}
\parbox{83mm}{\includegraphics[width=76mm]{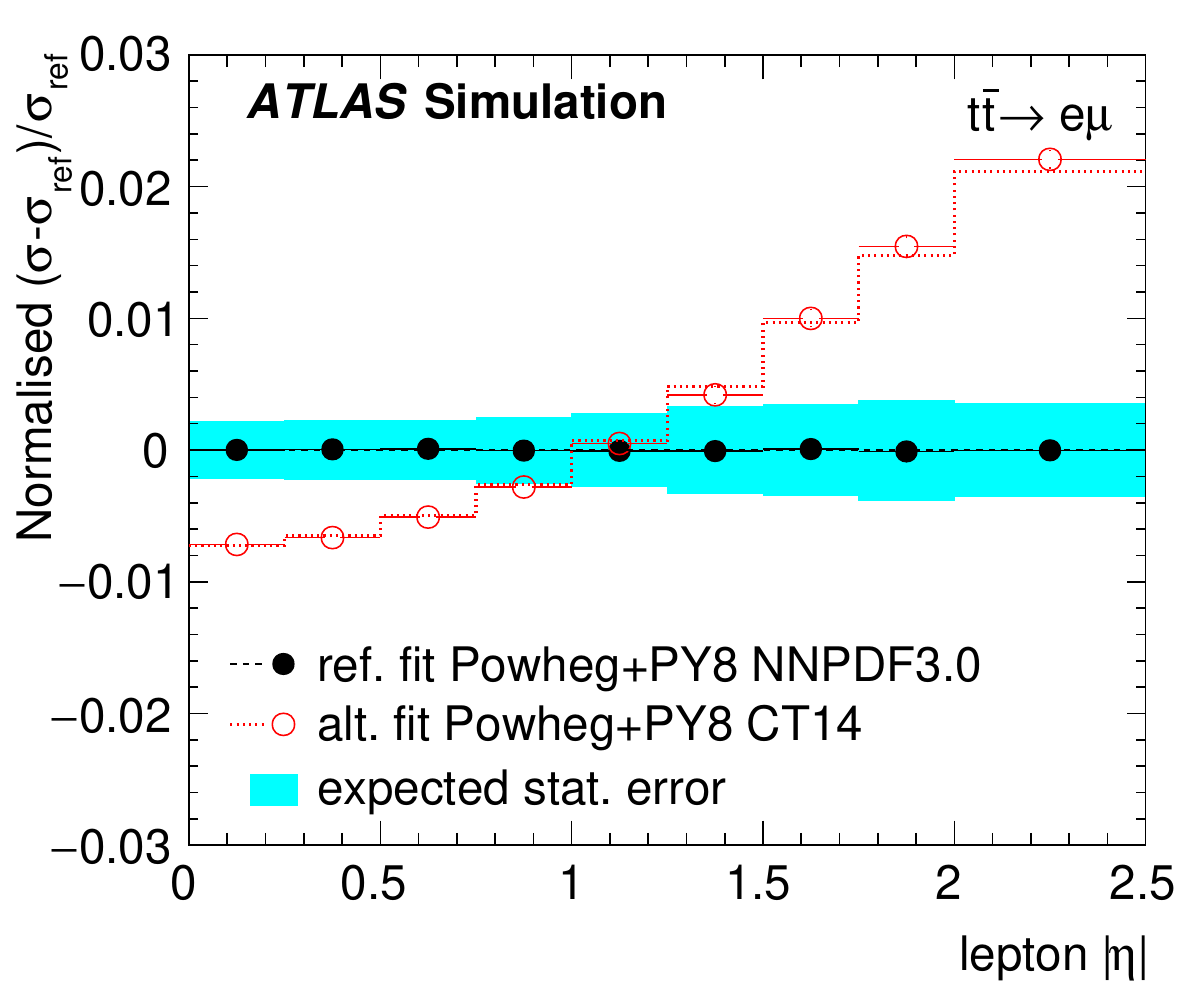}\vspace{-7mm}\center{(c)}}
\parbox{83mm}{\includegraphics[width=76mm]{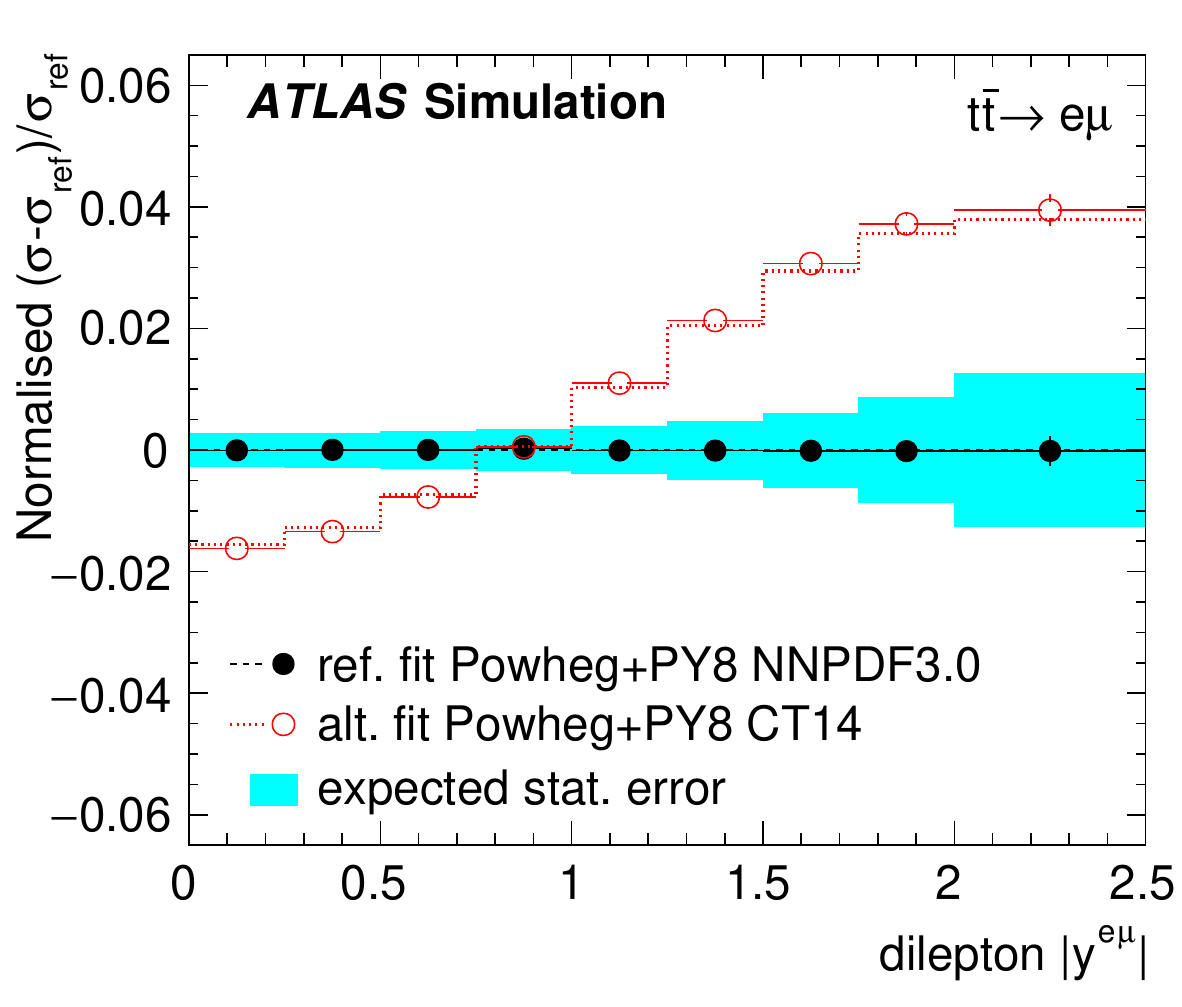}\vspace{-7mm}\center{(d)}}
\caption{\label{f:biastesta}Results of pseudo-experiment studies on simulated
events for the extraction of the normalised \ttem\ differential cross-section
distributions. The upper two plots show (a) \ptl\ and (b) \ptll\ using
{\scr Powheg\,+\,Pythia8} \ttbar\ events with
$\mtop=172.5$\,\GeV\ for the reference sample and \ttbar\ events
with $\mtop=176$\,\GeV\ for the alternative sample. The lower two plots show
(c) \etal\ and (d) \rapll\ using \ttbar\ events reweighted to the CT14 PDF set
instead of NNPDF3.0 for the alternative sample.
The black filled points show the mean deviations from the reference values
of the results from
pseudo-data samples generated with the reference simulation sample, with
error bars indicating the uncertainties due to the limited number of simulated
events. The cyan shaded bands indicate the expected statistical uncertainties
for a single sample corresponding to the data integrated luminosity.
The open red points show the mean deviations from the reference values obtained
from pseudo-experiments generated from the alternative simulation sample.
The red error bars represent the uncertainty due to the limited size
of these alternative samples, and the red dotted lines show the
true deviations from the reference in the alternative samples.}
\end{figure}

Pseudo-experiment tests were also performed for the \embb\ differential
cross-section analysis, using {\scr Powheg\,+\,Pythia8} \ttbar\ and
diagram-removal scheme $Wt$ events for the reference sample, and either \ttbar\
plus diagram-subtraction $Wt$ events, or the {\scr Powheg} $bb4l${\scr\,+Pythia8}
sample, for the alternative sample. Figure~\ref{f:biastestb} shows the
corresponding results for the \ptl\ and \ptll\ distributions, two of the
distributions that are
most sensitive to the modelling of $\ttbar/Wt$ interference.
The largest discrepancy in any of these tests
occurs for the last bin of the \ptll\ distribution when using
{\scr Powheg} $bb4l${\scr\,+\,Pythia8} for the alternative sample
(Figure~\ref{f:biastestb}(d)). This discrepancy has a statistical significance
of 3.5 standard deviations, which is comparable to the data statistical
uncertainty but less than half the total uncertainty in this bin when
systematic uncertainties are included. Other pseudo-experiment tests using
alternative values of \mtop\ or diagram subtraction $Wt$ events produced
larger changes in the differential cross-section in this bin but showed
smaller discrepancies, so no additional uncertainties were included.

\begin{figure}[tp]
\vspace{-5mm}
\parbox{83mm}{\includegraphics[width=76mm]{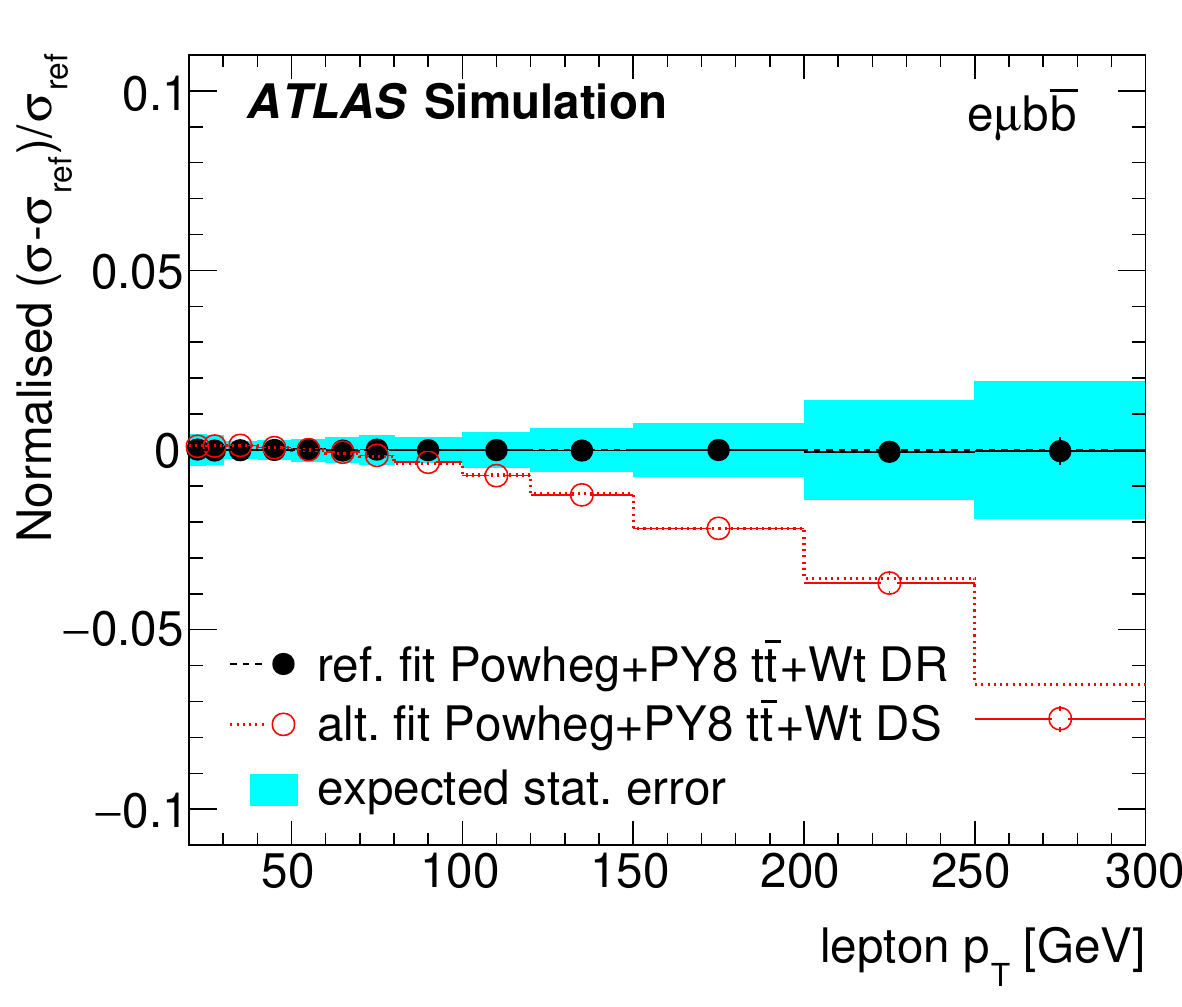}\vspace{-7mm}\center{(a)}}
\parbox{83mm}{\includegraphics[width=76mm]{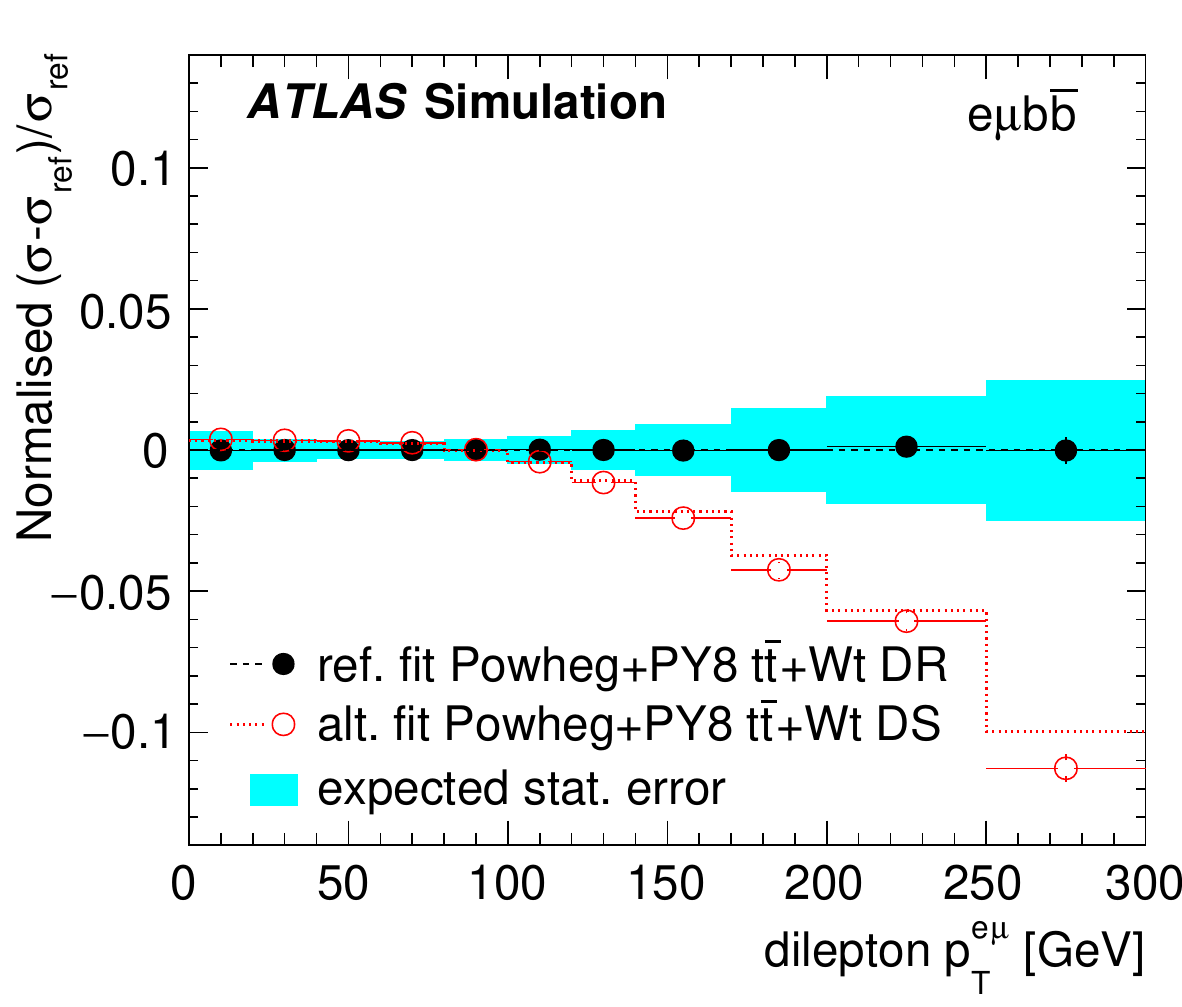}\vspace{-7mm}\center{(b)}}
\parbox{83mm}{\includegraphics[width=76mm]{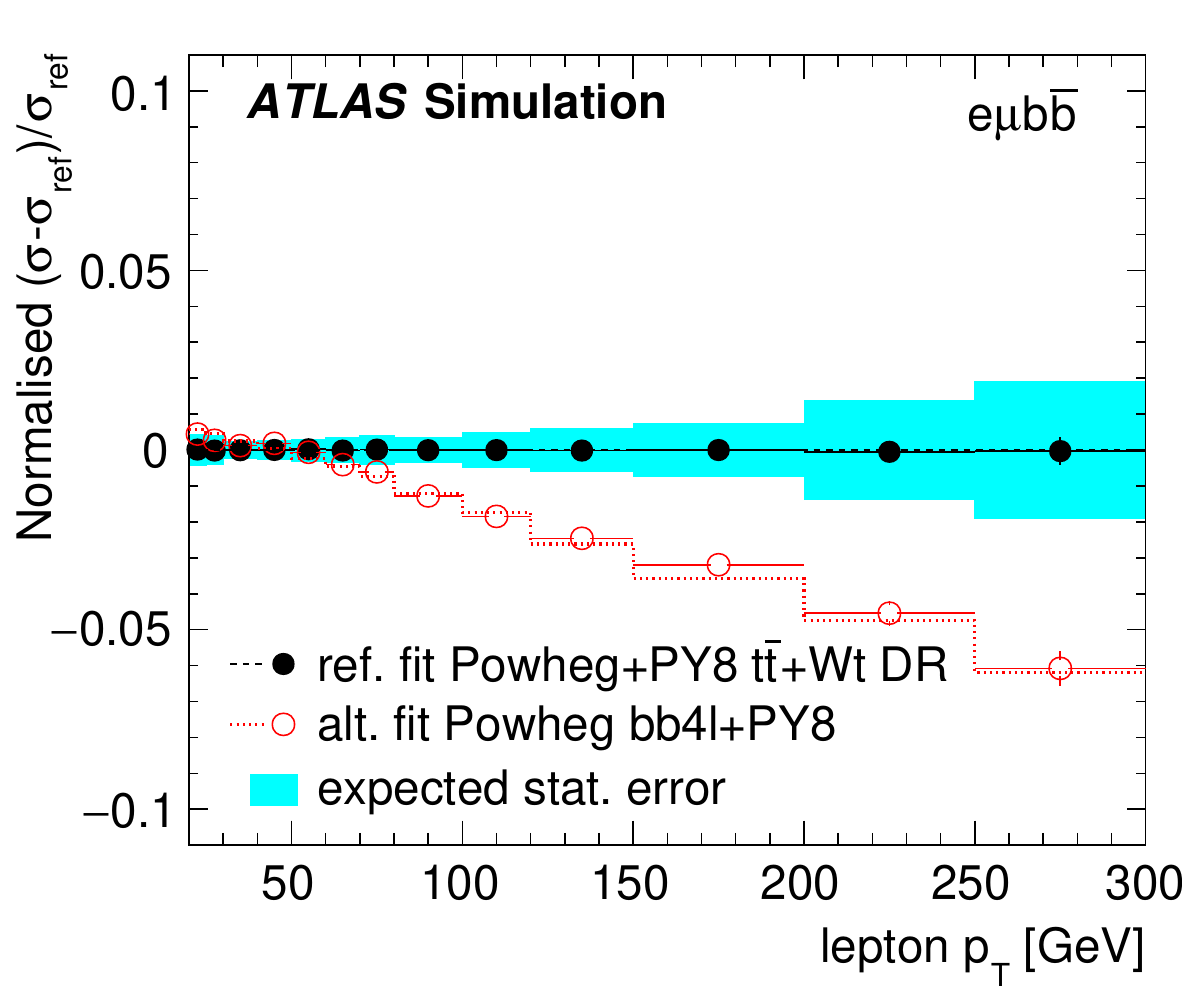}\vspace{-7mm}\center{(c)}}
\parbox{83mm}{\includegraphics[width=76mm]{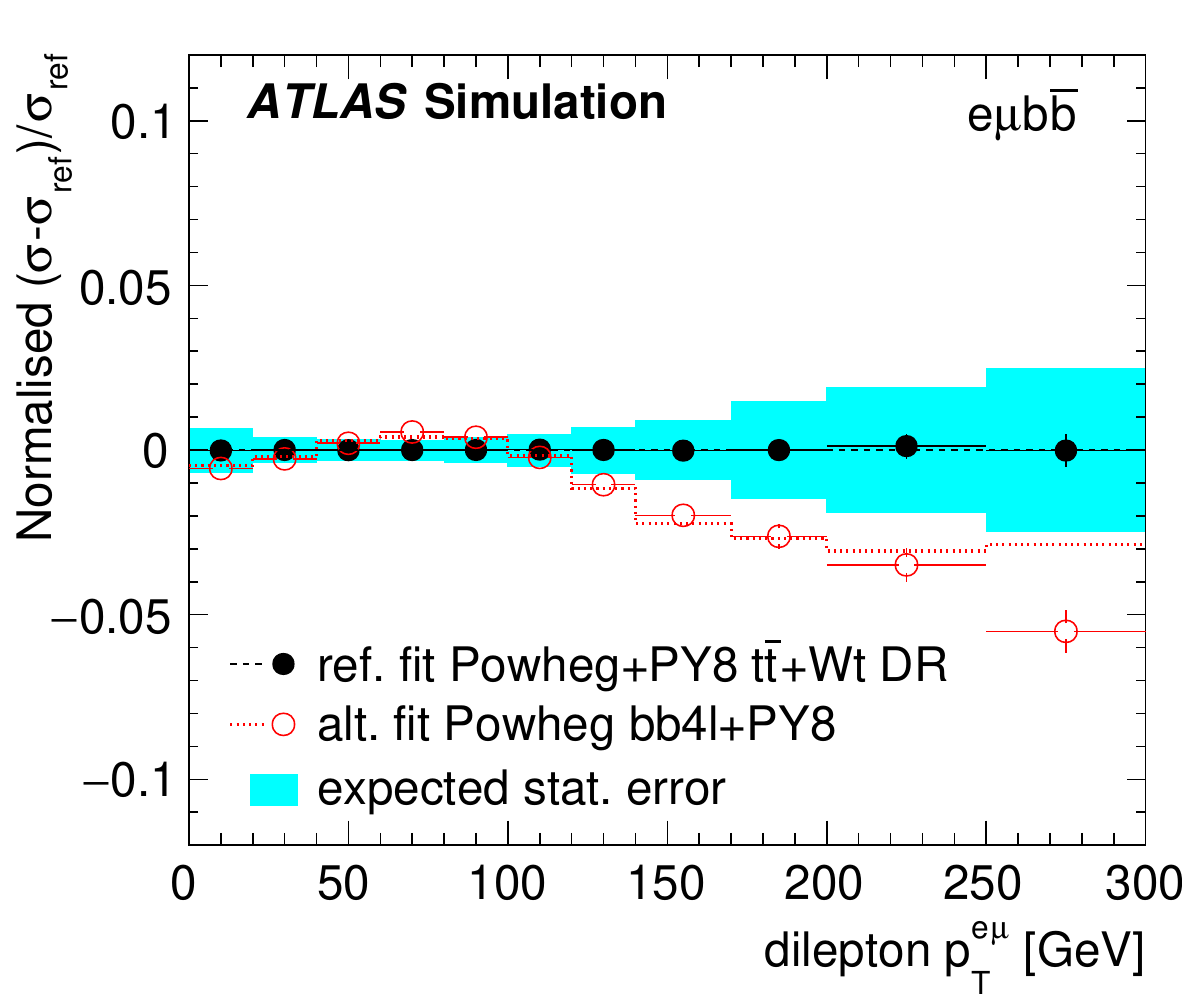}\vspace{-7mm}\center{(d)}}
\caption{\label{f:biastestb}Results of pseudo-experiment studies on simulated
events for the extraction of the normalised \embb\ differential cross-section
distributions, showing (a, c) \ptl\ and (b, d) \ptll.
Plots (a) and (b) use {\scr Powheg\,+\,Pythia8} \ttbar\ and $Wt$ events, with the
diagram removal (DR) scheme used for the reference sample and the diagram
subtraction (DS) scheme for the alternative sample.
Plots (c) and (d) use the same reference sample,
but {\scr Powheg} $bb4l${\scr\,+Pythia8} for the alternative sample.
The interpretation of the points and lines is the same as in
Figure~\ref{f:biastesta}.}
\end{figure}

\section{Systematic uncertainties}\label{s:syst}

Systematic uncertainties in the measured cross-sections arise from
uncertainties in the input quantities appearing in Eqs.~(\ref{e:tags}), namely
the $e\mu$ selection efficiency \epsem, the tagging correlation coefficient \cb,
the non-\ttbar\ background estimates \nib\ and  \niib, and the integrated
luminosity $L$, and from the corresponding quantities
in Eqs.~(\ref{e:fidtags}) and~(\ref{e:xembbi}) for the \ttem\ and
\embb\ differential cross-sections. Each uncertainty was evaluated
by changing all relevant input quantities coherently and re-solving
the tagging equations, thus taking into account correlations between
the different inputs, and between different bins in the differential
analyses. Partial correlations of systematic effects between bins in the
differential analyses (e.g.\ for PDF and lepton efficiency uncertainties)
were propagated to the final results. The sources of systematic uncertainty
are divided into the five groups discussed below, and are shown
for the inclusive and fiducial \ttbar\ cross-sections in Table~\ref{t:incsyst}.
The uncertainties for the normalised single-differential \ttem\ cross-sections
are shown graphically in Figure~\ref{f:ttemsyst}, and those for the
\embb\ cross-sections in Figure~\ref{f:embbsyst}.

\begin{table}[tp]
\caption{\label{t:incsyst}Breakdown of the relative systematic uncertainties
in \epsem, \gem\ and \cb, and the statistical, systematic (excluding
luminosity and beam energy) and total uncertainties in the inclusive
and fiducial \ttbar\ cross-section measurements. The five groups
of systematic uncertainties corresponding to the discussion in
Sections~\ref{ss:sytt} to~\ref{ss:sylumieb} are indicated in the leftmost
column.}
\centering

\begin{tabular}{llccccc}\toprule
Group & Uncertainty source & $\Delta\epsem/\epsem$ & $\Delta\gem/\gem$ &
$\Delta\cb/\cb$ & $\Delta\xtt/\xtt$ & $\Delta\xfid/\xfid$ \\
&  & (\%) & (\%) & (\%) & (\%) & (\%) \\
\midrule
&                        Data statistics &      &      &      & 0.16 & 0.16 \\
\ttbar\ mod. &       \ttbar\ matrix element matching & 0.01 & 0.02 & 0.02 & 0.01 & 0.00 \\
&                   QCD scale variation & 0.29 & 0.07 & 0.02 & 0.31 & 0.09 \\
&                 \ttbar\ hadronisation & 0.23 & 0.06 & 0.06 & 0.29 & 0.00 \\
&         Initial/final state radiation & 0.04 & 0.03 & 0.09 & 0.16 & 0.17 \\
&      \ttbar\ heavy-flavour production & 0.00 & 0.00 & 0.17 & 0.17 & 0.17 \\
&         Parton distribution functions & 0.47 & 0.05 &  -   & 0.47 & 0.08 \\
&                 Simulation statistics & 0.07 & 0.04 & 0.05 & 0.05 & 0.04 \\
Lept.&           Electron identification & 0.12 & 0.12 &  -   & 0.14 & 0.14 \\
&                   Muon identification & 0.13 & 0.13 &  -   & 0.14 & 0.14 \\
&                Electron charge mis-id & 0.05 & 0.05 &  -   & 0.06 & 0.06 \\
&                        Lepton trigger & 0.10 & 0.10 & 0.00 & 0.11 & 0.11 \\
&                 Electron energy scale & 0.04 & 0.04 & 0.00 & 0.05 & 0.05 \\
&            Electron energy resolution & 0.01 & 0.01 & 0.00 & 0.01 & 0.01 \\
&                   Muon momentum scale & 0.01 & 0.01 & 0.00 & 0.02 & 0.02 \\
&              Muon momentum resolution & 0.00 & 0.00 & 0.00 & 0.00 & 0.00 \\
&                    Electron isolation & 0.10 & 0.10 &  -   & 0.11 & 0.11 \\
&                        Muon isolation & 0.12 & 0.12 &  -   & 0.13 & 0.13 \\
Jet/$b$ &                      Jet energy scale &  -   &  -   & 0.05 & 0.02 & 0.02 \\
&                 Jet energy resolution &  -   &  -   & 0.03 & 0.06 & 0.06 \\
&                       Pileup jet veto &  -   &  -   & 0.00 & 0.02 & 0.02 \\
&                $b$-tagging efficiency &  -   &  -   & 0.01 & 0.13 & 0.13 \\
&                    $b$-tag mistagging &  -   &  -   & 0.02 & 0.02 & 0.02 \\
Bkg. &              Single-top cross-section &  -   &  -   &  -   & 0.40 & 0.40 \\
&       Single-top/\ttbar\ interference &  -   &  -   &  -   & 0.28 & 0.28 \\
&                  Single-top modelling &  -   &  -   &  -   & 0.30 & 0.30 \\
&                $Z$+jets extrapolation &  -   &  -   &  -   & 0.02 & 0.02 \\
&                Diboson cross-sections &  -   &  -   &  -   & 0.01 & 0.01 \\
&                     Diboson modelling &  -   &  -   &  -   & 0.05 & 0.05 \\
&                 Misidentified leptons &  -   &  -   &  -   & 0.18 & 0.18 \\
\midrule
&                  Analysis systematics & 0.66 & 0.29 & 0.22 & 0.97 & 0.74 \\
\midrule
$L/E_\mathrm{b}$ &                 Integrated luminosity &  -   &  -   &  -   & 0.89 & 0.89 \\
&                           Beam energy &  -   &  -   &  -   & 0.23 & 0.23 \\
\midrule
&                     Total uncertainty & 0.66 & 0.29 & 0.22 & 1.34 & 1.19 \\
\bottomrule
\end{tabular}
\end{table}

\begin{figure}[tp]
\vspace{-7mm}
\parbox{83mm}{\includegraphics[width=65mm]{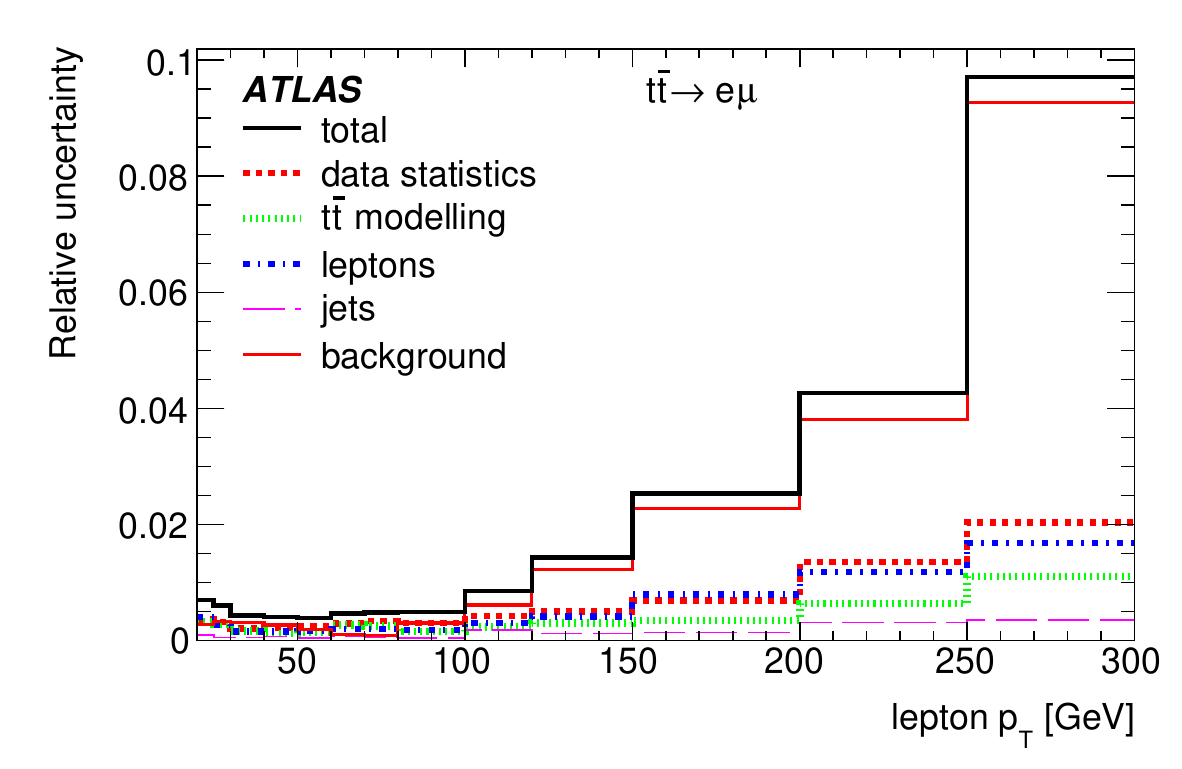}\vspace{-7mm}\center{(a)}}
\parbox{83mm}{\includegraphics[width=65mm]{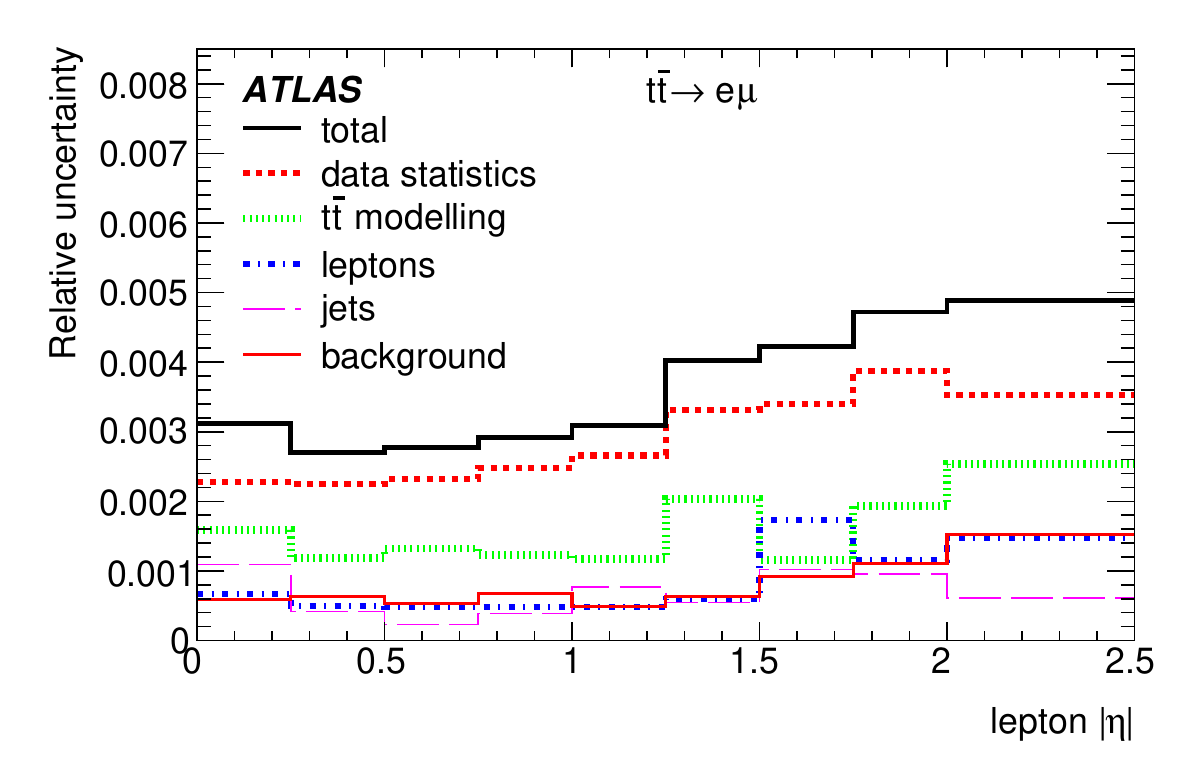}\vspace{-7mm}\center{(b)}}
\parbox{83mm}{\includegraphics[width=65mm]{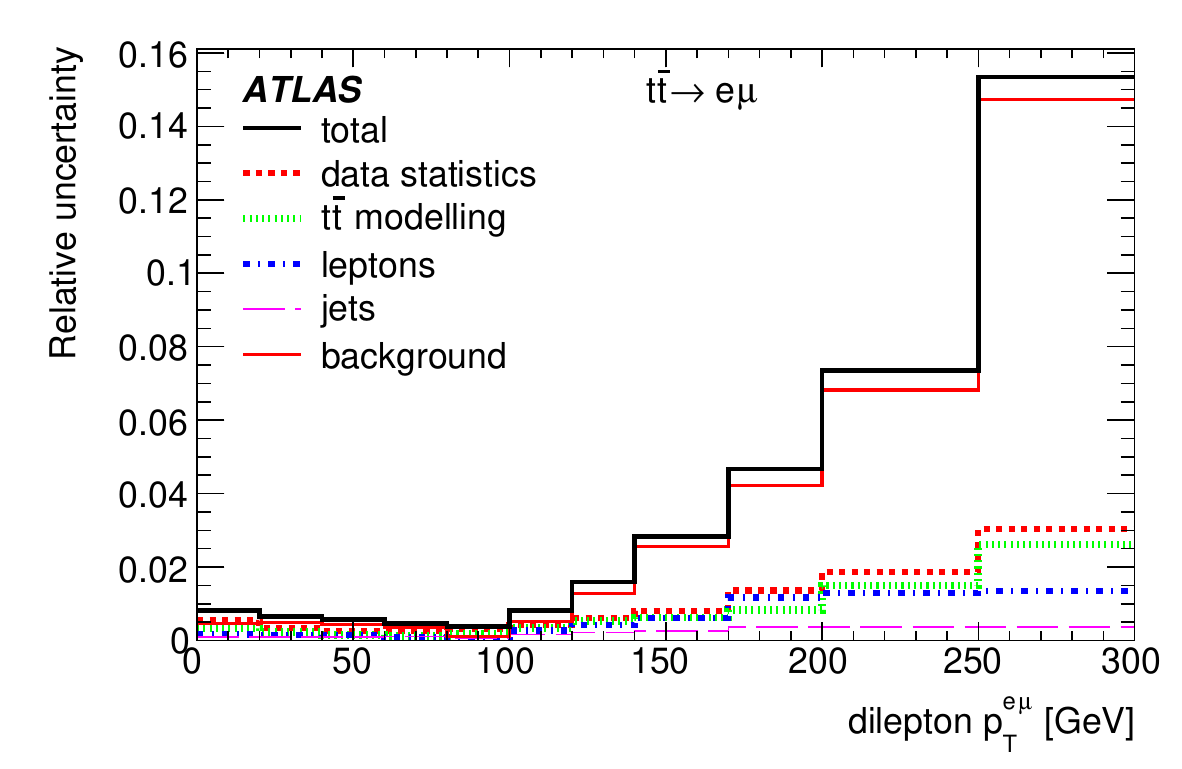}\vspace{-7mm}\center{(c)}}
\parbox{83mm}{\includegraphics[width=65mm]{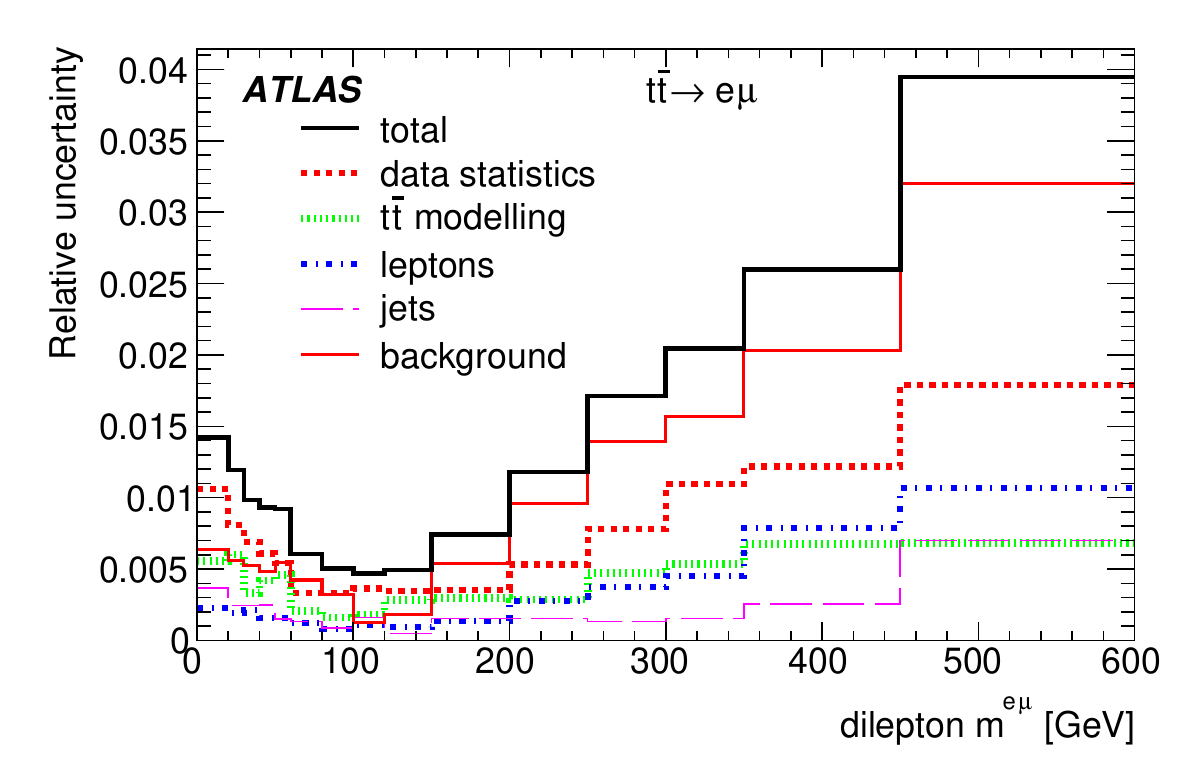}\vspace{-7mm}\center{(d)}}
\parbox{83mm}{\includegraphics[width=65mm]{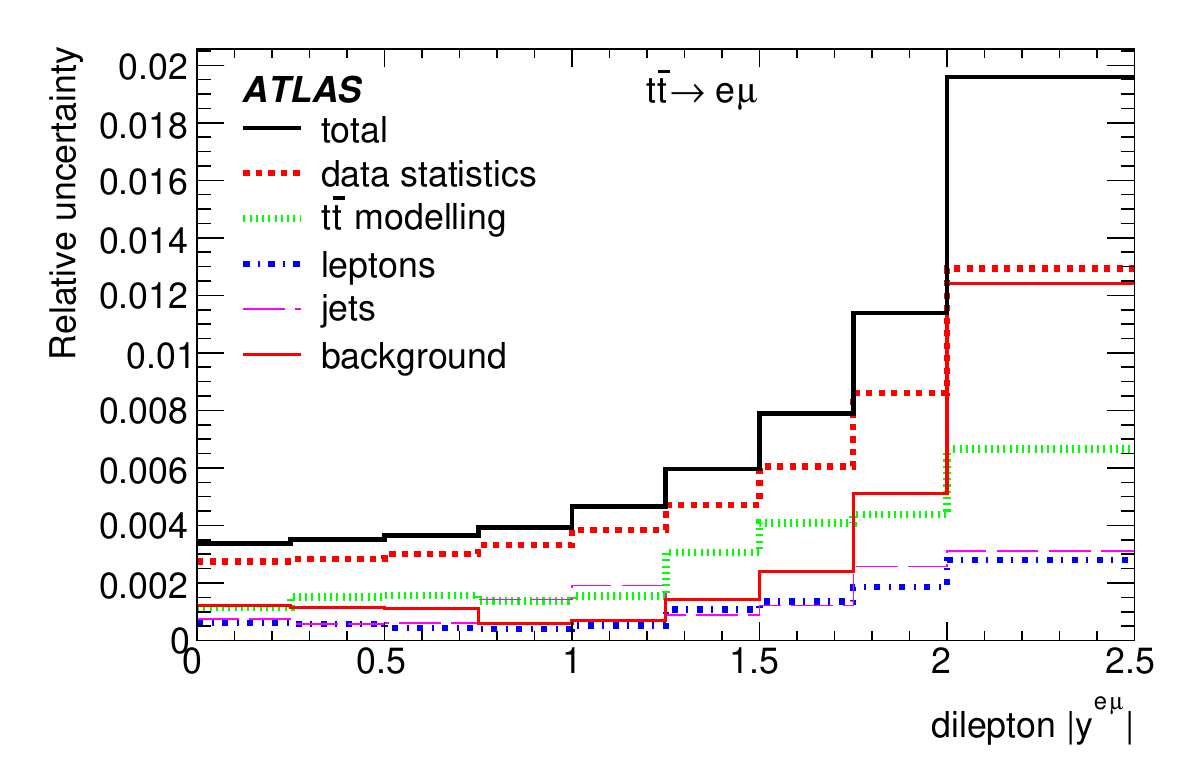}\vspace{-7mm}\center{(e)}}
\parbox{83mm}{\includegraphics[width=65mm]{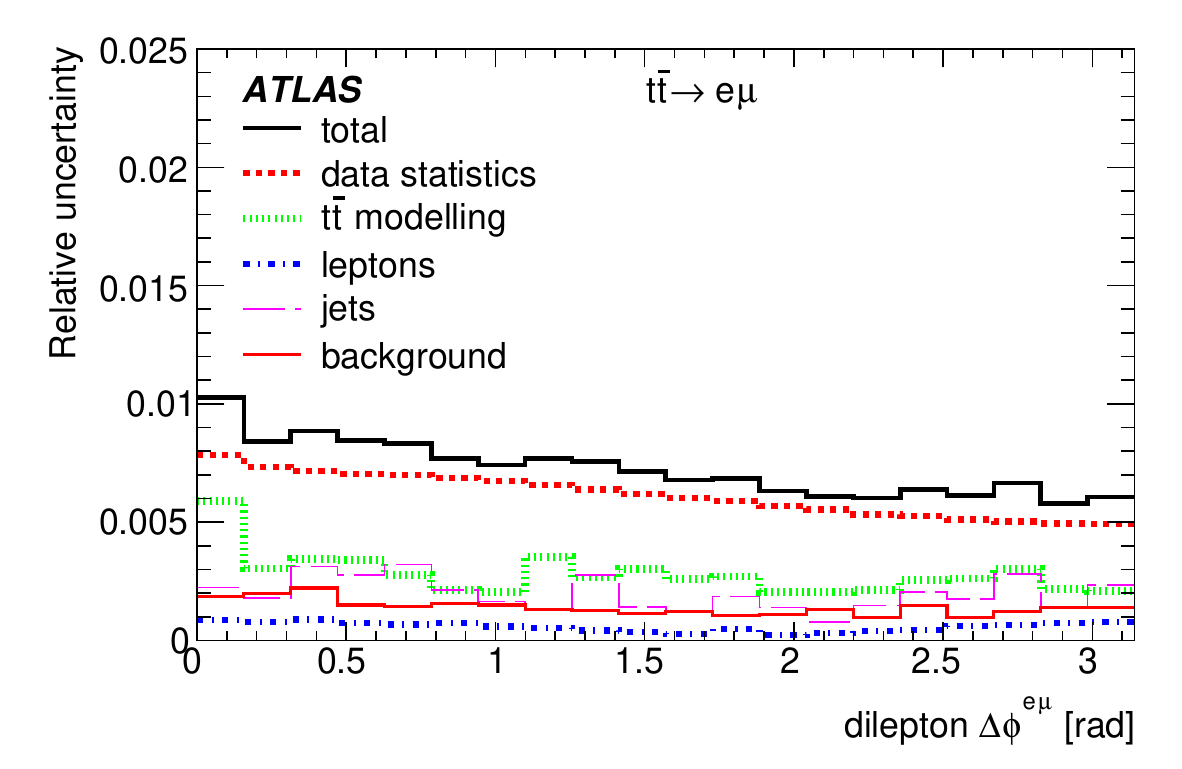}\vspace{-7mm}\center{(f)}}
\parbox{83mm}{\includegraphics[width=65mm]{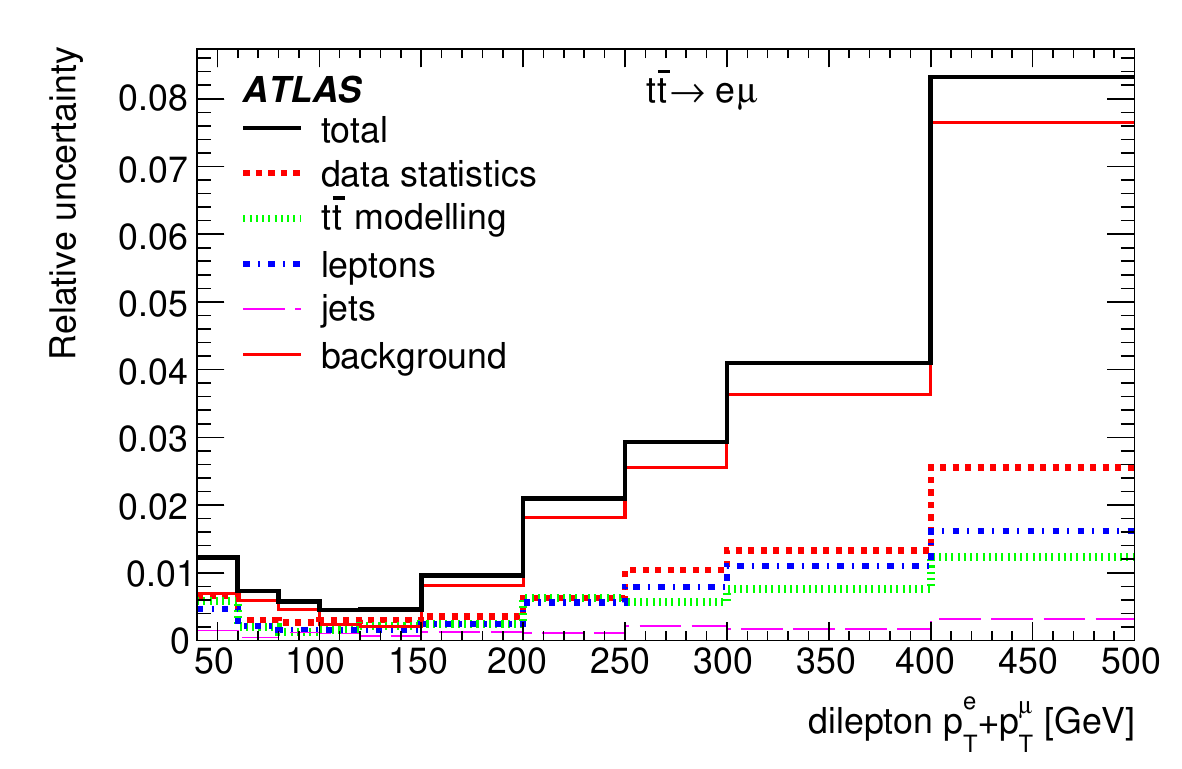}\vspace{-7mm}\center{(g)}}
\parbox{83mm}{\includegraphics[width=65mm]{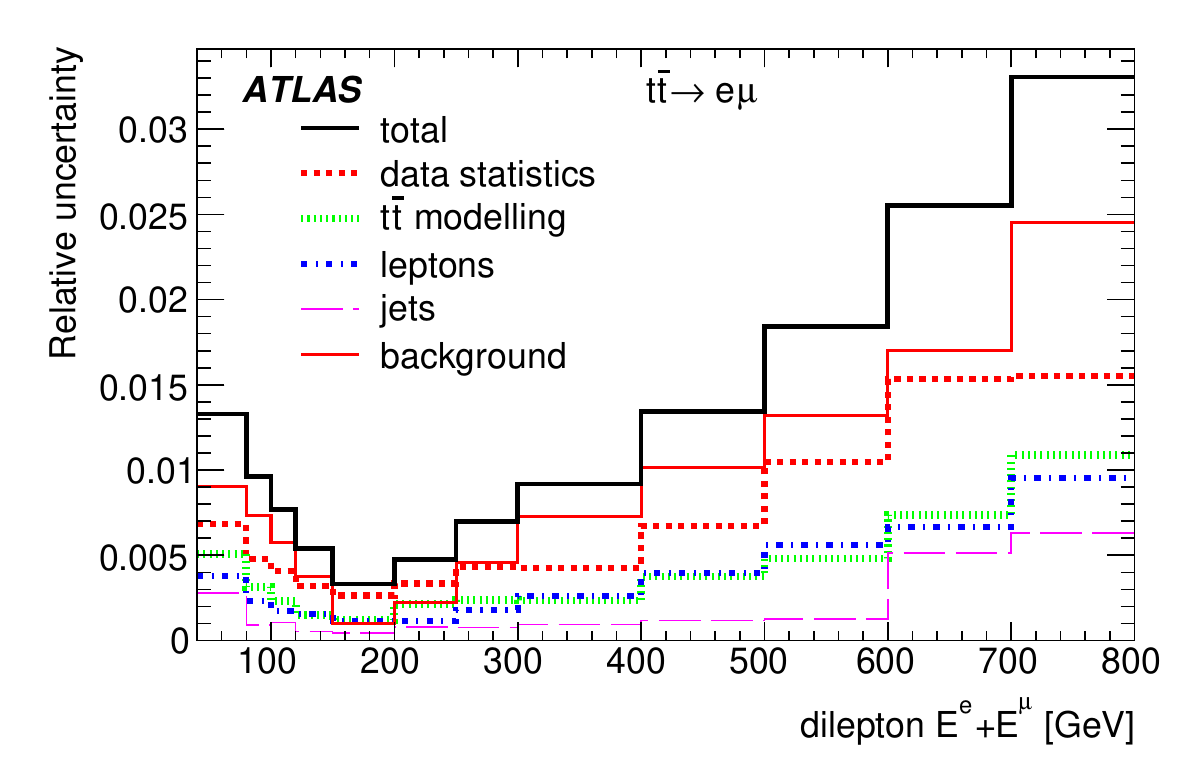}\vspace{-7mm}\center{(h)}}
\parbox{83mm}{\includegraphics[width=65mm]{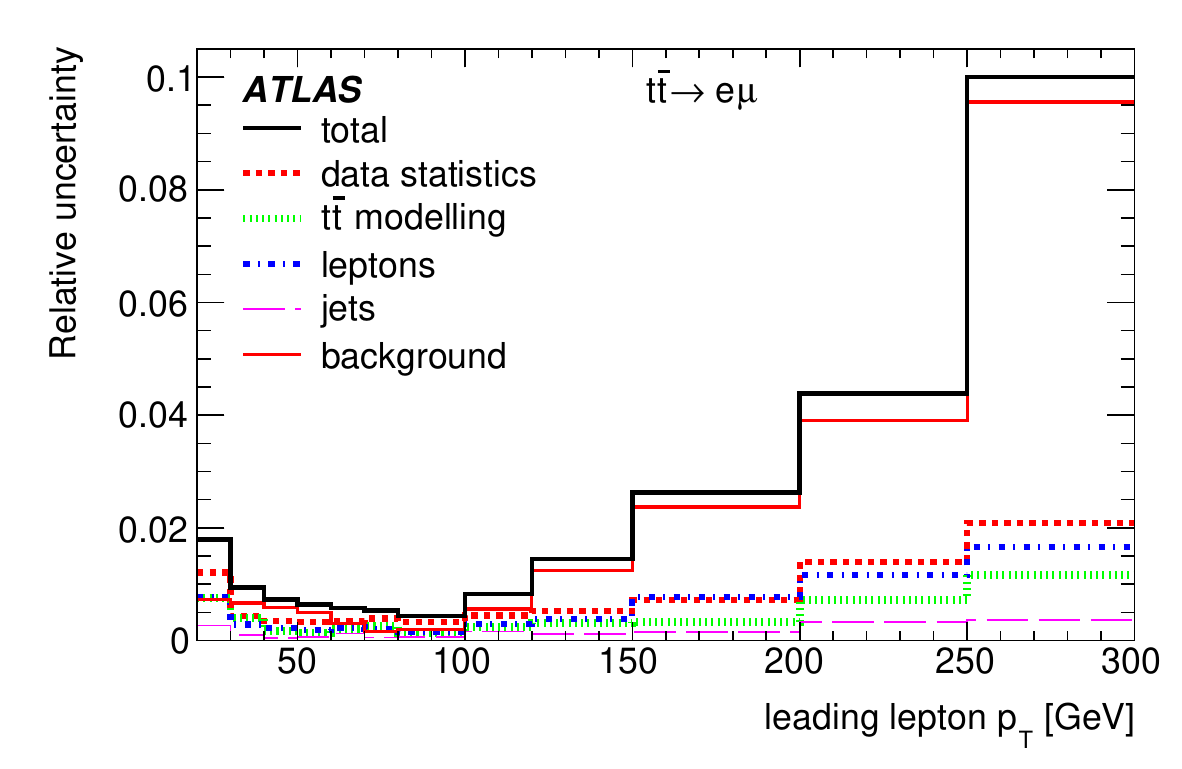}\vspace{-7mm}\center{(i)}}
\parbox{83mm}{\includegraphics[width=65mm]{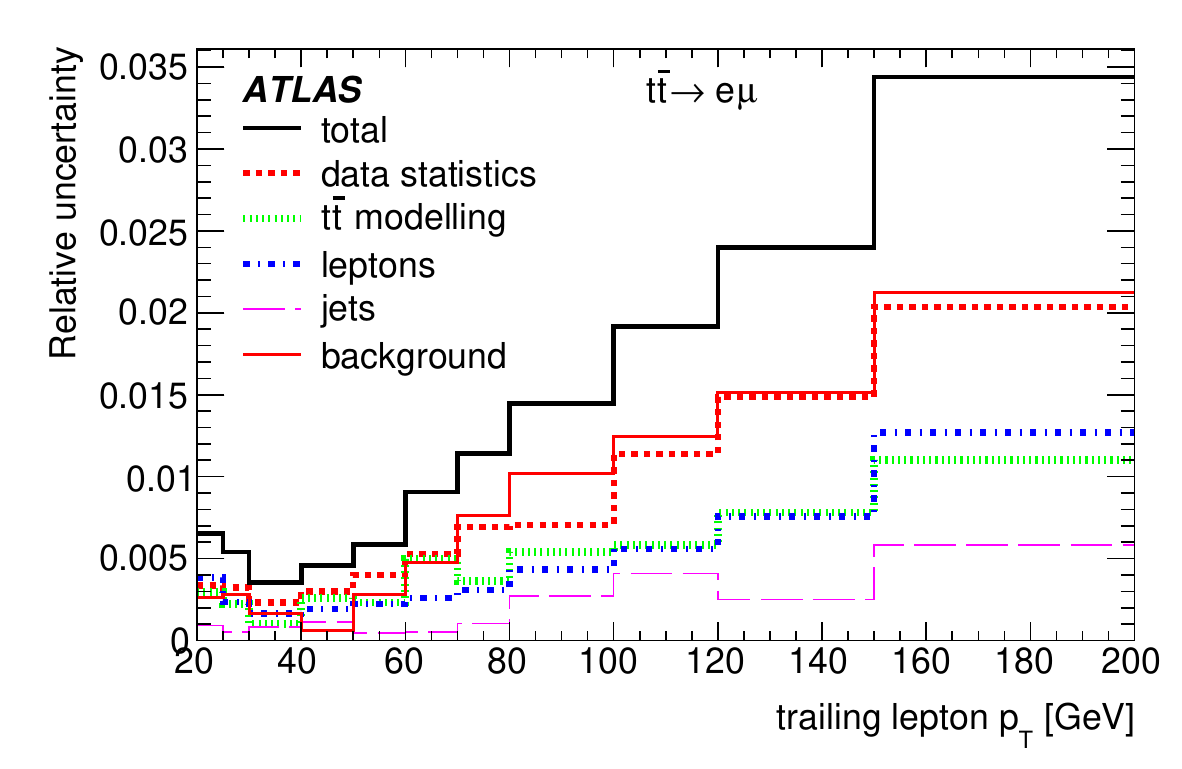}\vspace{-7mm}\center{(j)}}
\caption{\label{f:ttemsyst}Relative uncertainties in the measured
one-dimensional normalised \ttem\
differential cross-sections coming from data statistics,
\ttbar\ modelling, leptons, jets and background,  as a function of each lepton
or dilepton
differential variable. The total uncertainty is shown by the thick black lines,
and also includes small contributions from the integrated luminosity and
LHC beam energy uncertainties.}
\end{figure}

\begin{figure}[tp]
\vspace{-7mm}
\parbox{83mm}{\includegraphics[width=65mm]{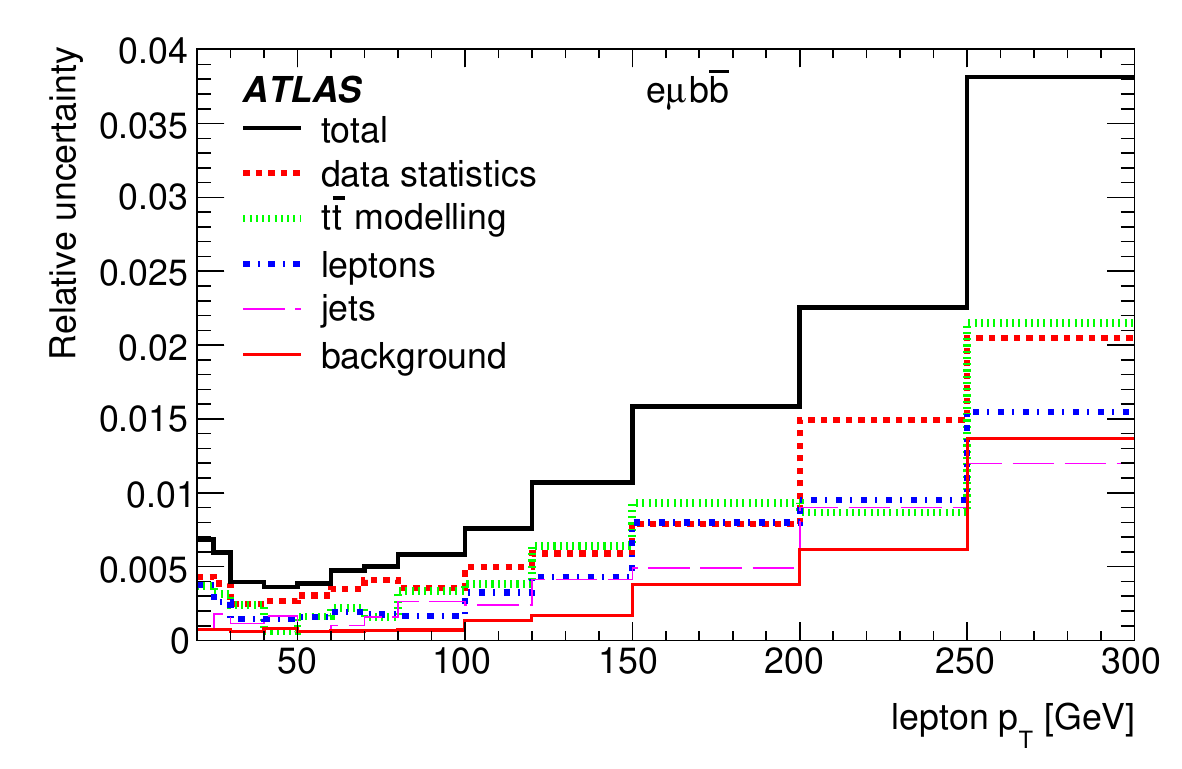}\vspace{-7mm}\center{(a)}}
\parbox{83mm}{\includegraphics[width=65mm]{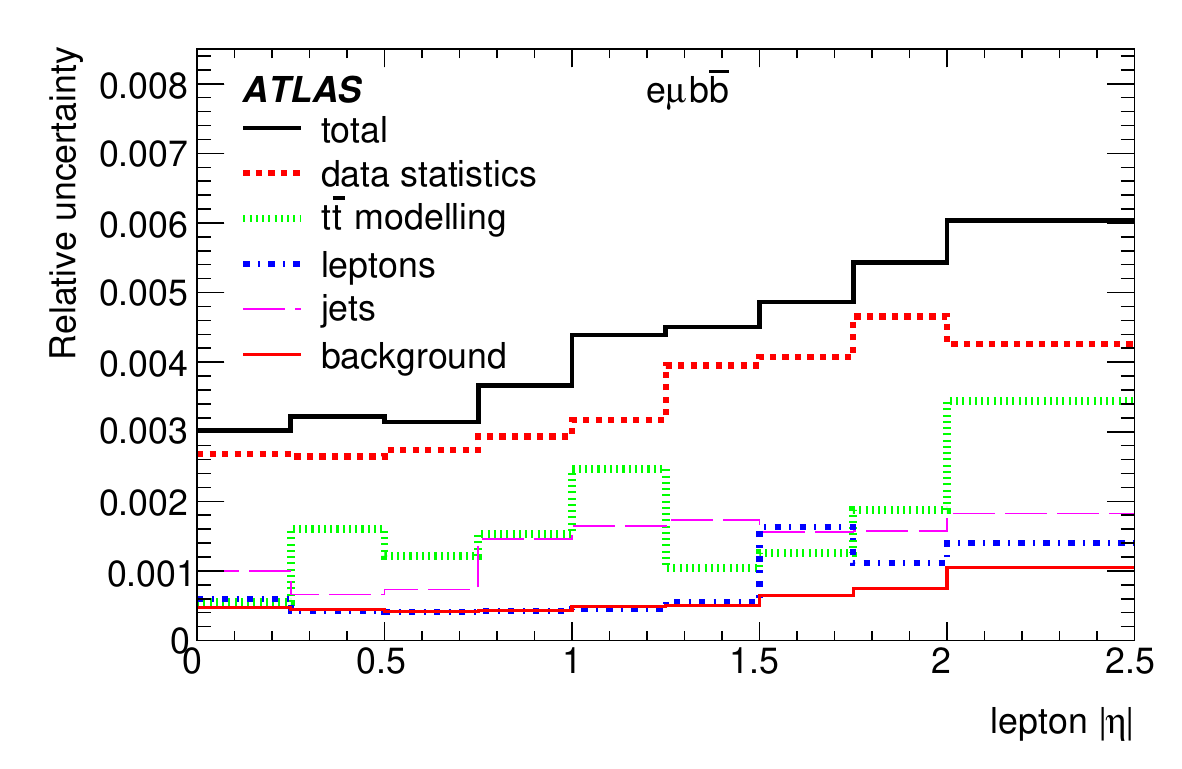}\vspace{-7mm}\center{(b)}}
\parbox{83mm}{\includegraphics[width=65mm]{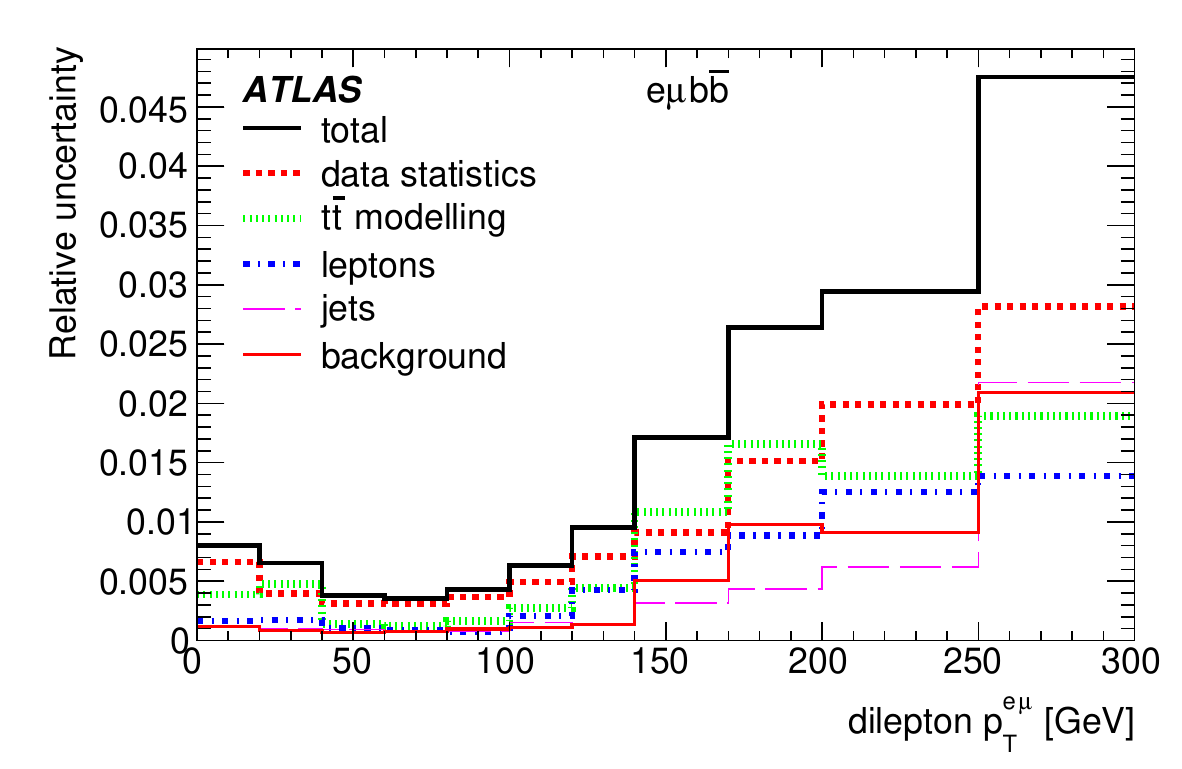}\vspace{-7mm}\center{(c)}}
\parbox{83mm}{\includegraphics[width=65mm]{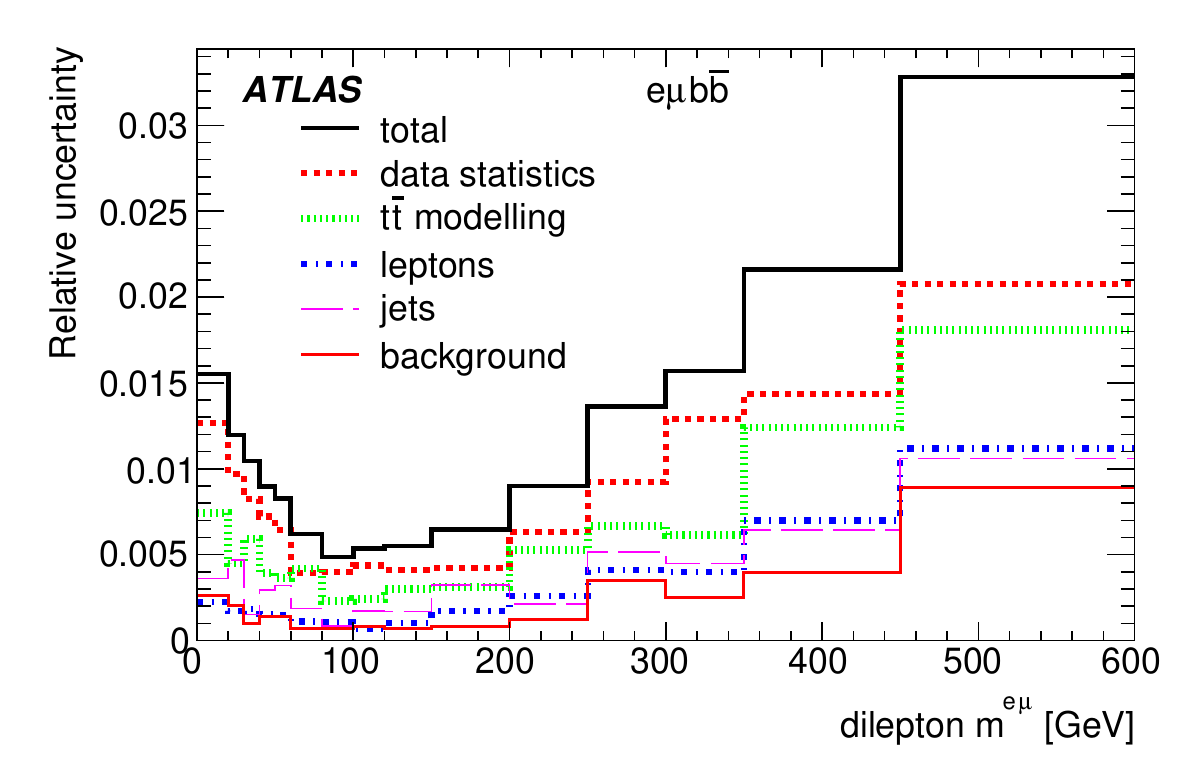}\vspace{-7mm}\center{(d)}}
\parbox{83mm}{\includegraphics[width=65mm]{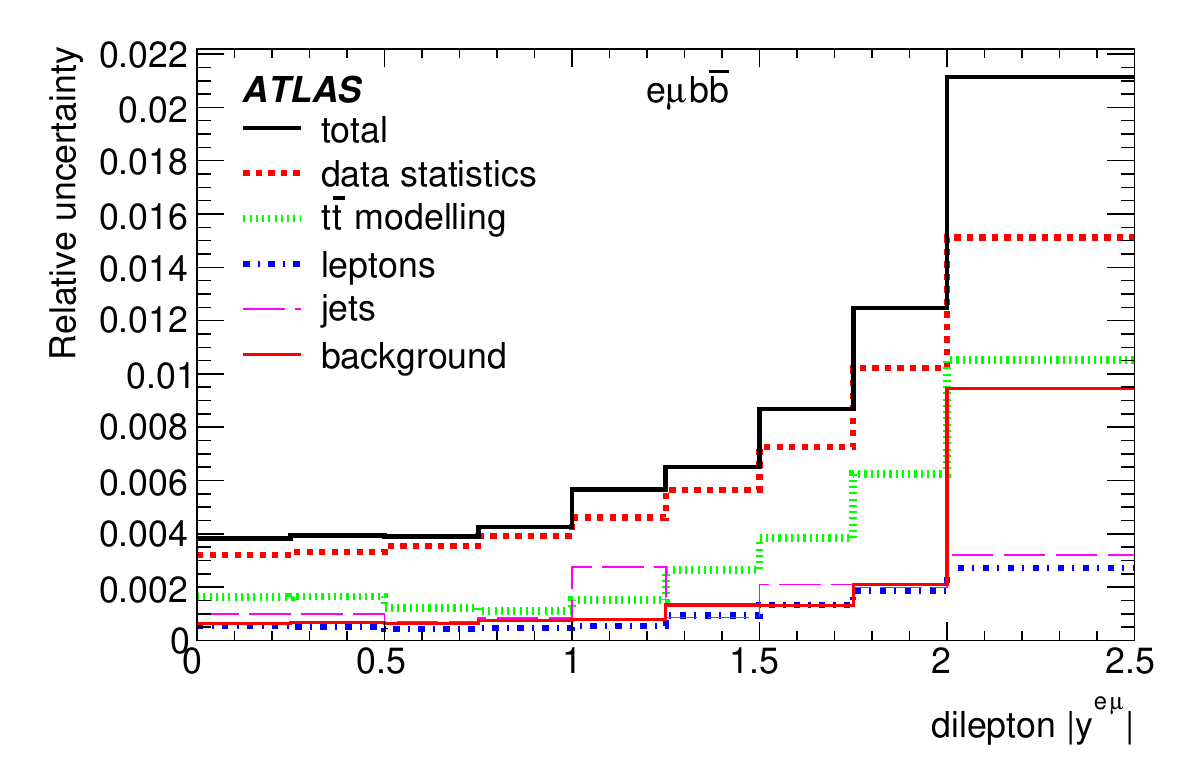}\vspace{-7mm}\center{(e)}}
\parbox{83mm}{\includegraphics[width=65mm]{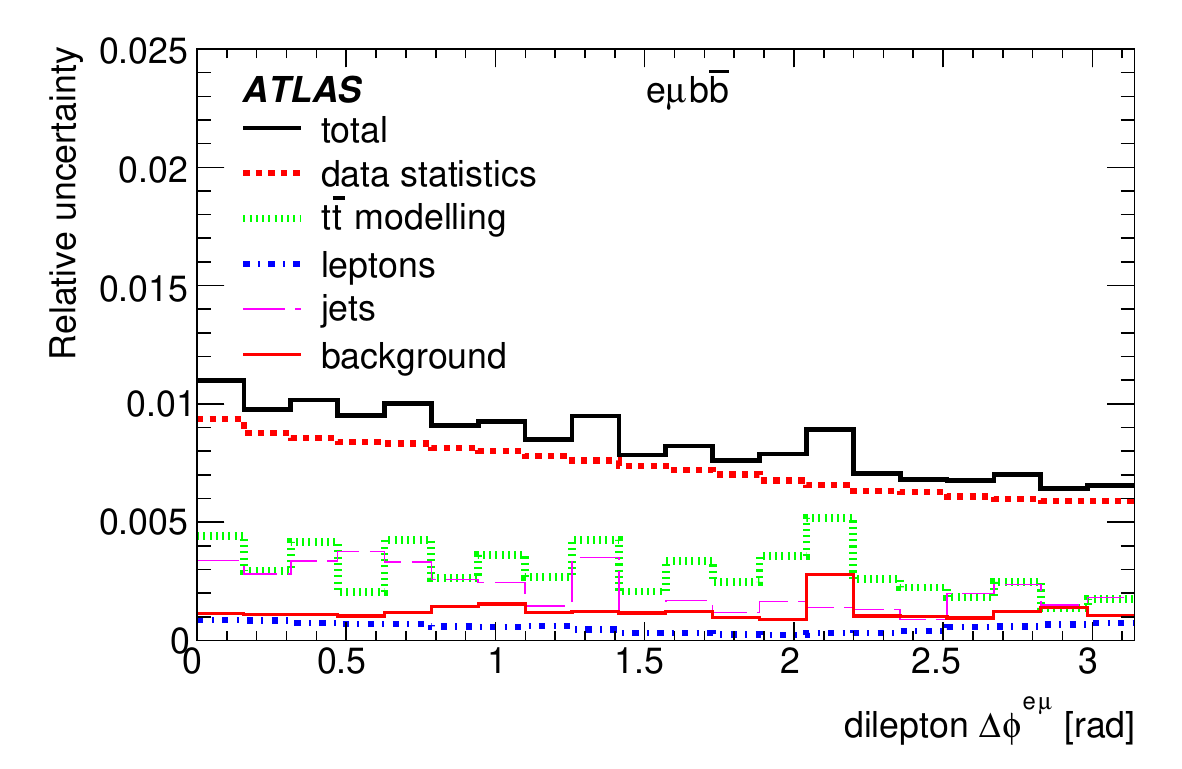}\vspace{-7mm}\center{(f)}}
\parbox{83mm}{\includegraphics[width=65mm]{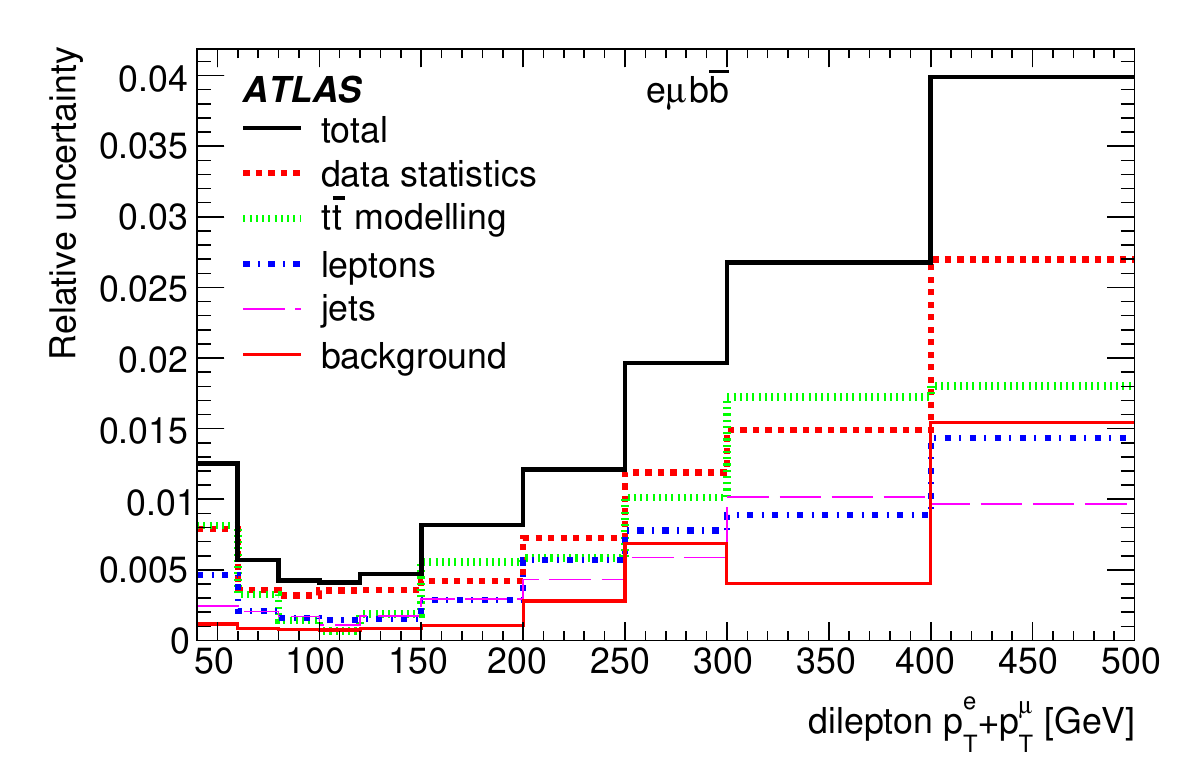}\vspace{-7mm}\center{(g)}}
\parbox{83mm}{\includegraphics[width=65mm]{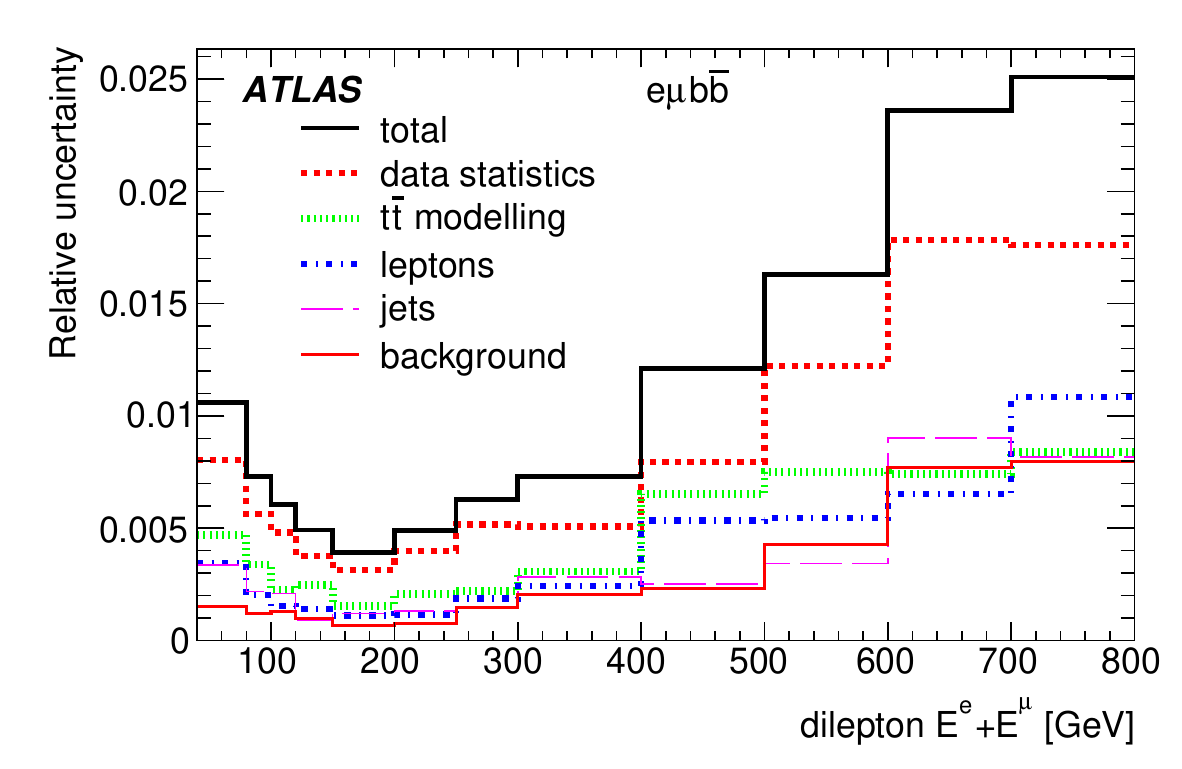}\vspace{-7mm}\center{(h)}}
\parbox{83mm}{\includegraphics[width=65mm]{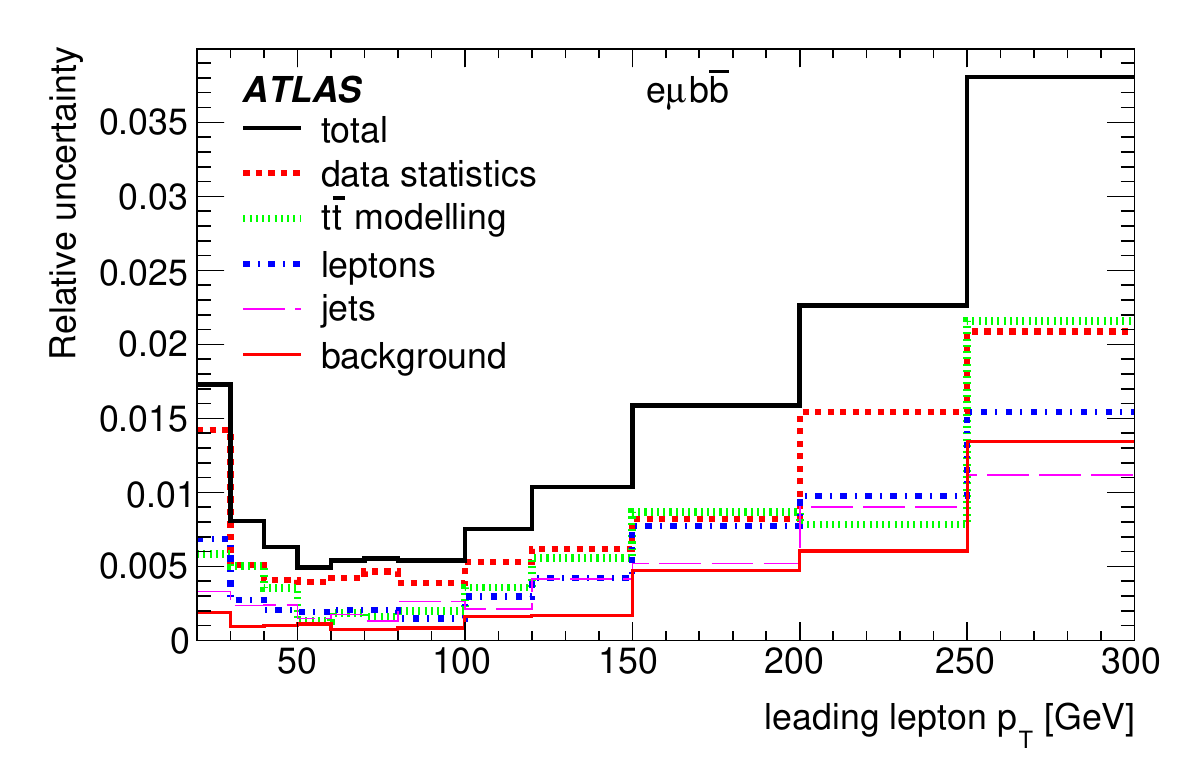}\vspace{-7mm}\center{(i)}}
\parbox{83mm}{\includegraphics[width=65mm]{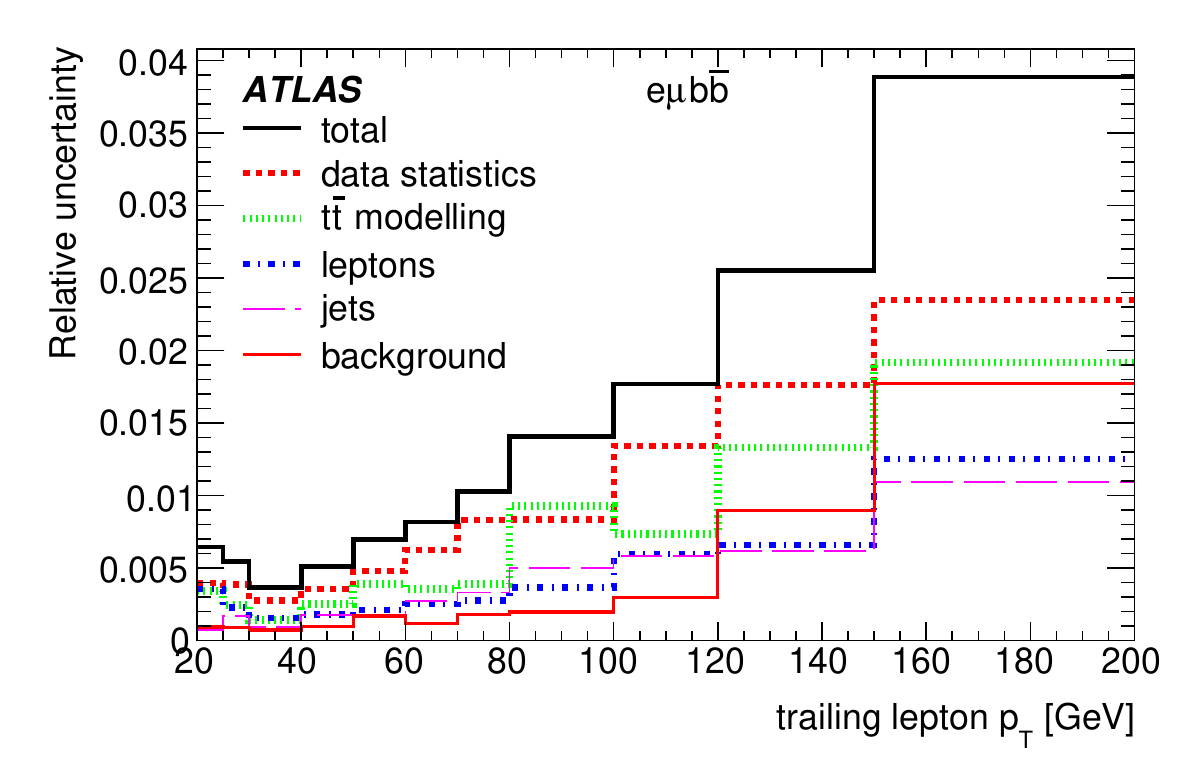}\vspace{-7mm}\center{(j)}}
\caption{\label{f:embbsyst}Relative uncertainties in the measured
one-dimensional normalised \embb\
differential cross-sections coming from data statistics,
\ttbar\ modelling, leptons, jets and background,  as a function of each lepton
or dilepton
differential variable. The total uncertainty is shown by the thick black lines,
and also includes small contributions from the integrated luminosity and
LHC beam energy uncertainties.}
\end{figure}

\subsection{\ttbar\ modelling}\label{ss:sytt}

The \ttbar\ modelling uncertainties in \epsem, \gem, \gemi, \gemembbi, \cb\ and
\cbi\ (and \fntaui\ for the $\tau$-corrected cross-sections) were evaluated by
reweighting the baseline \powhegminnlo\ \ttbar\ sample and by using the various
alternative samples described in Section~\ref{s:datmc}. Specifically, the \ttbar\
matrix element matching uncertainty was evaluated by comparing the
{\scr Powheg\,+\,Pythia8} sample generated with the setting $\pthard=1$,
which changes the definition of the shower veto region, with the nominal
$\pthard=0$ {\scr Powheg\,+\,Pythia8} sample.\footnote{Since only the baseline
sample with nominal settings was available for {\scr Powheg} MiNNLO,
uncertainty estimates requiring
comparison to alternative samples were assessed by evaluating variations
with respect to the {\scr Powheg\,+\,Pythia8} sample with nominal settings,
and applying
the same relative variation to the efficiencies and tagging correlations
obtained from the \powhegminnlo\ sample.} The uncertainties due to unknown
higher-order corrections in the matrix element were assessed by changing
the QCD renormalisation and factorisation scales through event weights
in the baseline \powhegminnlo\ sample, taking
half the difference between $(\mur,\muf)=(0.5,1)$ and $(2,1)$ variations
and half the difference between
$(\mur,\muf)=(1,0.5)$ and $(1,2)$ variations. These changes
were treated as being independent and added in quadrature. The uncertainty
due to the choice of hadronisation, underlying event and parton shower model was
assessed from the difference between {\scr Powheg} \ttbar\ samples
interfaced to {\scr Herwig7.1.3} instead of {\scr Pythia8}.
The initial/final-state radiation uncertainty was evaluated by reweighting
the \powhegminnlo\ sample according to the {\ttr var3c} A14 tune
variations~\cite{ATL-PHYS-PUB-2014-021} and by variations of the parameter
\murfsr\ controlling the amount of final-state radiation in the {\scr Pythia8}
shower by factors of 2 and 0.625 with respect to its default
value~\cite{TOPQ-2019-03}. All these uncertainties were evaluated without
applying the lepton isolation requirements, to avoid double-counting efficiency
differences absorbed in the lepton isolation efficiency scale factors
described in Section~\ref{ss:sylept} below.

The values of \cb\ and \cbi\ are sensitive to the fraction of \ttbar\
events with extra \bbbar\ pairs, which increase the probability of
$b$-tagging at least two jets in the event, and also contribute to the
rate of events with three or more $b$-tagged jets. As can be seen from the
purple dot-dashed line in
Figure~\ref{f:dmcjlept}(a), \powhegminnlo\ underestimates this rate,
as also seen for {\scr Powheg\,+\,Pythia8}
in dedicated analyses~\cite{TOPQ-2019-03,TOPQ-2017-12}. The discrepancy was
quantified through the ratio of event counts
$\rtt=N(\nbtag\geq 3)/N(\nbtag\geq 2)$, measured
to be $2.63\pm 0.03$\,\% in data compared with $2.02\pm 0.01$\,\% in
simulation, where the uncertainties are statistical only. The fraction of
simulated \ttbar\ events with at least three particle-level $b$-jets was
varied by reweighting, defining
such $b$-jets as particle level jets reconstructed as discussed in
Section~\ref{ss:embbdiff} with $\pt>10$\,\GeV\ and matched within $\Delta R<0.3$
to a weakly decaying $b$-hadron with $\pt>5$\,\GeV. These studies showed
a linear relationship between \rtt\ and \cb, and that increasing
the fraction of events with at least three particle-level $b$-jets by 50\%
brings \rtt\ into agreement with data. The reweighted simulation
models the kinematics of the third-highest \pt\ $b$-tagged jet well, and
increases the value of \cb\ for the inclusive cross-section analysis by
0.35\%. Half the size of this correction was taken as the corresponding
uncertainty (shown as `\ttbar\ heavy-flavour production'
in Table~\ref{t:incsyst}), based on the modelling of the kinematics of the
third- and forth-highest \pt\ $b$-tagged jets.
The shifts in \cb\ from all the \ttbar\ modelling
variations discussed above were also included, but are small in comparison.

Parton distribution function uncertainties were evaluated by reweighting
the \powhegminnlo\ \ttbar\ sample to the 100~variations (replicas)
of NNPDF3.0~\cite{nnpdf3} and calculating the RMS of the changes
induced in \epsem, \gem\, \gemi\ and \gemembbi. The resulting uncertainty
is 0.47\% in \epsem, but less than 0.1\% in \gem\ as PDF variations
mainly affect the acceptance rather than the reconstruction efficiency.
The central values from the CT18~\cite{ct18}, MSHT20~\cite{msht20} and
NNPDF3.1~\cite{nnpdf31} PDF sets lie within the NNPDF3.0 uncertainty.

The lepton kinematics, and hence the prediction for \epsem, also depend slightly
on the assumed value of \mtop. This effect was evaluated using five
{\scr Powheg\,+\,Pythia8} \ttbar\ samples with different \mtop\ values in the
range
169--176\,\GeV, and gives a relative change of $\pm 0.4$\% in \epsem\ for
a $\pm 1$\,\GeV\ change in \mtop. In the \xtt\ measurement, this
effect is partially counterbalanced by changes in the $Wt$ background
prediction, which decreases with increasing \mtop. Both effects were
parameterised using second-order polynomials as functions of \mtop,
and their combination gives a relative change of $\dmtopval$\% in \xtt\
for a 1\,\GeV increase in \mtop. By convention, the measured
\xtt\ is quoted at a fixed value of $\mtop=172.5$\,\GeV, but a $\pm 1$\,\GeV\
\mtop\ variation is included in the differential cross-section uncertainties.

The total \ttbar\ modelling uncertainties also include the small contributions
due to the limited size of the \powhegminnlo\ \ttbar\ sample
(`Simulation statistics' in Table~\ref{t:incsyst}) and are
shown for the differential distributions by the green hatched lines in
Figures~\ref{f:ttemsyst} and~\ref{f:embbsyst}.

\subsection{Lepton identification and measurement}\label{ss:sylept}

The lepton identification efficiencies were measured using tag-and-probe
techniques applied to \zee\ and \zmm\ events~\cite{EGAM-2021-01, MUON-2018-03},
as functions of lepton \pt\ and $\eta$ for electrons,
and $\eta$ and $\phi$ for muons.  The corresponding uncertainties
are only partially correlated across \pt, $\eta$ and $\phi$,
and this information was propagated to the final results
by generating multiple sets of scale factors whose
variations represent the full uncertainty model. The uncertainties
due to electron charge misidentification were studied using \zee\
events and taken into account using the same technique.
The lepton trigger efficiencies were also measured
in \zee\ and \zmm\ events using tag-and-probe
techniques~\cite{TRIG-2018-05,TRIG-2018-01}, giving per-lepton uncertainties of
0.2--1\% for electrons and 0.5--3\% for muons. Since a large fraction
of dilepton events are triggered redundantly by both leptons, the corresponding
uncertainty in \xtt\ is only 0.11\%.
The electron energy scale, muon momentum scale and corresponding resolution
were determined using \zee\ and \zmm\ events~\cite{EGAM-2018-01,MUON-2022-01},
and varied according to the corresponding uncertainties.

The lepton isolation efficiencies were measured directly in \ttbar-dominated
$e\mu$ plus $b$-tagged jet samples by determining the fraction of events
where either the electron or muon fails the isolation cut. After correcting
for background contamination~\cite{TOPQ-2023-28}, the results
as a function of lepton \pt\ for $|\eta|<1.5$ and $|\eta|>1.5$ are shown
in Figure~\ref{f:leptisol}, for data and simulation with \powhegminnlo\
\ttbar\ events. At low \pt, the simulation underestimates the electron
efficiency in data by about 1.5\%, and overestimates the muon efficiency
by 1\%. The ratios of data to simulation efficiencies were used to define
multiplicative efficiency corrections (scale factors) applied
to simulation on a per-lepton basis. The uncertainties in these scale factors
are dominated by the subtraction of the misidentified-lepton background for
the samples of leptons that fail the isolation requirement~\cite{TOPQ-2023-28};
such events are not used in the differential cross-section measurements.
The uncertainties
amount to about 0.1\% per lepton, partially correlated across the
differential distributions. The method was validated using various reweighted
and alternative \ttbar\ simulation samples that predict different isolation
efficiencies, also demonstrating that the sum of the measured electron and muon
efficiency shifts reproduce the isolation-related component of the shifts
in \epsem\ and \gem. The total lepton-related uncertainties are shown by
the blue dot-dashed lines in Figures~\ref{f:ttemsyst} and~\ref{f:embbsyst},
and include the effects on both \ttbar\ and $Wt$ events.

\begin{figure}[tp]
\parbox{83mm}{\includegraphics[width=76mm]{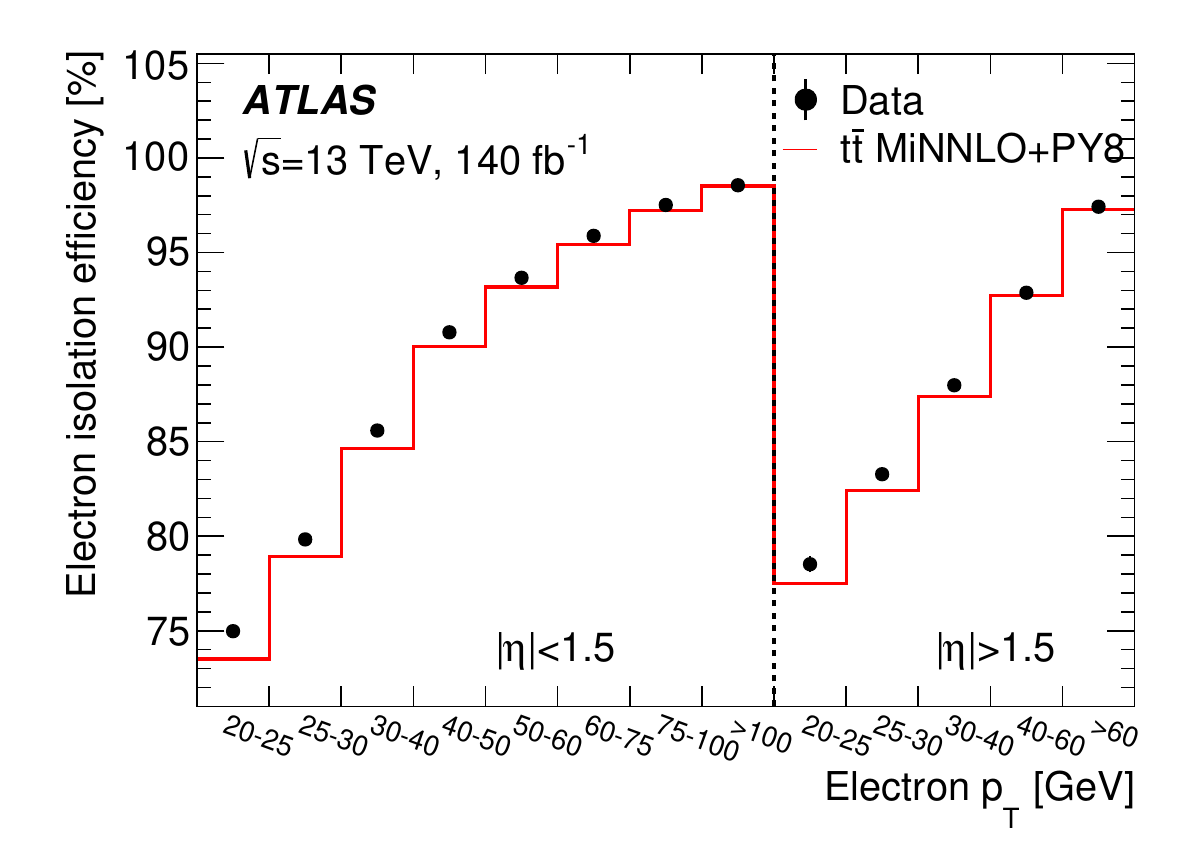}\vspace{-7mm}\center{(a)}}
\parbox{83mm}{\includegraphics[width=76mm]{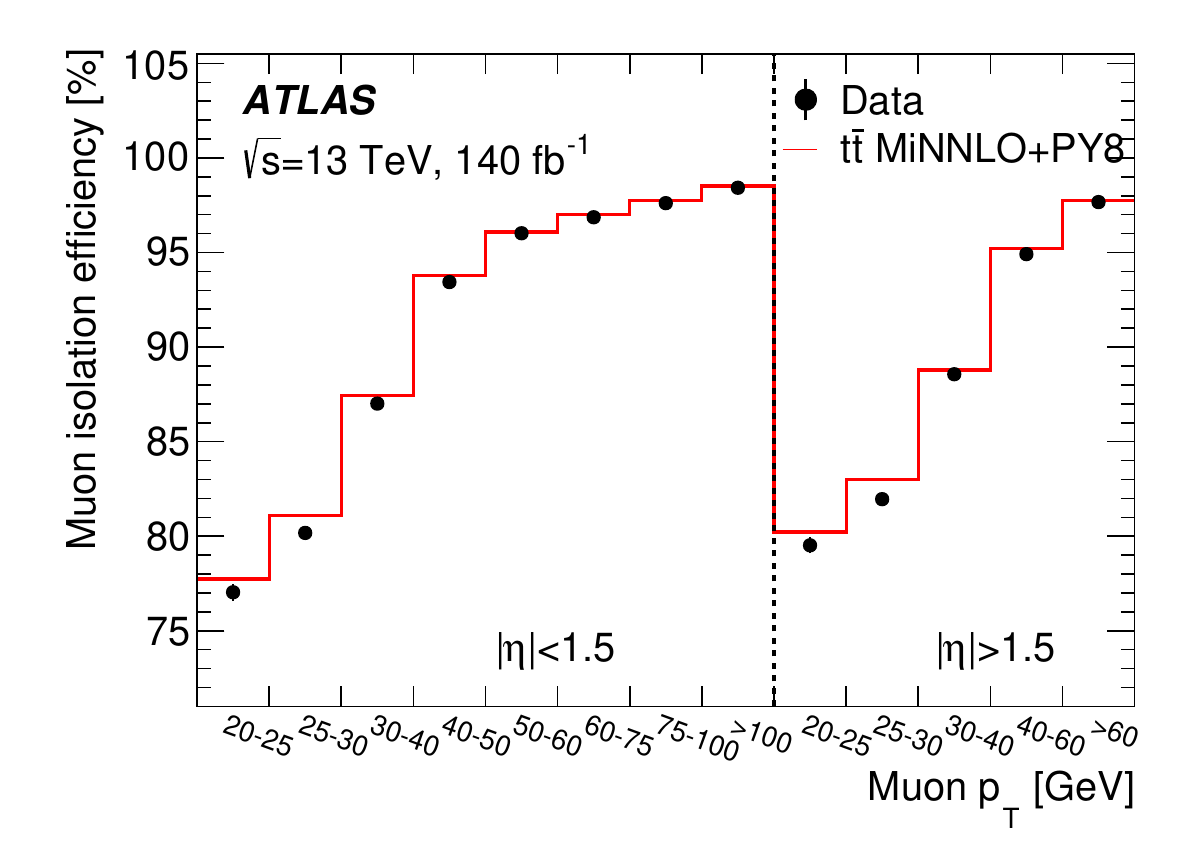}\vspace{-7mm}\center{(b)}}
\caption{\label{f:leptisol}Lepton isolation efficiencies measured
in \ttem\ events for (a)  electrons and (b) muons,
as functions of lepton \pt\ in two bins of $|\eta|$. The
data is shown by the points with error bars
and compared with simulation with \powhegminnlo\ \ttbar\
events without isolation efficiency scale factors, shown by the red histograms.}
\end{figure}

\subsection{Jet measurement and $b$-tagging}\label{ss:syjetb}

Uncertainties in jet reconstruction and calibration affect the estimates
of the background contributions from $Wt$ and diboson events, the values
of \cb\ and \cbi, and the values of \gemembbi\ and \epsbmc\
for the \embb\ differential cross-section analysis.
The jet energy scale was determined using a combination
of simulation, test beam and in-situ measurements, and the jet energy
resolution was evaluated using di-jet balance techniques~\cite{JETM-2018-05}.
The modelling of the pileup jet veto JVT requirement was studied using
jets in \zmm\ events~\cite{PERF-2014-03}.

The efficiency for $b$-tagging jets in \ttbar\ events was extracted from the
data via Eqs.~(\ref{e:tags}), but simulation was used to predict the number
of $b$-tagged jets in $Wt$ and diboson background events, as well as
\gemembbi\ and \epsbmc. The corresponding scale factors and uncertainties
for $b$, charm and light-flavour jets were
determined using \ttbar\ and $Z$+jets
events~\cite{FTAG-2018-01,FTAG-2020-08,FTAG-2019-02}. As these $b$-tagging
efficiency scale factors are used only for background determination in
the \ttbar\ cross-section measurements, the use of the same data sample to
determine them in Ref.~\cite{FTAG-2018-01} does not result in
significant correlation. The uncertainties
related to jets and $b$-tagging are shown by the purple long-dashed
lines in Figures~\ref{f:ttemsyst} and~\ref{f:embbsyst}.

\subsection{Background modelling}\label{ss:sybg}

The normalisation of the $Wt$ background was varied by 3.7\%, corresponding
to the PDF and QCD uncertainties in the prediction discussed
in Section~\ref{s:datmc}. The uncertainty due to $\ttbar/Wt$ interference
was assessed by comparing {\scr Powheg\,+\,Pythia8} $Wt$ samples with the
diagram removal and diagram subtraction schemes~\cite{wtinter1,wtinter2}.
Although small for the inclusive cross-section, this uncertainty is dominant
at the high ends of the lepton \pt\ and dilepton \ptll, \mll, \ptsum, \esum,
\ptlmax\ and \ptlmin\ distributions in the \ttem\ measurements.
The studies of Ref.~\cite{TOPQ-2020-14}
suggest that data lies between the diagram removal and diagram subtraction
predictions in the region where interference becomes important. Further
modelling uncertainties were evaluated by comparing the baseline $Wt$ sample
with samples generated with $\pthard=1$, with $\hdamp=3\mtop$, with
{\scr Powheg\,+\,Herwig7.1}, and by reweighting to change the values
of \mur\ and \muf. These uncertainties were taken to be uncorrelated with
the corresponding \ttbar\ modelling uncertainties. Initial/final state radiation
uncertainties were also evaluated based on {\ttr var3c} A14 tune and \murfsr\
variations, and taken to be correlated with the corresponding \ttbar\
uncertainties. The small background acceptance uncertainties
from NNPDF3.0 variations were also included.

The $Wt$ component is part of the signal in the \embb\ differential analysis,
so the $Wt$ modelling uncertainties were evaluated as changes in \gemembbi,
and correlated with the corresponding variations in the \ttbar\ signal
component where appropriate. In addition, \gemembbi\ depends on the relative
fractions of \ttbar\ and $Wt$ events in the signal, evaluated by separately
changing the \ttbar\ normalisation by 4.0\% and the $Wt$ normalisation by 3.7\%,
based on the predicted cross-sections given in
Sections~\ref{s:intro} and~\ref{s:datmc}.

The $Z(\rightarrow\tau\tau\rightarrow e\mu)$+jets background prediction was
normalised using the corresponding yield ratios in $Z(\rightarrow ee/\mu\mu)$
events as
described in Section~\ref{ss:bkgd}, and the corresponding uncertainties were
propagated to the cross-section results. For the differential analyses,
the changes in predictions as a function of lepton kinematics when performing
the normalisation in bins of $Z$-boson \pt\ rather than inclusively were also
taken into account. The inclusive diboson cross-sections were varied by
6\%, based on MCFM predictions~\cite{mcfm}, and the QCD factorisation
and normalisation scales were varied by reweighting the {\scr Sherpa}
samples. The uncertainties in the background from events with misidentified
leptons, evaluated using the same-charge $e\mu$ samples,
are discussed in Section~\ref{ss:bkgd}. The total background uncertainties
are shown by the red solid lines in Figures~\ref{f:ttemsyst}
and~\ref{f:embbsyst}, and are dominated by those in the $Wt$ background.

\subsection{Luminosity and beam energy}\label{ss:sylumieb}

The integrated luminosity of the dataset was evaluated using the LUCID2
detector~\cite{LUCID2}, complemented by measurements from the inner detector
and calorimeters, and has an uncertainty of 0.83\%~\cite{DAPR-2021-01}.
In the inclusive cross-section analysis, an optimised combination of results
from the separate 2015+16, 2017 and 2018 datasets is used as discussed in
Section~\ref{ss:ixres}, slightly reducing the effective luminosity uncertainty.
For
both inclusive and differential analyses, the luminosity-related uncertainty
in the cross-section values is around 10\% larger than that in the integrated
luminosity itself, as the simulation-based $Wt$ and diboson background
estimates depend on the integrated luminosity of the dataset.

The LHC beam energy is known to be within 0.1\% of the nominal value
at \sxyt~\cite{ebeam2}, which translates into an 0.23\% uncertainty in \xtt,
quoted as part of the experimental uncertainty as discussed in
Ref.~\cite{TOPQ-2018-17}. Similar uncertainties arise in the absolute
\ttem\ and \embb\ differential cross-sections, but almost completely cancel
in the normalised measurements.

\section{Inclusive cross-section results and interpretation}\label{s:ires}

The results of the inclusive cross-section analysis are given in
Section~\ref{ss:ixres}, followed by the top quark mass in Section~\ref{ss:mtop}.
The analysis results were initially blinded by multiplying the \xtt\ values
by a randomly chosen scale factor that was only removed after finalising
all systematic uncertainties and stability studies, and verifying that
consistent results were obtained from the 2015+16, 2017 and 2018 datasets.

\subsection{Total and fiducial cross-section results}\label{ss:ixres}

Table~\ref{t:incres} shows the results for \xtt\ and \xfid\ from the entire
dataset treated as a single sample, separate results from the 2015+16, 2017
and 2018 datasets, and the combination of these three datasets using
the best linear unbiased estimator technique~\cite{blue,valassiblue}, taking
into account correlations in the systematic uncertainties. Since the largest
uncertainties come from the luminosity measurement, and these uncertainties
are only partially correlated between years~\cite{DAPR-2021-01},
this combination gives a greater weight to the 2015+16 data despite its larger
statistical uncertainty, and is used for the final results:
\begin{eqnarray}
\xtt & = & \ttxval \pm \ttxstat \pm \ttxsyst \pm \ttxlumi \pm \ttxebeam\,\mathrm{pb},\ \mathrm{and} \nonumber \\
\xfid & = & \ttxfidval \pm \ttxfidstat \pm \ttxfidsyst \pm \ttxfidlumi \pm \ttxfidebeam\,\mathrm{pb},\nonumber
\end{eqnarray}
where the four uncertainties are due to data statistics, experimental and
theoretical  systematic effects internal to the analysis, the knowledge of
the integrated luminosity, and the knowledge of the LHC beam energy.
The total relative uncertainties are \ttxrel\ for \xtt\ and \ttxfidrel\
for \xfid. The three datasets have weights of 0.53, 0.23 and~0.24, and the
combination has a $\chi^2$ of 1.5 for two degrees of freedom, demonstrating
the good compatibility of the results. The values of \epsb\ are 0.5--1\%
lower in data than in simulation, well within the uncertainties in modelling
the $b$-tagging performance~\cite{FTAG-2018-01}. The result for \xtt\
is reported for a fixed top quark mass of $\mtop=172.5$\,\GeV, and depends on
the assumed value according to
$(1/\xtt)\, \mathrm{d}\xtt/\mathrm{d}\mtop=\dmtopval$\%/\GeV.
The \mtop\ dependence of the measured \xfid\ is negligible.
The fiducial cross-section
corrected to remove the contribution from leptonic decays of $\tau$-leptons
is $\xfidnotau=\ttxntfidval\pm\ttxntfidstat\pm\ttxntfidsyst\pm\ttxntfidlumi\pm\ttxntfidebeam$\,pb.

\begin{table}[tp]
\caption{\label{t:incres}Measurements of the inclusive total (\xtt) and
fiducial (\xfid) \ttbar\ production cross-sections at \sxyt\ using the
full dataset, the 2015+16, 2017 and 2018 datasets separately, and the combination
of the three separate measurements. The four uncertainties for each
measurement correspond to the statistical, experimental and theoretical
systematic, integrated luminosity, and beam energy uncertainties. The total
uncertainties are given in parentheses. The integrated luminosity $L$ of each
dataset is also shown.}

\hspace{-3mm}\begin{tabular}{lrcc}\toprule
Dataset & \multicolumn{1}{c}{$L$ [fb$^{-1}$] } & $\xtt$ [pb] & $\xfid$ [pb] \\
\midrule
All data & 140.1 & $ 827.5 \pm  1.1\pm  8.0\pm  7.6\pm  1.9$ (11.3) &
$14.01 \pm0.02\pm0.10\pm0.13\pm0.03$ (0.17) \\
\midrule
2015+16 data & 36.6 &$ 831.5 \pm  2.2\pm  8.1\pm  8.0\pm  1.9$ (11.7) &
$14.08 \pm0.04\pm0.10\pm0.14\pm0.03$ (0.18) \\
2017 data & 44.6 & $ 833.1 \pm  2.0\pm  8.1\pm 10.5\pm  1.9$ (13.5) &
$14.11 \pm0.03\pm0.10\pm0.18\pm0.03$ (0.21) \\
2018 data & 58.8 & $ 820.8 \pm  1.7\pm  8.0\pm 10.0\pm  1.9$ (13.1) &
$13.90 \pm0.03\pm0.10\pm0.17\pm0.03$ (0.21) \\
\midrule
Combination & 140.1 & $ 829.3 \pm  1.3\pm  8.0\pm  7.3\pm  1.9$ (11.1) &
$14.04 \pm0.02\pm0.10\pm0.12\pm0.03$ (0.17) \\
\bottomrule
\end{tabular}
\end{table}

The results are stable within statistical uncertainties when increasing
the jet \pt\ requirement from the nominal $\ptjet>25$\,\GeV\ to
$\ptjet>75$\,\GeV (where the tagging correlation increases to $\cb=1.17$),
when reducing the jet acceptance to $\etajet<1.5$ and when using the looser
77\% $b$-tagging efficiency working point. The cross-section decreases
slightly when increasing the lepton \pt\ requirement, e.g. \xtt\ changes
by $-0.77\pm 0.11$\% with $\ptl>35$\,\GeV instead of $\ptl>20$\,\GeV, where the
uncertainty represents the uncorrelated statistical component only. However,
the \ttbar\ modelling uncertainty from QCD scale variations, initial/final-state
radiation and hadronisation also increases from 0.43\% to 0.86\%, suggesting
this change in central value is reasonable given the 40\% reduction in
acceptance.

Table~\ref{t:incsyst} shows the breakdown of statistical and systematic
uncertainties in \xtt\ and \xfid, together with the average uncertainty
contributions to \epsem, \gem\ and \cb, weighted as in the combination.
The largest uncertainty comes from the calibration of the integrated luminosity,
followed by \ttbar\ modelling (in particular PDF uncertainties, hadronisation
and QCD scale variations, which are significantly smaller in the fiducial
cross-section measurement) and background uncertainties from the $Wt$
cross-section and modelling. The result is consistent with the previous
ATLAS measurement of \xtt\ at \sxyt~\cite{TOPQ-2018-26}, which it supersedes.
The total
uncertainty has been reduced from 1.8\% to \ttxrel, giving the most precise
measurement of \xtt\ to date. The new result benefits from the better modelling
of the top quark (and hence lepton) kinematics in \powhegminnlo\
compared with {\scr Powheg\,+\,Pythia8}, a more inclusive lepton selection
requiring  $\pt>20$\,\GeV rather than $\pt>25$\,\GeV,
a more precise understanding of lepton identification
and isolation efficiencies, refined modelling of the $Wt$ background
and an optimal treatment of the data given the luminosity uncertainties.

The inclusive cross-section result, together with other ATLAS measurements
in dilepton and lepton+jets final states at \sxut~\cite{TOPQ-2018-40},
\sxwt~\cite{TOPQ-2013-04,TOPQ-2017-08}, \sxvt~\cite{TOPQ-2013-04,TOPQ-2016-08},
\sxyt~\cite{TOPQ-2020-02} and \sxzt~\cite{TOPQ-2023-21}, is compared in
Figure~\ref{f:xsvse}
with the NNLO+NNLL QCD prediction described in Section~\ref{s:intro},
calculated with {\ttr Top++}~\cite{toppp} with the PDF4LHC21~\cite{pdf4lhc3}
PDF combination. The result is in good agreement with the prediction, and has
about one third of the uncertainty. The result also agrees with measurements
at \sxyt\ from the CMS Collaboration~\cite{CMS-TOP-16-006,CMS-TOP-17-001}
but again has significantly higher precision.

\begin{figure}[tp]
\centering

\includegraphics[width=120mm]{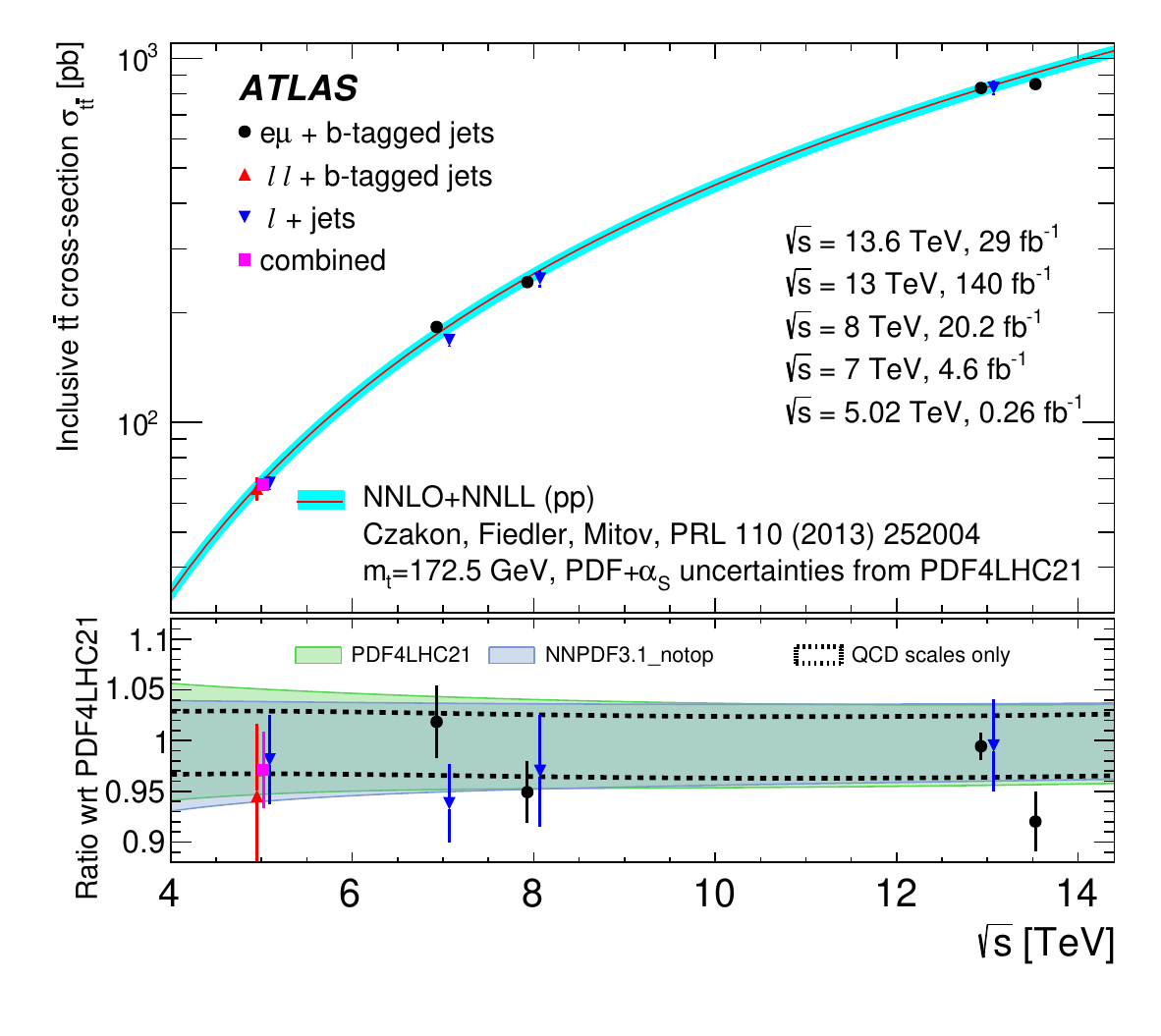}
\caption{\label{f:xsvse}The upper panel shows the inclusive \ttbar\ cross-section
as a function of centre-of-mass energy $\sqrt{s}$, comparing ATLAS results
from the $e\mu$, dilepton ($\ell\ell$) and lepton+jets final states (and the
combination of $\ell\ell$ and lepton+jets at $\sqrt{s}=5.02$\,\TeV) with
NNLO+NNLL theoretical predictions~\cite{topxtheo5,toppp} using the
PDF4LHC21 PDF set~\cite{pdf4lhc3}. The lower panel shows the ratios of
measurements and predictions to the central value of the prediction with
PDF4LHC21. The total uncertainties when using the PDF4LHC21 and NNPDF3.1\_notop
PDF sets are shown as the overlapping coloured bands, and the dotted lines show
the QCD scale uncertainties alone.}
\end{figure}

\subsection{Extraction of the top quark pole mass}\label{ss:mtop}

The strong dependence of the prediction for \xtt\ on the top quark pole
mass \mtpole\ can be exploited to interpret the precise measurement of \xtt\
as a measurement of \mtpole. This interpretation was performed
for two recent PDF sets that do not use any LHC \ttbar\ differential
cross-section data as input, as the use of such data in PDF fits can introduce a
dependence on the assumed value of \mtop. The baseline result is based on
the NNPDF3.1\_notop NNLO PDF set~\cite{nnpdf31} with $\alphas=0.1180\pm 0.0010$,
as used in Ref.~\cite{TOPQ-2018-39}. This prediction is shown in the lower
panel of Figure~\ref{f:xsvse}.
The CT14~\cite{ct14} NNLO PDF
set was also considered, assuming $\alphas=0.1180\pm 0.0012$ and
scaling the uncertainties to correspond to 68\% confidence levels.

Figure~\ref{f:mtpole} shows the dependence of the predicted \xtt\ on \mtpole\
for the NNPDF3.1\_notop PDF set, calculated using
{\ttr Top++}~\cite{topxtheo5,toppp}.
It also shows the experimental measurement from Section~\ref{ss:ixres} with
its dependence on the top quark mass in simulation \mtmc\ obtained from the
polynomial parameterisation described in Section~\ref{ss:sytt},
assuming $\mtmc\approx\mtpole$. The value of
\mtpole\ maximising the compatibility with the measured \xtt\ was extracted
using the Bayesian likelihood formulation detailed in Ref.~\cite{TOPQ-2018-17},
giving a result of:
\[
\mtpole = \mtpval^{+\mtpeup}_{-\mtpedn}\,\GeV\ .
\]
The breakdown of uncertainties is shown in Table~\ref{t:mtopsyst}. The high
precision of the experimental measurement of \xtt\ results in the uncertainty
being dominated by those on the theoretical  prediction. The result for the CT14
PDF set is also given; this PDF set gives a slightly higher central value
for \mtop\ and a larger uncertainty. Compared with these uncertainties,
the assumption that the top quark mass used in simulation for the \xtt\
measurement is close to \mtpole\ has a very small effect, corresponding
e.g.\ to a 0.12\,\GeV\ shift for a 1\,\GeV\ difference between \mtmc\ and
\mtpole.

\begin{figure}[tp]
\centering

\includegraphics[width=120mm]{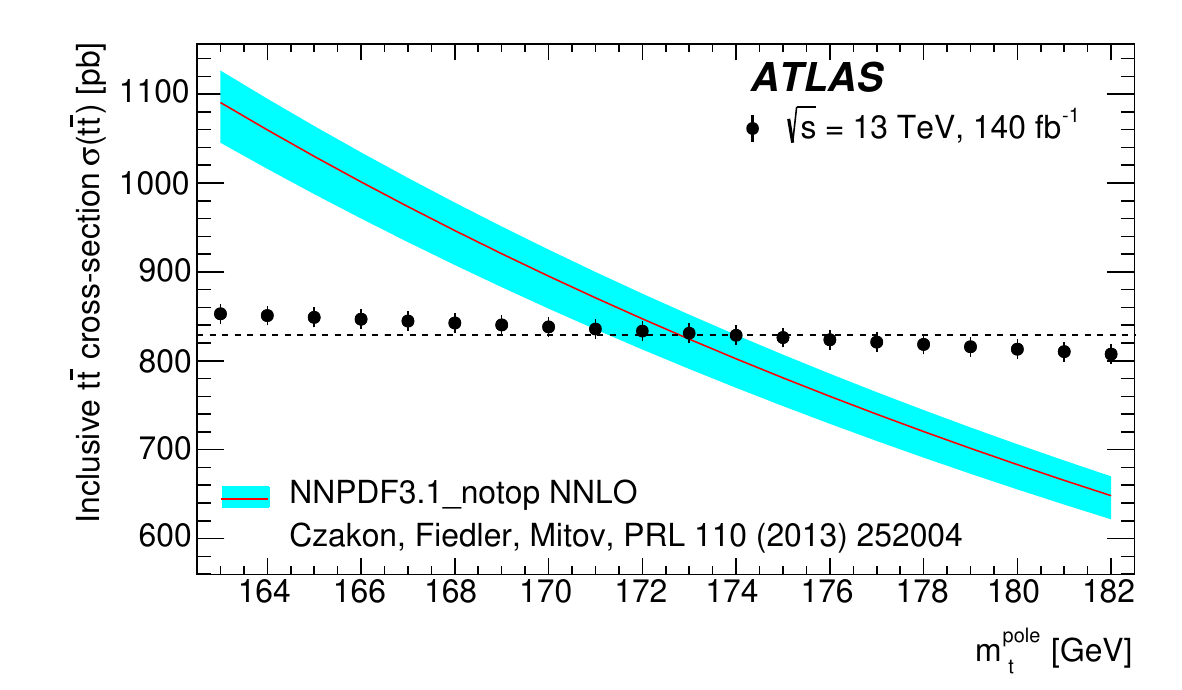}
\caption{\label{f:mtpole}The red line shows the predicted inclusive \ttbar\
cross-section at \sxyt\ as a function of \mtpole\ for the NNPDF3.1\_notop PDF
set, with the cyan band indicating the total uncertainty in the prediction from
PDF+$\alphas$ and QCD scale uncertainties. The black points show the experimental
measurement with its uncertainty and dependence on the assumed value of
\mtop\ through acceptance and background corrections. The dotted line shows
the experimental central value at a fixed top quark mass of 172.5\,\GeV.}
\end{figure}

\begin{table}[tp]
\caption{\label{t:mtopsyst}Extraction of the top quark pole mass from the
\ttbar\ production cross-section measurement at \sxyt\ using the NNPDF3.1\_notop
and CT14 PDF sets. The first row gives the result with the total
uncertainty, and the subsequent rows show the individual uncertainty
components.}
\centering

\begin{tabular}{lrr}\toprule
PDF set & NNPDF3.1\_notop & CT14 \\
\midrule
Result [\GeV\hspace{-1mm}] &  $172.79^{+1.52}_{-1.70}$ & $173.00^{+1.84}_{-2.04}$\\
\midrule
Experimental &  0.55 & 0.55 \\
PDF+$\alphas$ & $^{+1.00}_{-0.89}$ & $^{+1.50}_{-1.48}$\rule[-2mm]{0mm}{7mm}\\
QCD scales & $^{+1.00}_{-1.48}$ & $^{+1.01}_{-1.49}$\rule[-2mm]{0mm}{7mm}\\
\bottomrule
\end{tabular}
\end{table}

The result is compatible with the value of $172.9\pm 1.7$\,\GeV\ derived
from the ATLAS measurement of \xtt\ with 36\,\ifb of \sxyt\
data~\cite{TOPQ-2018-17},
and with the value of $173.4^{+1.8}_{-2.0}$\,\GeV\ obtained from the
combination of ATLAS and CMS \xtt\ measurements at
$\sqrt{s}=7$--8\,\TeV~\cite{TOPQ-2018-39}, both using NNPDF3.1\_notop,
as well as with measurements
from CMS at \sxyt~\cite{CMS-TOP-17-001} and from D0 at the Tevatron
$pp$ collider~\cite{d0pole}. It is also compatible with the average
of direct measurements of the top quark mass from its decay products,
$\mtop=172.57\pm 0.29$\,\GeV~\cite{pdg}, which has much higher precision but
makes stronger assumptions on the interpretation of the mass parameter in \ttbar\
event generators.

\section{Differential cross-section results}\label{s:dres}

The single-lepton and dilepton \ttem\ differential cross-sections
were obtained from the full dataset by solving the tagging
equations Eqs.~(\ref{e:fidtags}) for each bin $i$ of each distribution.
The results were found to be stable when varying the jet \pt, $|\eta|$ and
$b$-tagging requirements. The single-lepton \ptl\ and \etal\ distributions
were also measured separately for electrons and muons, instead of combining
them into lepton distributions with two entries per event, and found to be
compatible. The distributions of bin-by-bin differences between the electron
and muon differential cross-sections have $\chi^2$ per degree of freedom
of $8/12$ for \ptl\ and $10/8$ for \etal, taking statistical and uncorrelated
systematic uncertainties into account. The normalised distributions
were also found to be consistent with the measurements from the 36\,\ifb\
data sample~\cite{TOPQ-2018-17}, in most cases within the uncorrelated
statistical uncertainties.

The \embb\ differential cross-sections were obtained from Eq.~(\ref{e:xembbi}).
The value of \epsb\ obtained from solving Eqs.~(\ref{e:tags}) with the inclusive
data event counts corresponding to the 77\% $b$-tagging efficiency working point
is slightly lower than that in simulation, leading to a correction factor
$\stag=1.0086\pm 0.0007$, where the uncertainty is statistical only. The
correction \stagsq\ was applied to all bins of all absolute \embb\ differential
cross-sections. The integrals of the ten one-dimensional differential
distributions each give the integrated fiducial cross-section \xembbfid, and
agree at the 0.1\% level.\footnote{These integrals do not agree exactly
because of statistical fluctuations and the variation of \gemembbi\ with
each differential variable.}
The cross-section for the complete \embb\ fiducial region was calculated from
the unweighted average of the individual distribution integrals, giving:
\[
\xembbfid = \embbfidval \pm \embbfidstat \pm \embbfidsyst\ \mathrm{pb},
\]
where the first uncertainty is statistical, and the second systematic
(including the luminosity and beam energy uncertainties), giving a relative
uncertainty of \embbfidrel. A breakdown of the
uncertainties in broad categories is given in Table~\ref{t:xembbfid},
This fiducial cross-section is more inclusive than that measured
in Ref.~\cite{TOPQ-2020-14}, which requires $\ptl>28$\,\GeV\ rather
than $\ptl>20$\,\GeV, with overlap removal requirements between
the $b$-jets and leptons, and has a relative precision of 4.8\%. The
better precision of the present result is due primarily to the in situ
measurement of the $b$-tagging and jet reconstruction efficiency via \stag.

\begin{table}[tp]
\caption{\label{t:xembbfid}Breakdown of uncertainties in the measured
\embb\ fiducial cross-section in broad categories.}
\centering

\begin{tabular}{lc}\toprule
Uncertainty & $\Delta\xembbfid/\xembbfid$ (\%) \\
\midrule
Data statistics &   0.16 \\
$\ttbar+Wt$ modelling &   1.34 \\
Leptons &   0.29 \\
Jets/$b$-tagging &   0.26 \\
Backgrounds &   0.36 \\
Luminosity/$E_\mathrm{beam}$ &   0.92 \\
\midrule
Total uncertainty &   1.72 \\
\bottomrule
\end{tabular}
\end{table}

\subsection{Results for $\boldmath\ttbar\rightarrow e\mu$ and $\boldmath e\mu b\bar{b}$ differential cross-sections}\label{ss:diffres}

The measured absolute one-dimensional \ttem\ and \embb\ differential
cross-sections are shown in the upper panels of
Figures~\ref{f:distresa}--\ref{f:distrese},
and the two-dimensional cross-sections are shown
in Figures~\ref{f:distresf}--\ref{f:distresh}. The latter Figures show
`unrolled' distributions with the four groups of \etal, \rapll\ or \dphill\
bins corresponding to each \mll\ bin shown consecutively on the $x$-axis.
Numerical values for both the absolute and normalised differential
cross-sections, including a set of results corrected to remove the
contributions from $W\rightarrow\tau\rightarrow e/\mu$ decays, can be found
at Ref.~\cite{ttx8aux}, and are available in the
HEPData repository~\cite{hepdata}.

\begin{figure}[tp]
\parbox{83mm}{\includegraphics[width=76mm]{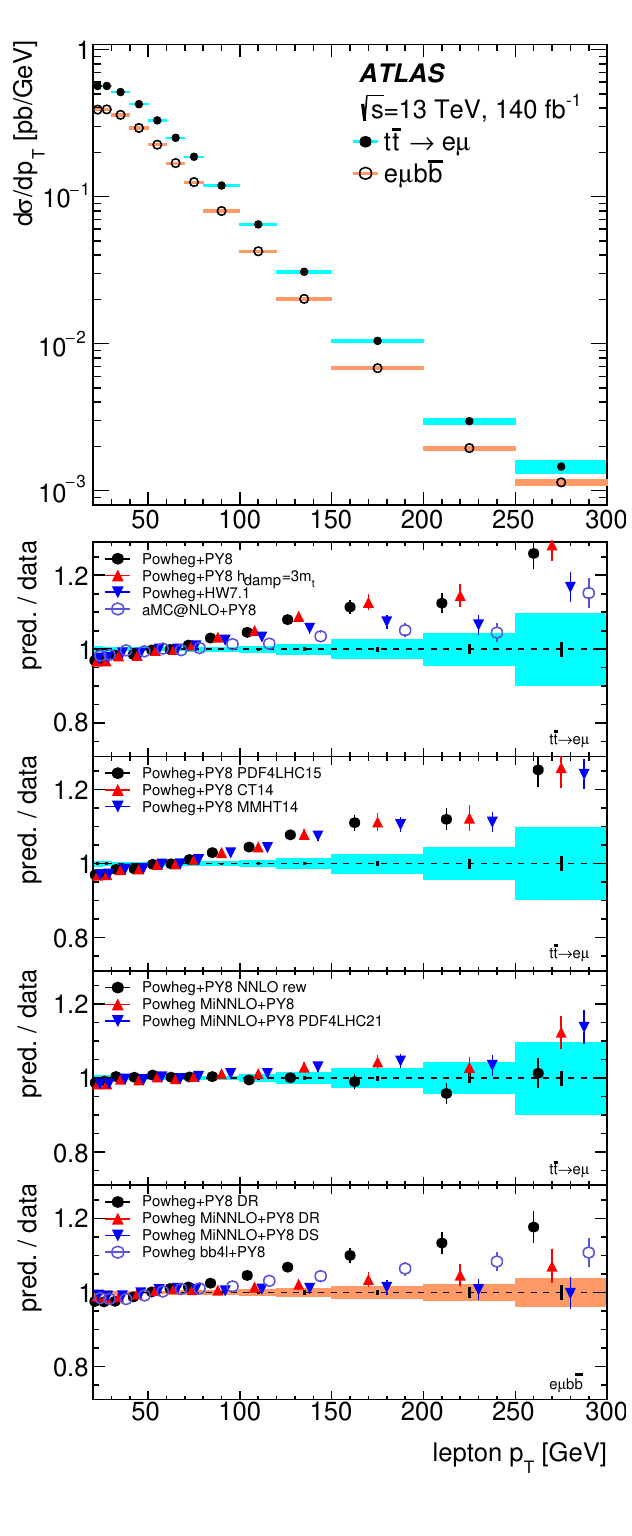}\vspace{-7mm}\center{(a)}}
\parbox{83mm}{\includegraphics[width=76mm]{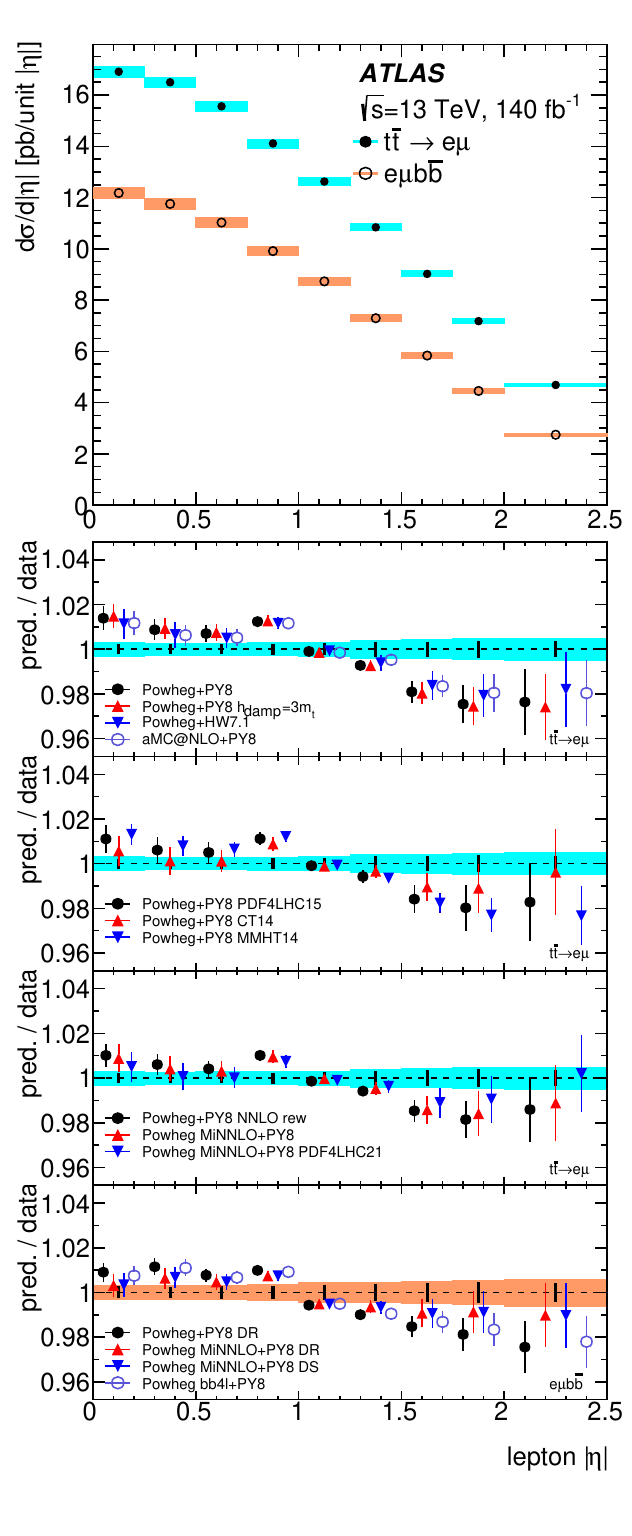}\vspace{-7mm}\center{(b)}}
\caption{\label{f:distresa}Differential cross-sections as functions of (a) \ptl\
and (b) \etal. The upper panels show the measured absolute \ttem\ (filled points)
and \embb\ (open points) cross-sections, including the contributions from
$W\rightarrow\tau\rightarrow e/\mu$ decays, with the total uncertainties shown
by the shaded bands. The other panels show ratios of various predictions to the
measured normalised differential cross-sections, for \ttem\ (middle three panels)
and \embb\ (lower panels). The markers showing the ratios for each prediction
are offset from the bin centres for better visibility, and the total
uncertainty in the prediction is shown by the error bar. The data statistical
uncertainty is shown by the black error bars and the total uncertainty by the
cyan (\ttem) or orange (\embb) band around unity.}
\end{figure}

\begin{figure}[tp]
\parbox{83mm}{\includegraphics[width=76mm]{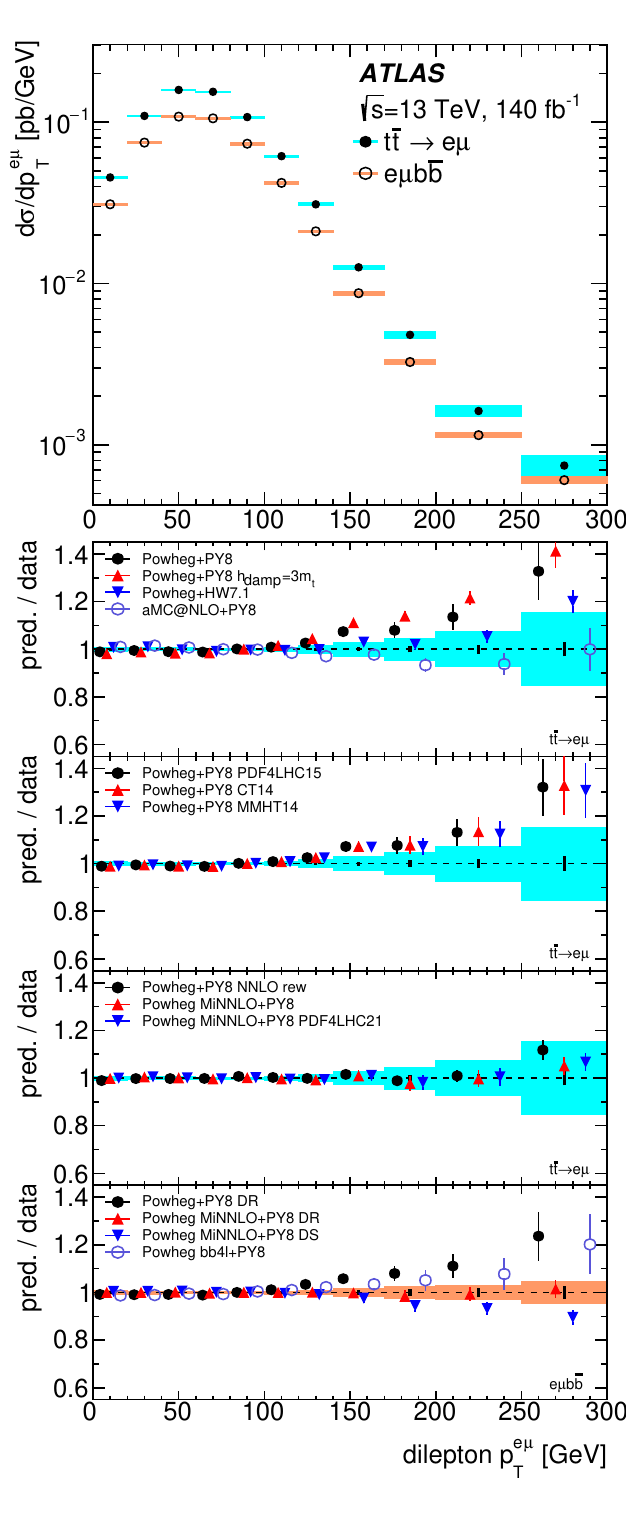}\vspace{-7mm}\center{(a)}}
\parbox{83mm}{\includegraphics[width=76mm]{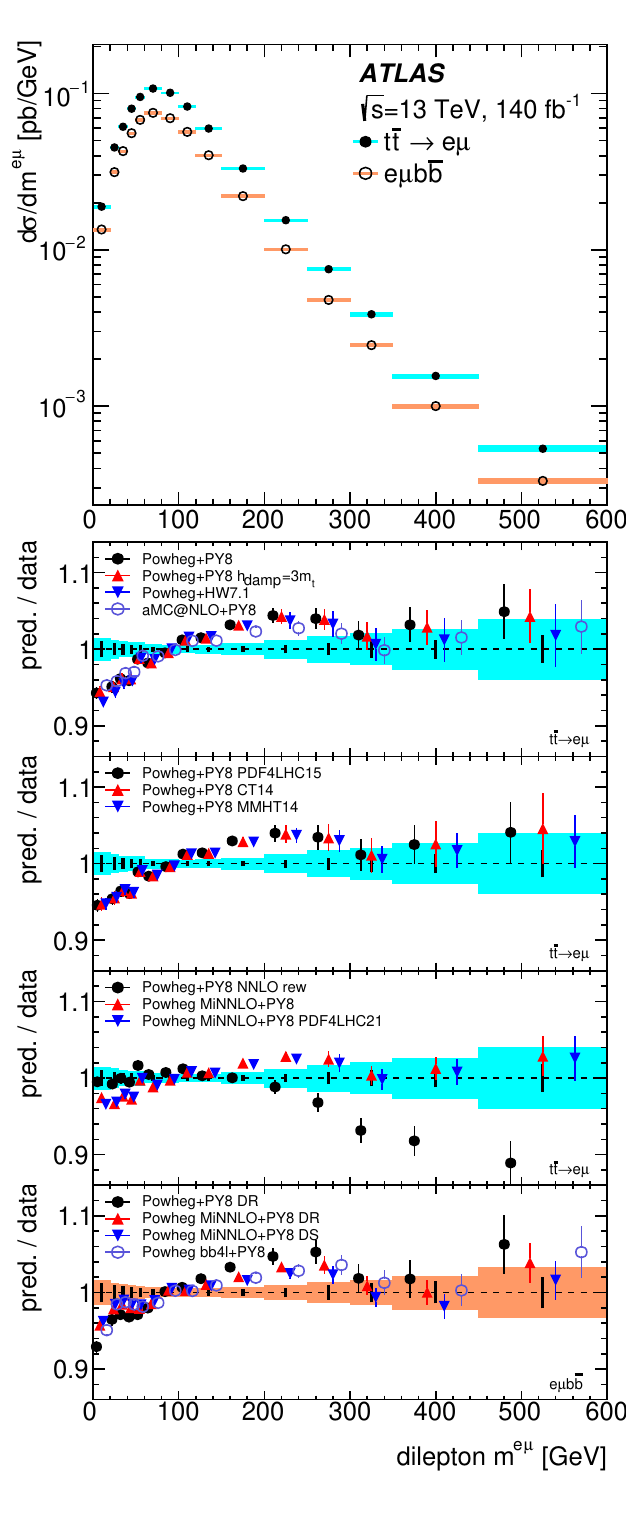}\vspace{-7mm}\center{(b)}}
\caption{\label{f:distresb}Differential cross-sections as functions of (a) \ptll\
and (b) \mll. The upper panels show the measured absolute \ttem\ (filled points)
and \embb\ (open points) cross-sections, including the contributions from
$W\rightarrow\tau\rightarrow e/\mu$ decays, with the total uncertainties shown
by the shaded bands. The other panels show ratios of various predictions to the
measured normalised differential cross-sections, for \ttem\ (middle three panels)
and \embb\ (lower panels). The markers showing the ratios for each prediction
are offset from the bin centres for better visibility, and the total
uncertainty in the prediction is shown by the error bar. The data statistical
uncertainty is shown by the black error bars and the total uncertainty by the
cyan (\ttem) or orange (\embb) band around unity.}
\end{figure}

\begin{figure}[tp]
\parbox{83mm}{\includegraphics[width=76mm]{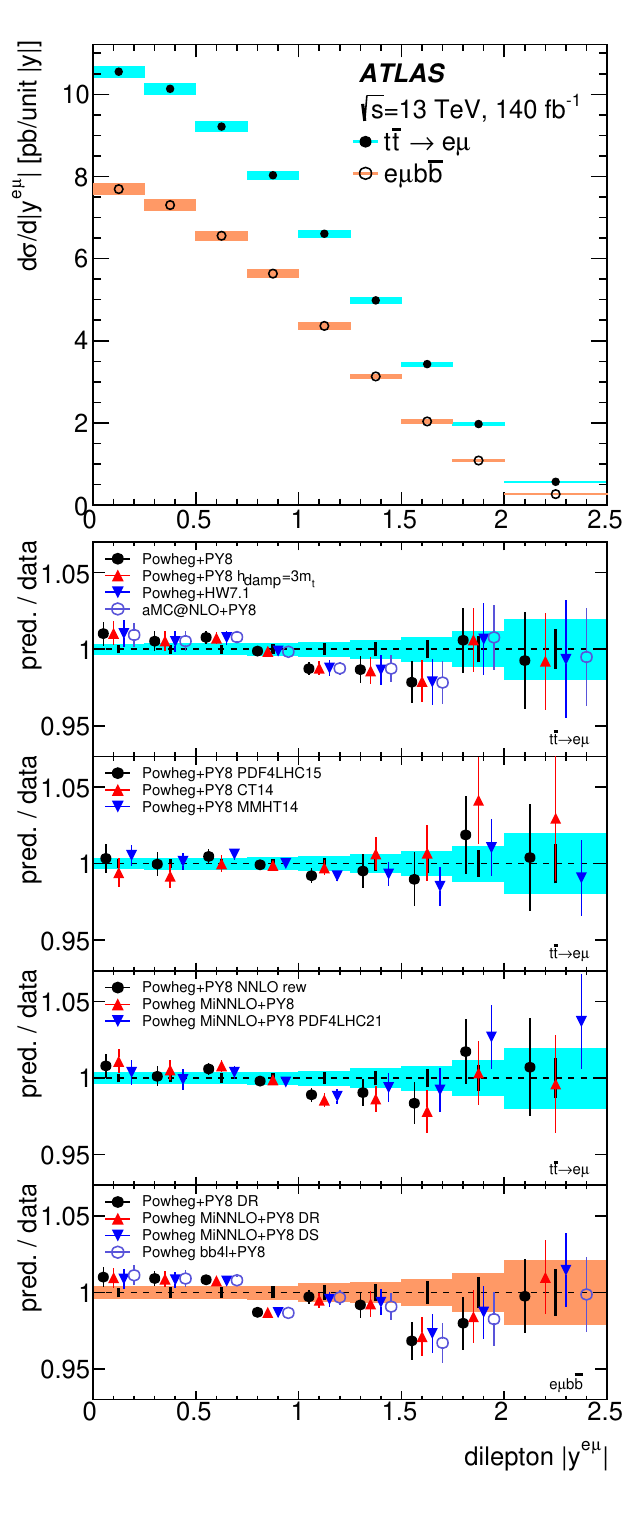}\vspace{-7mm}\center{(a)}}
\parbox{83mm}{\includegraphics[width=76mm]{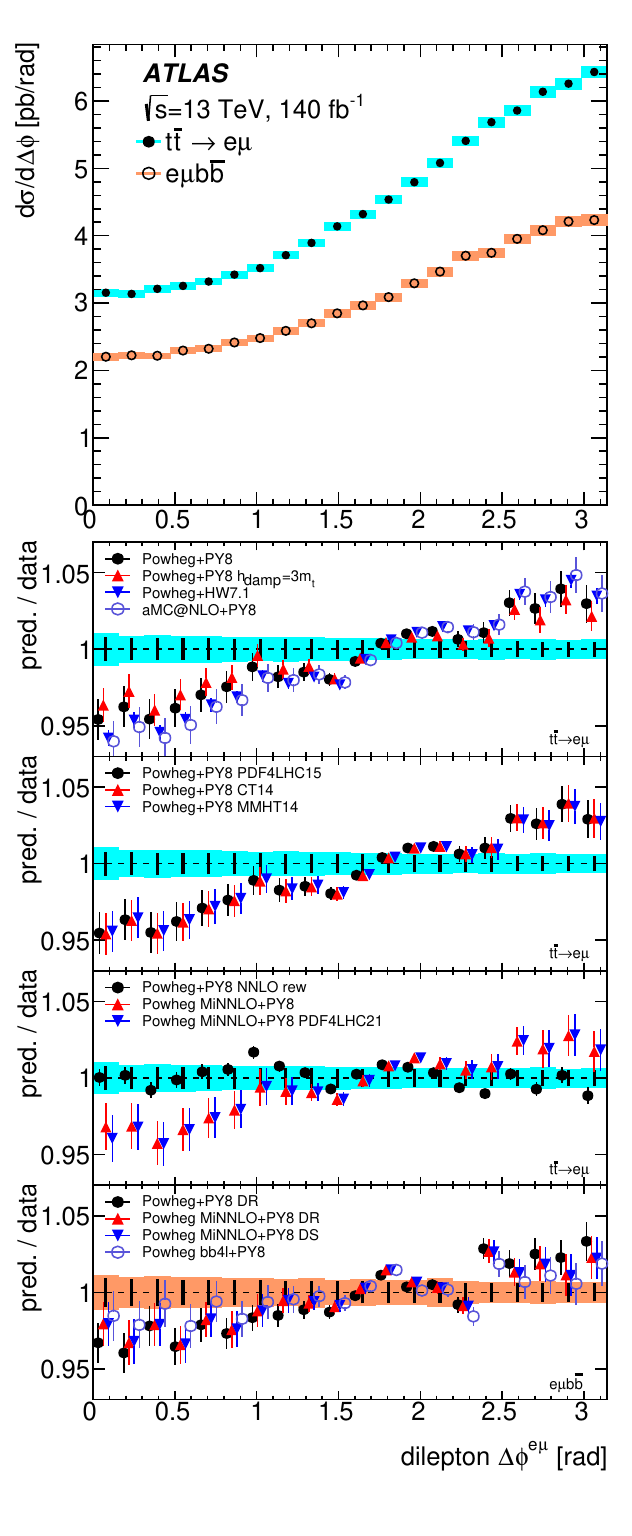}\vspace{-7mm}\center{(b)}}
\caption{\label{f:distresc}Differential cross-sections as functions of (a)
\rapll\ and (b) \dphill. The upper panels show the measured absolute \ttem\
(filled points)
and \embb\ (open points) cross-sections, including the contributions from
$W\rightarrow\tau\rightarrow e/\mu$ decays, with the total uncertainties shown
by the shaded bands. The other panels show ratios of various predictions to the
measured normalised differential cross-sections, for \ttem\ (middle three panels)
and \embb\ (lower panels). The markers showing the ratios for each prediction
are offset from the bin centres for better visibility, and the total
uncertainty in the prediction is shown by the error bar. The data statistical
uncertainty is shown by the black error bars and the total uncertainty by the
cyan (\ttem) or orange (\embb) band around unity.}
\end{figure}

\begin{figure}[tp]
\parbox{83mm}{\includegraphics[width=76mm]{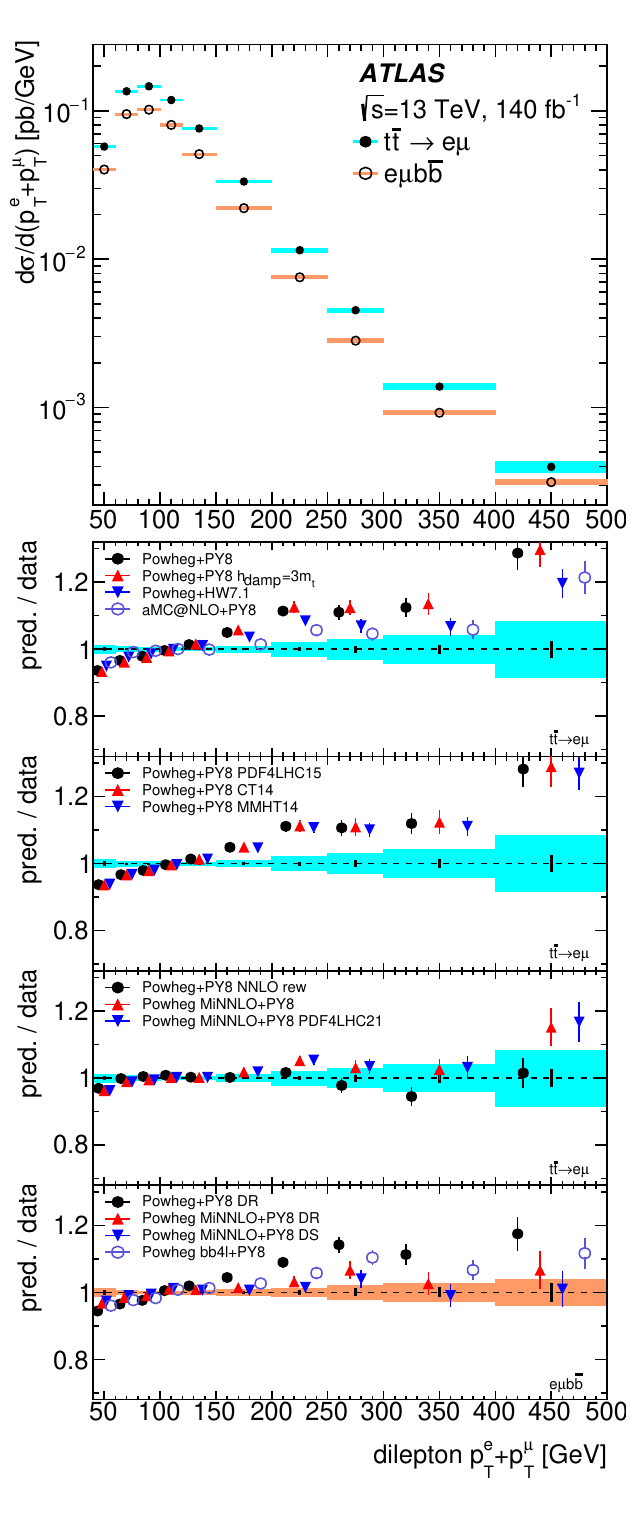}\vspace{-7mm}\center{(a)}}
\parbox{83mm}{\includegraphics[width=76mm]{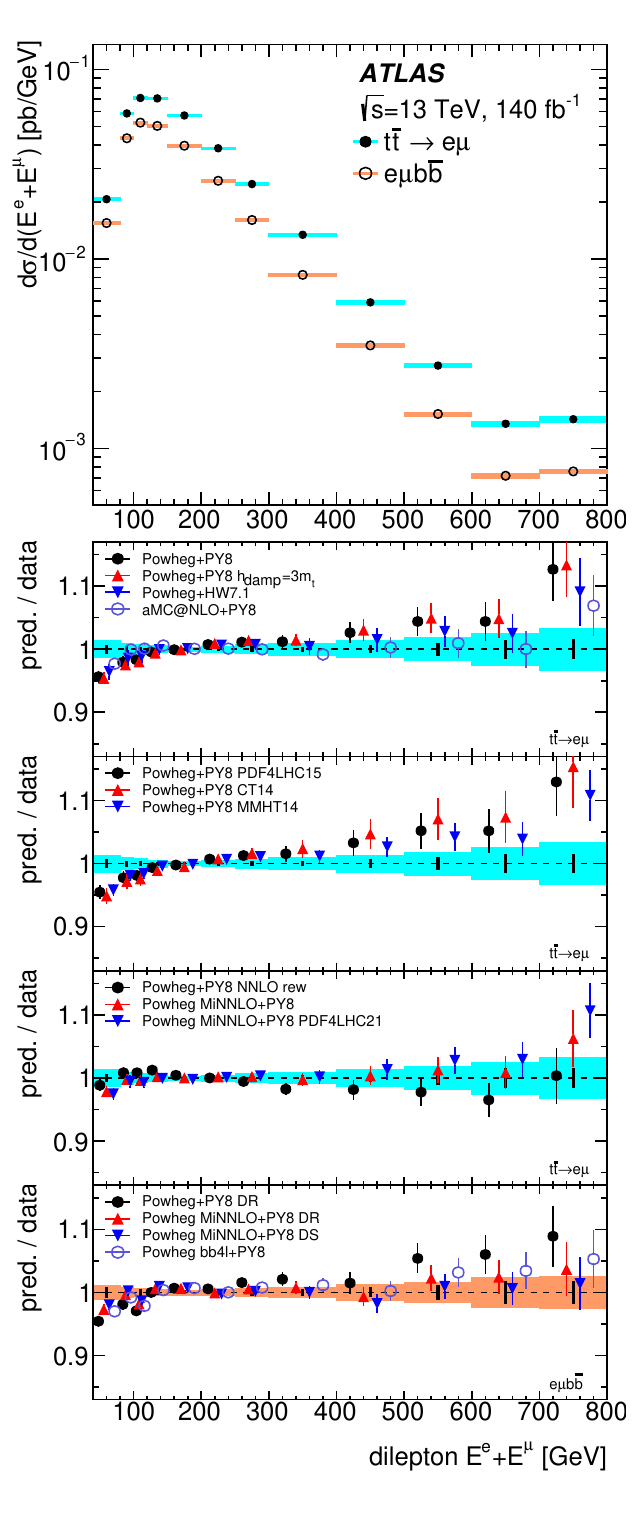}\vspace{-7mm}\center{(b)}}
\caption{\label{f:distresd}Differential cross-sections as functions of (a)
\ptsum\ and (b) \esum. The upper panels show the measured absolute \ttem\
(filled points)
and \embb\ (open points) cross-sections, including the contributions from
$W\rightarrow\tau\rightarrow e/\mu$ decays, with the total uncertainties shown
by the shaded bands. The other panels show ratios of various predictions to the
measured normalised differential cross-sections, for \ttem\ (middle three panels)
and \embb\ (lower panels). The markers showing the ratios for each prediction
are offset from the bin centres for better visibility, and the total
uncertainty in the prediction is shown by the error bar. The data statistical
uncertainty is shown by the black error bars and the total uncertainty by the
cyan (\ttem) or orange (\embb) band around unity.}
\end{figure}

\begin{figure}[tp]
\parbox{83mm}{\includegraphics[width=76mm]{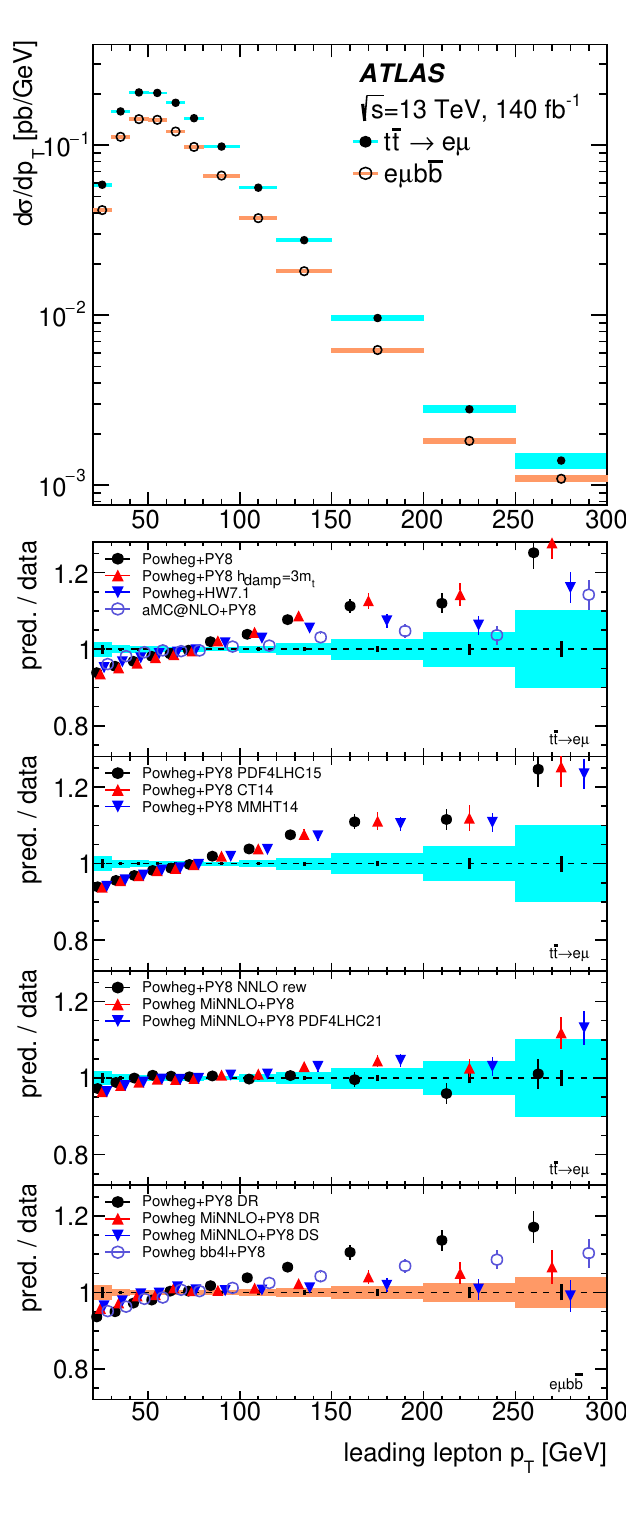}\vspace{-7mm}\center{(a)}}
\parbox{83mm}{\includegraphics[width=76mm]{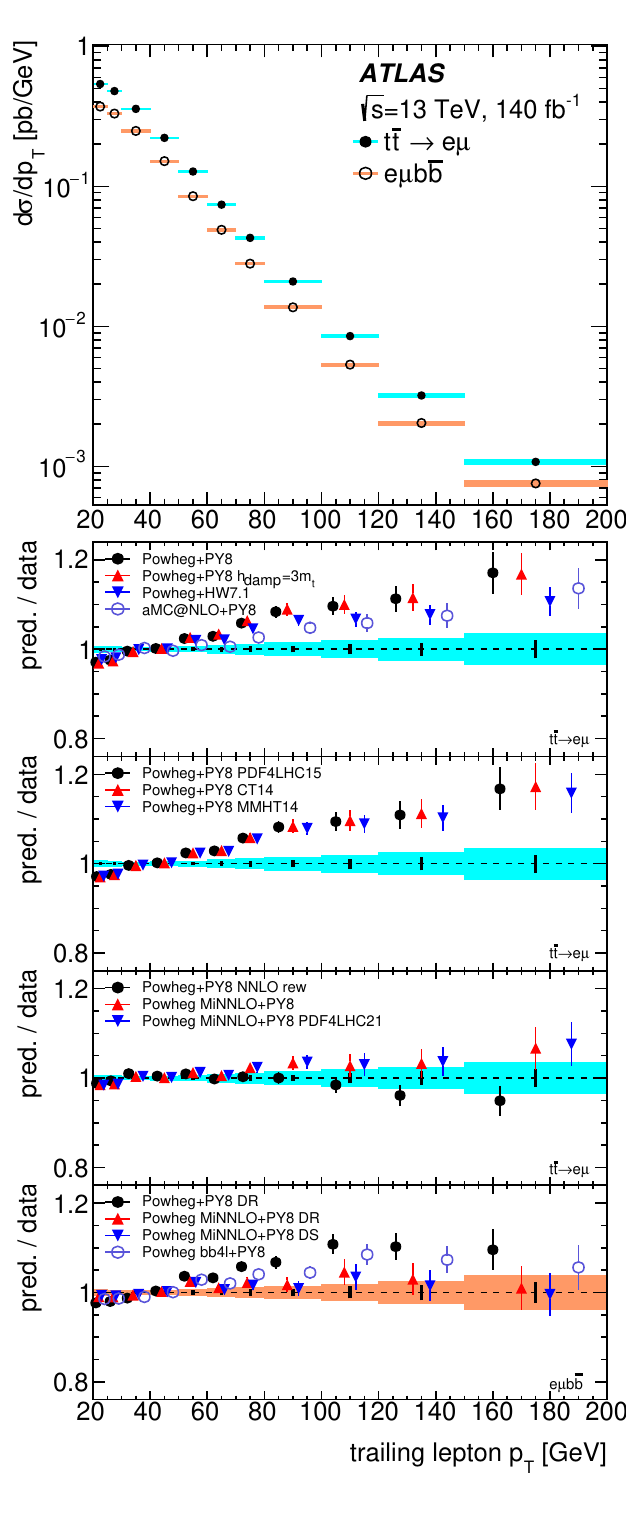}\vspace{-7mm}\center{(b)}}
\caption{\label{f:distrese}Differential cross-sections as functions of (a)
\ptlmax\ and (b) \ptlmin. The upper panels show the measured absolute \ttem\
(filled points)
and \embb\ (open points) cross-sections, including the contributions from
$W\rightarrow\tau\rightarrow e/\mu$ decays, with the total uncertainties shown
by the shaded bands. The other panels show ratios of various predictions to the
measured normalised differential cross-sections, for \ttem\ (middle three panels)
and \embb\ (lower panels). The markers showing the ratios for each prediction
are offset from the bin centres for better visibility, and the total
uncertainty in the prediction is shown by the error bar. The data statistical
uncertainty is shown by the black error bars and the total uncertainty by the
cyan (\ttem) or orange (\embb) band around unity.}
\end{figure}

\begin{figure}[tp]
\centering

\includegraphics[width=160mm]{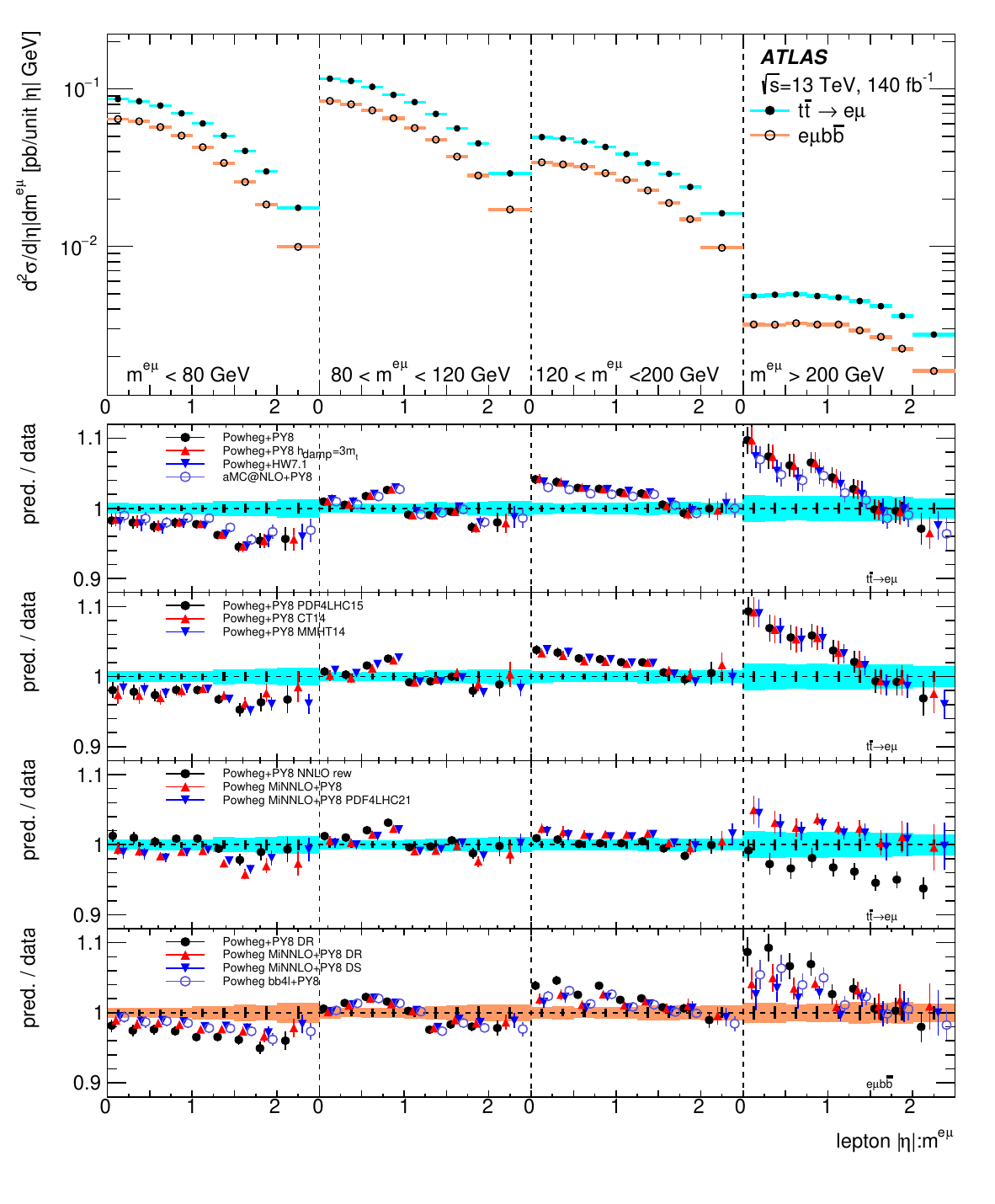}
\caption{\label{f:distresf}Double-differential cross-section as a function
of \etal\ and \mll, showing the unrolled cross-sections as a function of \etal\ in
four bins of \mll. The upper panel shows the measured absolute \ttem\
(filled points)
and \embb\ (open points) cross-sections, including the contributions from
$W\rightarrow\tau\rightarrow e/\mu$ decays, with the total uncertainties shown
by the shaded bands. The other panels show ratios of various predictions to the
measured normalised differential cross-sections, for \ttem\ (middle three panels)
and \embb\ (lower panel). The markers showing the ratios for each prediction
are offset from the bin centres for better visibility, and the total
uncertainty in the prediction is shown by the error bar. The data statistical
uncertainty is shown by the black error bars and the total uncertainty by the
cyan (\ttem) or orange (\embb) band around unity.}
\end{figure}

\begin{figure}[tp]
\centering

\includegraphics[width=160mm]{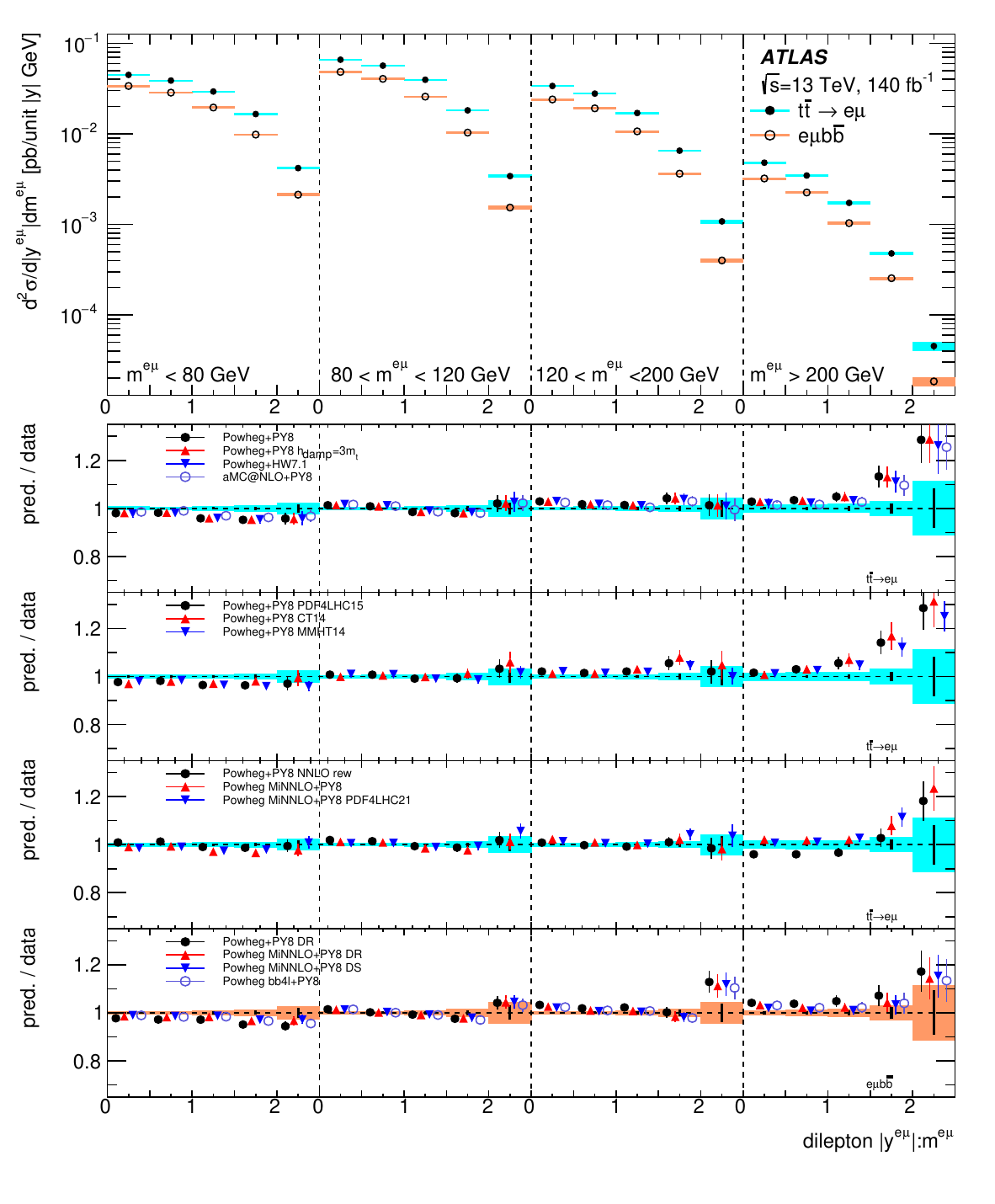}
\caption{\label{f:distresg}Double-differential cross-section as a function
of \rapll\ and \mll, showing the unrolled cross-sections as a function of \rapll\
in four bins of \mll. The upper panel shows the measured absolute \ttem\
(filled points)
and \embb\ (open points) cross-sections, including the contributions from
$W\rightarrow\tau\rightarrow e/\mu$ decays, with the total uncertainties shown
by the shaded bands. The other panels show ratios of various predictions to the
measured normalised differential cross-sections, for \ttem\ (middle three panels)
and \embb\ (lower panel). The markers showing the ratios for each prediction
are offset from the bin centres for better visibility, and the total
uncertainty in the prediction is shown by the error bar. The data statistical
uncertainty is shown by the black error bars and the total uncertainty by the
cyan (\ttem) or orange (\embb) band around unity.}
\end{figure}

\begin{figure}[tp]
\centering

\includegraphics[width=160mm]{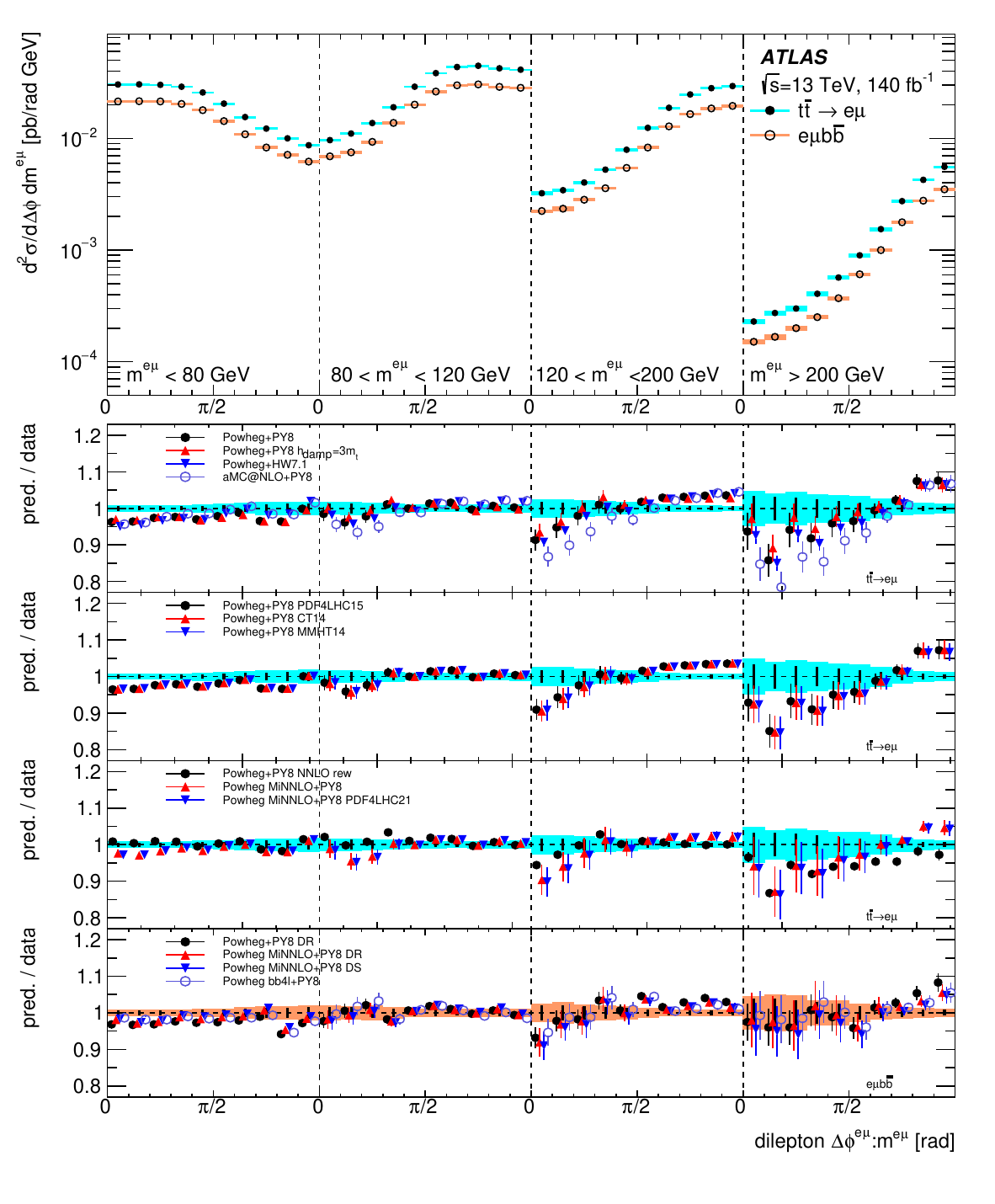}
\caption{\label{f:distresh}Double-differential cross-section as a function
of \dphill\ and \mll, showing the unrolled cross-sections as a function of
\dphill\ in four bins of \mll. The upper panel shows the measured absolute \ttem\
(filled points)
and \embb\ (open points) cross-sections, including the contributions from
$W\rightarrow\tau\rightarrow e/\mu$ decays, with the total uncertainties shown
by the shaded bands. The other panels show ratios of various predictions to the
measured normalised differential cross-sections, for \ttem\ (middle three panels)
and \embb\ (lower panel). The markers showing the ratios for each prediction
are offset from the bin centres for better visibility, and the total
uncertainty in the prediction is shown by the error bar. The data statistical
uncertainty is shown by the black error bars and the total uncertainty by the
cyan (\ttem) or orange (\embb) band around unity.}
\end{figure}

The uncertainties in the one-dimensional normalised \ttem\ cross-sections
are shown in Figure~\ref{f:ttemsyst}, and range from 0.3\% to around
10\% at the upper ends of the \pt-related distributions. For the
\etal, \rapll\ and \dphill\ distributions, the largest uncertainties are
typically statistical, whereas \ttbar\ modelling also plays an important role
at the lower ends of the \pt-related distributions. At the upper ends of those
distributions, the background modelling, in particular the $\ttbar/Wt$
interference uncertainty, becomes dominant. The normalised differential
cross-sections benefit from substantial cancellations in the systematic
uncertainties across bins, which are absent in the absolute differential
cross-section measurements; the latter also suffer from the integrated
luminosity uncertainty, which contributes 0.9-1.1\% in most bins.
The corresponding uncertainties in the one-dimensional \embb\ cross-sections
are shown in Figure~\ref{f:embbsyst}, and exhibit many similar features.
However, the background uncertainties in the \pt-related
distributions do not suffer as much from the $\ttbar/Wt$ interference
uncertainty, being e.g.\ 2.3\% and 3.8\% in the $200<\ptl<250$\,\GeV\ and
$\ptl>250$\,\GeV\ bins in the \embb\ measurement compared with
4.3\% and 9.7\% in the \ttem\ measurement.

As an additional validation of the analysis procedure, the measured \ttem\
and \embb\ differential cross-sections were used to reweight the
{\scr Powheg\,+\,Pythia8} simulation at particle level. This sample was then
used in pseudo-experiment studies as described in Section~\ref{ss:diffval},
taking unweighted simulation as the reference for the calculation
of \gemi, \cbi\ and \gemembbi. The mean values of the differential cross-sections
obtained from the pseudo-experiments were found to be consistent with the
values extracted from data, and no significant biases were observed.

\subsection{Comparison with event generator predictions}\label{ss:diffcomp}

The measured normalised \ttem\ and \embb\ differential cross-sections are
compared with a set of particle-level predictions from different
Monte Carlo event generators in the ratio panels of
Figures~\ref{f:distresa}--\ref{f:distresh}. For each distribution, the first
three ratio panels show comparisons of the \ttem\ cross-sections to
\ttbar\ event generators, and the last ratio panel shows comparisons
of the \embb\ distributions to combined samples of \ttbar\ and $Wt$ events,
or the \powhegbbfl\ sample. The samples considered are summarised in
Table~\ref{t:mcsam}. They include \ttbar\ event generator configurations
discussed in Section~\ref{s:datmc}, as well as a sample generated
with {\scr aMC@NLO\,+\,Pythia8}. The \powhegminnlo\ and {\scr Powheg\,+\,Pythia8}
samples generated with NNPDF3.0 were also reweighted to the predictions of the
PDF4LHC15~\cite{pdf4lhc2}, CT14~\cite{ct14}, MMHT14~\cite{mmht14} and
PDF4LHC21~\cite{pdf4lhc3} NLO PDF sets. The \ttbar\ NNLO reweighting procedure
described in Ref.~\cite{TOPQ-2020-14} was also applied to the
{\scr Powheg\,+\,Pythia8} \ttbar\ sample, in order to bring the top quark
kinematics into agreement with the predictions of the NNLO QCD\,+\,NLO
electroweak calculation of Ref.~\cite{ttnnlocalc}. In these studies,
no reweighting of the fraction of events with three or more $b$-jets was
applied to any of the samples.

\begin{table}[tp]
\caption{\label{t:mcsam}Summary of particle-level simulation samples used in the
comparison with measured \ttem\ and \embb\ differential cross-sections,
giving the matrix-element generator, PDF set, parton shower and other relevant
settings. The top quark mass was set to $\mtop=172.5$\,\GeV in all samples.
The four groups shown correspond to the four ratio panels on
Figures~\ref{f:distresa}--\ref{f:distresh}.}
\centering

\begin{tabular}{lllll}\toprule
Group & Matrix element & PDF set & Parton shower & Comments \\
\midrule
\ttbar-1 & {\scr Powheg} & NNPDF3.0 & {\scr Pythia8} & \\
& {\scr Powheg} & NNPDF3.0 & {\scr Pythia8} & \hdamp=3\mtop \\
& {\scr Powheg} & NNPDF3.0 & {\scr Herwig7.1.3} & \\
& {\scr aMC@NLO} & NNPDF3.0 & {\scr Pythia8} & \\
\midrule
\ttbar-2 & {\scr Powheg} & PDF4LHC15 & {\scr Pythia8} & \\
& {\scr Powheg} & CT14 & {\scr Pythia8} & \\
& {\scr Powheg} & MMHT14 & {\scr Pythia8} & \\
\midrule
\ttbar-3 & {\scr Powheg}& NNPDF3.0 & {\scr Pythia8} & NNLO reweighting ($\pt(t)$,
$\pt(\ttbar)$, $m(\ttbar)$) \\
& {\scr Powheg} MiNNLO & NNPDF3.0 & {\scr Pythia8} & Baseline \\
& {\scr Powheg} MiNNLO & PDF4LHC21 & {\scr Pythia8} & \\
\midrule
\embb & {\scr Powheg} & NNPDF3.0 & {\scr Pythia8} & $Wt$ diagram removal \\
& {\scr Powheg} MiNNLO & NNPDF3.0 & {\scr Pythia8} & $Wt$ diagram removal, baseline \\
& {\scr Powheg} MiNNLO & NNPDF3.0 & {\scr Pythia8} & $Wt$ diagram subtraction \\
& {\scr Powheg} $bb4l$ & NNPDF3.0 & {\scr Pythia8} & Complete $e\mu\nu\bar{\nu}\bbbar$ \\
\bottomrule
\end{tabular}
\end{table}

In previous comparisons of ATLAS \ttbar\ lepton differential distributions with
predictions~\cite{TOPQ-2015-02,TOPQ-2018-17,TOPQ-2018-26}, the predictions
were treated without uncertainties due to e.g.\ QCD scale or PDF variations,
making the comparisons harder to interpret. In this analysis, such prediction
uncertainties have been explicitly considered, including separate variations
of the QCD renormalisation and factorisation scales by factors of two
up and down about their default values, and PDF variations according to the
eigenvector or replica variations provided with each PDF set. These variations
were calculated using event weights in each sample. In addition,
the effects of the {\ttr var3c} A14 tune variations, changes in the
\murfsr\ parameter by factors of~2 and 0.625, and variations
of the top quark mass by $\pm 0.5$\,\GeV\ around the default value of
$\mtop=172.5$\,\GeV\
were evaluated using {\scr Powheg\,+\,Pythia8} and applied to all samples.
The PDF eigenvectors or replicas were used to calculate the correlations
between bins of each distribution, whilst other variations were taken to be
fully correlated between bins. The resulting uncertainties are shown as
error bars on the predictions in Figures~\ref{f:distresa}--~\ref{f:distresh},
and are typically dominated by PDF uncertainties for \etal\ and \rapll,
and QCD scale uncertainties for the other distributions.

The compatibility of each prediction with each measured normalised
distribution was tested using a $\chi^2$ calculated as
\begin{eqnarray}\label{e:redchi}
\chi^2 = {\mathbf\Delta}_{(n-1)}^T {\mathbf S}^{-1}_{(n-1)} {\mathbf\Delta}_{(n-1)}\,,
\end{eqnarray}
where ${\mathbf\Delta}_{(n-1)}$ is the vector of differences between the measured
and predicted normalised differential cross-section in each of the $n$
bins, excluding the last one, and ${\mathbf S}^{-1}_{(n-1)}$ is the inverse of
the corresponding
covariance matrix, including both the experimental uncertainties in the
measurement and the uncertainties in the predictions, and their
bin-to-bin correlations through off-diagonal terms. The last bin of each
distribution was excluded to account for the normalisation condition.
The resulting $\chi^2$ and associated $p$ values (for $n-1$ degrees of freedom)
are shown for the single-differential distributions in Table~\ref{t:reschioned},
and for the double-differential distributions in Table~\ref{t:reschitwod}.

\begin{table}[tp]
\caption{\label{t:reschioned}$\chi^2$ values (top block) and associated
probabilities
(bottom block) for comparison of normalised measured single-differential
fiducial cross-sections with various simulation
samples .Within each block, the last four rows correspond to comparisons
of the \embb\ cross-sections with combined \ttbar+$Wt$ samples, and the
other rows to \ttem\ cross-sections compared with \ttbar\ samples.
Probabilities smaller than $10^{-10}$ are shown as zero.}
\small

\hspace{-20mm}
\begin{tabular}{lrrrrrrrrrr}\toprule
Generator & \ptl & \etal & \ptll & \mll & \rapll & \dphill & \ptsum & \esum & \ptlmax & \ptlmin \\
$N_\mathrm{dof}$ & 12 & 8 & 10 & 14 & 8 & 19 & 9 & 11 & 11 & 10 \\ \midrule
$\chi^2$ values & \\ \cmidrule{1-1}{\scr Powheg\,+\,PY8}&  35.2 &  19.7 &  12.5 &  49.8 &  11.6 &  22.7 &  34.9 &  14.7 &  32.0 &  27.4 \\
{\scr Powheg\,+\,PY8} $\hdamp=3\mtop$&  38.9 &  20.1 &  20.2 &  50.1 &  11.0 &  24.0 &  41.0 &  15.3 &  36.3 &  30.6 \\
{\scr Powheg\,+\,HW7.1}&  32.5 &  15.9 &  20.4 &  92.9 &  10.2 &  71.3 &  27.6 &  12.2 &  26.6 &  20.9 \\
{\scr aMC@NLO\,+\,PY8}&  16.2 &  16.7 &  15.4 &  35.2 &  12.0 &  30.5 &  27.5 &  13.5 &  12.5 &  17.4 \\
{\scr Powheg\,+\,PY8} PDF4LHC15&  33.7 &  15.2 &  12.0 &  47.3 &   8.7 &  22.3 &  33.8 &  13.3 &  30.3 &  26.0 \\
{\scr Powheg\,+\,PY8} CT14&  32.8 &  12.6 &  11.8 &  44.6 &   8.0 &  22.4 &  33.0 &  12.7 &  29.4 &  25.3 \\
{\scr Powheg\,+\,PY8} MMHT14&  33.0 &  19.0 &  11.5 &  48.2 &   9.5 &  22.0 &  33.1 &  13.5 &  29.6 &  25.5 \\
{\scr Powheg\,+\,PY8} NNLO rew&  16.4 &  14.9 &  11.7 &  33.6 &  12.8 &  16.5 &  24.6 &  11.3 &  14.3 &  14.6 \\
{\scr Powheg MiNNLO\,+\,PY8}&  15.4 &  12.3 &   7.2 &  25.9 &  11.7 &  23.6 &  20.9 &   7.5 &  12.1 &  12.5 \\
{\scr Powheg MiNNLO\,+\,PY8} PDF4LHC21&  15.6 &  12.8 &   7.5 &  25.6 &  12.6 &  25.2 &  21.3 &  10.2 &  11.8 &  13.1 \\
\cmidrule{1-1}{\scr Powheg\,+\,PY8} DR&  26.9 &  12.0 &  10.1 &  48.6 &  14.7 &  23.8 &  34.5 &  28.0 &  32.5 &  35.3 \\
{\scr Powheg MiNNLO\,+\,PY8} DR&  12.4 &   6.4 &   1.4 &  31.4 &  14.4 &  26.7 &  15.3 &  17.9 &  16.4 &  14.9 \\
{\scr Powheg MiNNLO\,+\,PY8} DS&  10.8 &   6.6 &   6.0 &  25.8 &  14.4 &  26.1 &  14.0 &  17.1 &  14.3 &  13.1 \\
{\scr Powheg} $bb4l$\,+\,{\scr PY8}&  15.5 &  11.2 &   3.1 &  20.5 &  14.7 &  21.4 &  20.5 &  18.6 &  19.2 &  21.8 \\
\midrule[\heavyrulewidth]
$\chi^2$ probabilities & \\
\cmidrule{1-1}{\scr Powheg\,+\,PY8}& $4\ 10^{-4}$ & 0.011 & 0.25 & $7\ 10^{-6}$ & 0.17 & 0.25 & $6\ 10^{-5}$ & 0.20 & $8\ 10^{-4}$ & $2\ 10^{-3}$ \\
{\scr Powheg\,+\,PY8} $\hdamp=3\mtop$& $1\ 10^{-4}$ & 0.010 & 0.027 & $6\ 10^{-6}$ & 0.20 & 0.20 & $5\ 10^{-6}$ & 0.17 & $1\ 10^{-4}$ & $7\ 10^{-4}$ \\
{\scr Powheg\,+\,HW7.1}& $1\ 10^{-3}$ & 0.044 & 0.026 & 0& 0.25 & $6\ 10^{-8}$ & $1\ 10^{-3}$ & 0.35 & $5\ 10^{-3}$ & 0.022 \\
{\scr aMC@NLO\,+\,PY8}& 0.18 & 0.033 & 0.12 & $1\ 10^{-3}$ & 0.15 & 0.046 & $1\ 10^{-3}$ & 0.26 & 0.33 & 0.065 \\
{\scr Powheg\,+\,PY8} PDF4LHC15& $8\ 10^{-4}$ & 0.056 & 0.29 & $2\ 10^{-5}$ & 0.37 & 0.27 & $1\ 10^{-4}$ & 0.27 & $1\ 10^{-3}$ & $4\ 10^{-3}$ \\
{\scr Powheg\,+\,PY8} CT14& $1\ 10^{-3}$ & 0.13 & 0.30 & $5\ 10^{-5}$ & 0.44 & 0.27 & $1\ 10^{-4}$ & 0.32 & $2\ 10^{-3}$ & $5\ 10^{-3}$ \\
{\scr Powheg\,+\,PY8} MMHT14& $1\ 10^{-3}$ & 0.015 & 0.32 & $1\ 10^{-5}$ & 0.30 & 0.28 & $1\ 10^{-4}$ & 0.26 & $2\ 10^{-3}$ & $4\ 10^{-3}$ \\
{\scr Powheg\,+\,PY8} NNLO rew& 0.18 & 0.061 & 0.31 & $2\ 10^{-3}$ & 0.12 & 0.62 & $3\ 10^{-3}$ & 0.41 & 0.22 & 0.15 \\
{\scr Powheg MiNNLO\,+\,PY8}& 0.22 & 0.14 & 0.71 & 0.026 & 0.16 & 0.21 & 0.013 & 0.76 & 0.35 & 0.25 \\
{\scr Powheg MiNNLO\,+\,PY8} PDF4LHC21& 0.21 & 0.12 & 0.68 & 0.029 & 0.13 & 0.15 & 0.012 & 0.51 & 0.38 & 0.22 \\
\cmidrule{1-1}{\scr Powheg\,+\,PY8} DR& $8\ 10^{-3}$ & 0.15 & 0.44 & $1\ 10^{-5}$ & 0.065 & 0.20 & $7\ 10^{-5}$ & $3\ 10^{-3}$ & $6\ 10^{-4}$ & $1\ 10^{-4}$ \\
{\scr Powheg MiNNLO\,+\,PY8} DR& 0.41 & 0.60 & 1.00 & $5\ 10^{-3}$ & 0.071 & 0.11 & 0.084 & 0.083 & 0.13 & 0.13 \\
{\scr Powheg MiNNLO\,+\,PY8} DS& 0.54 & 0.58 & 0.82 & 0.027 & 0.072 & 0.13 & 0.12 & 0.11 & 0.22 & 0.22 \\
{\scr Powheg} $bb4l$\,+\,{\scr PY8}& 0.22 & 0.19 & 0.98 & 0.11 & 0.064 & 0.32 & 0.015 & 0.069 & 0.057 & 0.016 \\
\bottomrule
\end{tabular}
\end{table}

\begin{table}[tp]
\caption{\label{t:reschitwod}$\chi^2$ values (top block) and associated
probabilities
(bottom block) for comparison of normalised measured double-differential
fiducial cross-sections with various simulation
samples. Within each block, the last four rows correspond to comparisons
of the \embb\ cross-sections with combined \ttbar+$Wt$ samples, and the
other rows to \ttem\ cross-sections compared with \ttbar\ samples.
Probabilities smaller than $10^{-10}$ are shown as zero.}
\small\centering

\begin{tabular}{lrrr}\toprule
Generator & \etalvmll & \rapllvmll & \dphillvmll \\
$N_\mathrm{dof}$ & 35 & 19 & 39 \\ \midrule
$\chi^2$ values & \\ \cmidrule{1-1}{\scr Powheg\,+\,PY8}&  72.7 &  46.0 &  79.9 \\
{\scr Powheg\,+\,PY8} $\hdamp=3\mtop$&  74.2 &  44.9 &  78.3 \\
{\scr Powheg\,+\,HW7.1}& 122.2 &  67.4 & 150.8 \\
{\scr aMC@NLO\,+\,PY8}&  63.0 &  34.2 &  63.2 \\
{\scr Powheg\,+\,PY8} PDF4LHC15&  69.9 &  43.7 &  77.2 \\
{\scr Powheg\,+\,PY8} CT14&  68.8 &  42.2 &  74.9 \\
{\scr Powheg\,+\,PY8} MMHT14&  71.6 &  47.3 &  78.5 \\
{\scr Powheg\,+\,PY8} NNLO rew&  71.0 &  33.8 &  71.0 \\
{\scr Powheg MiNNLO\,+\,PY8}&  44.8 &  26.7 &  50.2 \\
{\scr Powheg MiNNLO\,+\,PY8} PDF4LHC21&  44.3 &  27.0 &  49.7 \\
\cmidrule{1-1}{\scr Powheg\,+\,PY8} DR&  63.7 &  37.0 &  75.3 \\
{\scr Powheg MiNNLO\,+\,PY8} DR&  49.1 &  29.3 &  60.9 \\
{\scr Powheg MiNNLO\,+\,PY8} DS&  41.9 &  24.5 &  54.5 \\
{\scr Powheg} $bb4l$\,+\,{\scr PY8}&  47.8 &  23.7 &  66.0 \\
\midrule[\heavyrulewidth]
$\chi^2$ probabilities & \\
\cmidrule{1-1}{\scr Powheg\,+\,PY8}& $2\ 10^{-4}$ & $5\ 10^{-4}$ & $1\ 10^{-4}$ \\
{\scr Powheg\,+\,PY8} $\hdamp=3\mtop$& $1\ 10^{-4}$ & $7\ 10^{-4}$ & $2\ 10^{-4}$ \\
{\scr Powheg\,+\,HW7.1}& 0& $2\ 10^{-7}$ & 0\\
{\scr aMC@NLO\,+\,PY8}& $3\ 10^{-3}$ & 0.018 & $8\ 10^{-3}$ \\
{\scr Powheg\,+\,PY8} PDF4LHC15& $4\ 10^{-4}$ & $1\ 10^{-3}$ & $3\ 10^{-4}$ \\
{\scr Powheg\,+\,PY8} CT14& $6\ 10^{-4}$ & $2\ 10^{-3}$ & $5\ 10^{-4}$ \\
{\scr Powheg\,+\,PY8} MMHT14& $3\ 10^{-4}$ & $3\ 10^{-4}$ & $2\ 10^{-4}$ \\
{\scr Powheg\,+\,PY8} NNLO rew& $3\ 10^{-4}$ & 0.020 & $1\ 10^{-3}$ \\
{\scr Powheg MiNNLO\,+\,PY8}& 0.12 & 0.11 & 0.11 \\
{\scr Powheg MiNNLO\,+\,PY8} PDF4LHC21& 0.13 & 0.10 & 0.12 \\
\cmidrule{1-1}{\scr Powheg\,+\,PY8} DR& $2\ 10^{-3}$ & $8\ 10^{-3}$ & $4\ 10^{-4}$ \\
{\scr Powheg MiNNLO\,+\,PY8} DR& 0.057 & 0.062 & 0.014 \\
{\scr Powheg MiNNLO\,+\,PY8} DS& 0.20 & 0.18 & 0.050 \\
{\scr Powheg} $bb4l$\,+\,{\scr PY8}& 0.073 & 0.21 & $4\ 10^{-3}$ \\
\bottomrule
\end{tabular}
\end{table}

The measurements are precise and discriminating, and many of the predictions
disagree strongly with the data, even after taking the prediction uncertainties
into account. Predictions based on the {\scr Powheg} {\ttr hvq} process
with the {\scr Pythia8} or {\scr Herwig7} parton shower generally model
distributions sensitive to the top quark \pt\ distribution poorly,
but this can be improved by reweighting the top quark kinematics to
the NNLO prediction of Ref.~\cite{ttnnlocalc}. The \powhegminnlo\
and \powhegbbfl\ samples provide a much better description of the data,
and {\scr aMC@NLO\,+\,Pythia8} is also better than {\scr Powheg\,+\,Pythia8}.
For \embb, some distributions e.g. \ptl\ and \ptll\ offer potential
discrimination between the diagram removal and diagram subtraction
models of $Wt$ events (as suggested by Figure~\ref{f:biastestb}),
but the uncertainties in both measurements and predictions mean that definite
conclusions cannot be drawn. Here, more discrimination can be achieved by
dedicated measurements in the interference region~\cite{TOPQ-2020-14}.

In more detail, the \ptl, \ptlmax\ and \ptlmin\ \ttem\ distributions
(Figures~\ref{f:distresa}(a) and~\ref{f:distrese}) predicted by
{\scr Powheg\,+\,Pythia8} and its $\hdamp=3\mtop$ variation are significantly
harder than the data.
{\scr Powheg\,+\,Herwig7} and {\scr aMC@NLO\,+\,Pythia8} are closer but
still harder than the data, whereas \powhegminnlo\ and \powhegbbfl\ are
compatible with the data. In the \ptlmin\ distribution, which is most relevant
for modelling the acceptance in the inclusive \xtt\ analysis, \powhegminnlo\
is still slightly harder than the data (which is the origin of the dependence
of \xtt\ on the lepton \pt\ requirement discussed in Section~\ref{ss:ixres}),
but is still compatible within uncertainties.
The best modelling of the corresponding \embb\ distributions is provided
by \powhegminnlo\ combined with {\scr Powheg\,+\,Pythia8} diagram subtraction
$Wt$ events. The diagram removal $Wt$ sample predicts a harder \pt\ spectrum,
as does \powhegbbfl\, but these are still consistent with the data
within uncertainties.

The data are more forward than all the predictions for the \ttem\
\etal\ distribution (Figure~\ref{f:distresa}(b)) and to a lesser extent for
\rapll\ (Figure~\ref{f:distresc}(a)). For {\scr Powheg\,+\,Pythia8},
the PDF sets PDF4LHC15 and CT14 give a better description than NNPDF3.0
or MMHT14, demonstrating the sensitivity to PDFs.
For \powhegminnlo, both NNPDF3.0 and PDF4LHC21 give similar predictions.
Similar features are seen in the corresponding \embb\ distributions.

All the predictions model the \ttem\ and \embb\ \ptll\ distributions well
at low \ptll\ (Figure~\ref{f:distresb}(a)), but the {\scr Powheg\,+\,Pythia8}
samples give a too-hard distribution at high \ptll.  No prediction models
the \mll\ distributions (Figure~\ref{f:distresb}(b)) at low \mll, where there
is a clear data excess. \powhegminnlo\ provides the best modelling
of the intermediate $100<\ptll<200$\,\GeV\ range, whilst the NNLO-reweighted
{\scr Powheg\,+\,Pythia8} sample undershoots the data at high \mll.
The poor modelling of \mll\ is also seen in the double-differential
cross-sections shown in Figures~\ref{f:distresf}--\ref{f:distresh}, where
all predictions except \powhegminnlo\ have very poor $\chi^2$ values in
Table~\ref{t:reschitwod}.

Most of the predictions show a clear slope with respect to the data for
\dphill\ (Figure~\ref{f:distresc}(b)), but bin-to-bin correlations in the
uncertainties in the predictions (in particular from QCD renormalisation scale
variations) result
in the $\chi^2$ values still being acceptable, with the exception
of {\scr Powheg\,+\,Herwig7}. The double-differential
cross-section of \dphill\ in bins of \mll\ (Figure~\ref{f:distresh}) suggests
that the slope between data and predictions increases with \mll.
The NNLO-reweighted {\scr Powheg\,+\,Pythia8} sample models this distribution
well, without a significant slope except for $\mll>200$\,\GeV.
However, the \powhegminnlo\ sample does not model this distribution significantly
better than {\scr Powheg\,+\,Pythia8}.

The \ttem\ data is softer than all predictions for \ptsum\
(Figure~\ref{f:distresd}(a)), with \powhegminnlo\ coming closest, but still with
a $\chi^2$ probability of only around 1\%. These predictions are in better
agreement with the \embb\ \ptsum\ data distribution, where \powhegbbfl\
shows some tension. For \esum\ (Figure~\ref{f:distresd}(b)), all the
predictions except \powhegminnlo\ and NNLO-reweighted {\scr Powheg\,+\,Pythia8}
are harder than the data at low \esum, but the $\chi^2$ values are still
acceptable, except for the \embb\ prediction from {\scr Powheg\,+\,Pythia8}
with the diagram removal scheme for $Wt$ events.

\section{Conclusions}\label{s:conc}

The inclusive \ttbar\ production cross-section \xtt\ has been measured in
$pp$ collisions at \sxyt\ using the \intlumi\,\ifb\ ATLAS dataset recorded
at the LHC in 2015--18. Using events with an opposite-charge $e\mu$ pair
and $b$-tagged jets, the result is:
\begin{eqnarray}\nonumber
\xtt & = & \ttxval \pm \ttxstat \pm \ttxsyst \pm \ttxlumi \pm \ttxebeam\,\mathrm{pb},
\end{eqnarray}
where the four uncertainties reflect the limited size of the data sample,
experimental and theoretical systematic effects, uncertainty in the integrated
luminosity and the knowledge of the LHC beam energy, giving a total uncertainty
of \ttxrel. The result is the most precise achieved to date, and is consistent
with NNLO+NNLL QCD predictions. Fiducial cross-sections corresponding
to the lepton acceptance have also been measured, with or without the
contribution from leptonic decays of tau leptons. The dependence of predictions
for \xtt\ on the top quark pole mass has been exploited to determine
\[
\mtpole = \mtpval^{+\mtpeup}_{-\mtpedn}\,\GeV\ ,
\]
using the NNPDF3.1\_notop PDF set. This result is consistent with other
determinations of \mtop\ using a variety of techniques.

The same dataset has been used to measure ten single-differential and
three double-differential cross-sections as functions of lepton and dilepton
kinematic variables, for both the \ttem\ process and the \embb\ final state
including contributions from both \ttbar\ and $Wt$ processes. Uncertainties
as small as 0.3\% have been achieved for normalised distributions in some
parts of the fiducial regions. Comparisons with event generator predictions
show that state-of-the-art generators such as {\scr Powheg} MiNNLO or
{\scr Powheg} {\em bb4l} better model the lepton kinematics than the
{\scr Powheg} {\ttr hvq} process traditionally used for LHC physics analyses.
These precise measurements
provide input that can be used to further refine the modelling of
top quark production at hadron colliders.

\section*{Acknowledgements}
%

%

%
%

%
%

We thank CERN for the very successful operation of the LHC and its injectors, as well as the support staff at
CERN and at our institutions worldwide without whom ATLAS could not be operated efficiently.

The crucial computing support from all WLCG partners is acknowledged gratefully, in particular from CERN, the ATLAS Tier-1 facilities at TRIUMF/SFU (Canada), NDGF (Denmark, Norway, Sweden), CC-IN2P3 (France), KIT/GridKA (Germany), INFN-CNAF (Italy), NL-T1 (Netherlands), PIC (Spain), RAL (UK) and BNL (USA), the Tier-2 facilities worldwide and large non-WLCG resource providers. Major contributors of computing resources are listed in Ref.~\cite{ATL-SOFT-PUB-2026-001}.

We gratefully acknowledge the support of ANPCyT, Argentina; YerPhI, Armenia; ARC, Australia; BMWFW and FWF, Austria; ANAS, Azerbaijan; CNPq and FAPESP, Brazil; NSERC, NRC and CFI, Canada; CERN; ANID, Chile; CAS, MOST and NSFC, China; Minciencias, Colombia; MEYS CR, Czech Republic; DNRF and DNSRC, Denmark; IN2P3-CNRS and CEA-DRF/IRFU, France; SRNSFG, Georgia; BMFTR, HGF and MPG, Germany; GSRI, Greece; RGC and Hong Kong SAR, China; ICHEP and Academy of Sciences and Humanities, Israel; INFN, Italy; MEXT and JSPS, Japan; CNRST, Morocco; NWO, Netherlands; RCN, Norway; MNiSW, Poland; FCT, Portugal; MNE/IFA, Romania; MSTDI, Serbia; MSSR, Slovakia; ARIS and MVZI, Slovenia; DSI/NRF, South Africa; MICIU/AEI, Spain; SRC and Wallenberg Foundation, Sweden; SERI, SNSF and Cantons of Bern and Geneva, Switzerland; NSTC, Taipei; TENMAK, T\"urkiye; STFC/UKRI, United Kingdom; DOE and NSF, United States of America.

Individual groups and members have received support from BCKDF, CANARIE, CRC and DRAC, Canada; CERN-CZ, FORTE and PRIMUS, Czech Republic; COST, ERC, ERDF, Horizon 2020, ICSC-NextGenerationEU and Marie Sk{\l}odowska-Curie Actions, European Union; Investissements d'Avenir Labex, Investissements d'Avenir Idex and ANR, France; DFG and AvH Foundation, Germany; Herakleitos, Thales and Aristeia programmes co-financed by EU-ESF and the Greek NSRF, Greece; BSF-NSF and MINERVA, Israel; NCN and NAWA, Poland; La Caixa Banking Foundation, CERCA Programme Generalitat de Catalunya and PROMETEO and GenT Programmes Generalitat Valenciana, Spain; G\"{o}ran Gustafssons Stiftelse, Sweden; The Royal Society and Leverhulme Trust, United Kingdom; United States of America.

In addition, individual members wish to acknowledge support from CERN: European Organization for Nuclear Research (CERN DOCT); Chile: Agencia Nacional de Investigaci\'on y Desarrollo (ANID FONDECYT reg. 1230987, FONDECYT 1230812, FONDECYT 1240864, Fondecyt 3240661, Fondecyt Regular 1240721); China: Chinese Ministry of Science and Technology (MOST-2023YFA1605700, MOST-2023YFA1609300), National Natural Science Foundation of China (NSFC - 12175119, NSFC 12275265); Czech Republic: Czech Science Foundation (GACR - 24-11373S), Ministry of Education Youth and Sports (ERC-CZ-LL2327, FORTE CZ.02.01.01/00/22\_008/0004632), PRIMUS Research Programme (PRIMUS/21/SCI/017); EU: H2020 European Research Council (ERC - 101002463); European Union: European Research Council (BARD No. 101116429, ERC - 948254, ERC 101089007), European Regional Development Fund (HE COFUND GA No.101081355, ERDF), European Union, Future Artificial Intelligence Research (FAIR-NextGenerationEU PE00000013), Italian Center for High Performance Computing, Big Data and Quantum Computing (ICSC, NextGenerationEU), Marie Sklodowska-Curie Actions (GAP-101168829); France: Agence Nationale de la Recherche (ANR-21-CE31-0013, ANR-21-CE31-0022, ANR-22-EDIR-0002, ANR-24-CE31-0504-01); Germany: Deutsche Forschungsgemeinschaft (DFG - 469666862, DFG - CR 312/5-2); China: Research Grants Council (GRF); Italy: Istituto Nazionale di Fisica Nucleare (ICSC, NextGenerationEU), Ministero dell'Università e della Ricerca (NextGenEU 153D23001490006 M4C2.1.1, NextGenEU I53D23000820006 M4C2.1.1, NextGenEU I53D23001490006 M4C2.1.1, SOE2024\_0000023); Japan: Japan Society for the Promotion of Science (JSPS KAKENHI JP22H01227, JSPS KAKENHI JP22H04944, JSPS KAKENHI JP22KK0227, JSPS KAKENHI JP24K23939, JSPS KAKENHI JP24KK0251, JSPS KAKENHI JP25H00650, JSPS KAKENHI JP25H01291, JSPS KAKENHI JP25K01023); Norway: Research Council of Norway (RCN-314472); Poland: Ministry of Science and Higher Education (IDUB AGH, POB8, D4 no 9722), Polish National Science Centre (NCN 2021/42/E/ST2/00350, NCN OPUS 2023/51/B/ST2/02507, NCN OPUS nr 2022/47/B/ST2/03059, NCN UMO-2019/34/E/ST2/00393, UMO-2022/47/O/ST2/00148, UMO-2023/49/B/ST2/04085, UMO-2023/51/B/ST2/00920, UMO-2024/53/N/ST2/00869); Portugal: Foundation for Science and Technology (FCT); Spain: Agencia de Gestión de Ayudas Universitarias y de Investigación (AGAUR - 2023 BP 00141), Generalitat Valenciana (ASFAE/2022/008), Ministry of Science and Innovation (MCIN \& NextGenEU PCI2022-135018-2, MICIN \& FEDER PID2021-125273NB, RYC2019-028510-I, RYC2020-030254-I, RYC2021-031273-I, RYC2022-038164-I), Ministerio de Ciencia, Innovación y Universidades/Agencia Estatal de Investigaci\'on (PID2022-142604OB-C22); Sweden: Carl Trygger Foundation (Carl Trygger Foundation CTS 22:2312), Swedish Research Council (Swedish Research Council 2023-04654, VR 2021-03651, VR 2022-03845, VR 2022-04683, VR 2023-03403, VR 2024-05451), Knut and Alice Wallenberg Foundation (KAW 2018.0458, KAW 2022.0358, KAW 2023.0366); Switzerland: Swiss National Science Foundation (SNSF - PCEFP2\_194658); United Kingdom: Royal Society (NIF-R1-231091); United States of America: U.S. Department of Energy (ECA DE-AC02-76SF00515), Neubauer Family Foundation.

%
%


%
\clearpage
\printbibliography
\clearpage

%

\section{Addendum: quasi-bound-state formation near the $\boldmath\ttbar$
threshold}

The recent ATLAS measurement of the normalised $e\mu$ dilepton invariant mass
and azimuthal angle distributions in $\sqrt{s}=13$\,\TeV\ $pp$ collisions at
the Large Hadron Collider is extended to study the sensitivity
to the formation of quasi-bound states near the \ttbar\ threshold.
The measured differential distributions are compared with \ttbar\ models
incorporating perturbative QCD predictions for the hard process, with or
without the addition of colour-singlet quasi-bound-states. The
data are better described by the predictions incorporating quasi-bound-state
formation. Fits to the dilepton invariant mass distribution show
evidence for the latter process with an observed significance exceeding
three standard deviations, and a measured cross-section in agreement with
dedicated studies of the threshold region.

\subsection{Introduction}\label{sa:intro}

The study of top-quark--antiquark (\ttbar) production at the CERN Large Hadron
Collider (LHC) allows quantum chromodynamics to be probed at some of the
highest accessible energy scales, and forms a key part of the LHC experimental
programme. Recently, the modelling of the \ttbar\ threshold region, where the
\ttbar\ invariant mass \mtt\ is close to twice the top quark mass of
$\mtop\approx 172.5$\,\GeV, has received particular attention.
Using combinations of \mtt\ and angular variables, both
CMS~\cite{CMS-TOP-24-007} and ATLAS~\cite{TOPQ-2025-11} have reported excesses
at low \mtt\ compared with the predictions of standard \ttbar\ event generators
based on a perturbative QCD (pQCD) description of the hard-scattering process,
and that appear compatible with the formation of a quasi-bound state
close to threshold.

Precise measurements of lepton and dilepton kinematic distributions in \ttbar\
events provide important information for developing and tuning event generator
models of \ttbar\ production. The recent ATLAS measurement of lepton differential
cross-sections in $\ttbar\rightarrow e\mu\nu\bar{\nu}\bbbar$ events from
140\,\ifb\ of $pp$ collisions at \sxyt ~\cite{TOPQ-2024-12} provides a
wide range of precise differential measurements including the dilepton invariant
mass \mll\ and dilepton azimuthal angular difference \dphill, as well as the
measurement of \dphill\ in coarse bins of \mll. The invariant mass distribution
shows a significant data excess in the region $\mll<100$\,\GeV\ when compared
with pQCD predictions from \ttbar\ event generators.
This excess was also present in previous
measurements of \mll\ with a 36\,\ifb\ dataset at \sxyt~\cite{TOPQ-2018-17}, and
at \sxvt~\cite{TOPQ-2015-02}. In Ref.~\cite{oniastat}, the \mll, \dphill\
and two-dimensional \dphillvmll\ distributions
measured in Ref.~\cite{TOPQ-2018-17}, along with related measurements
from CMS~\cite{CMS-TOP-17-014}, were fitted to a combination of pQCD and
quasi-bound-state predictions, showing that the data can be better described
by predictions including the latter contribution, and claiming evidence for
the formation of toponium. These studies were performed
using the HEPData~\cite{hepdatattx5} records of the measured distributions,
and did not take into account bin-to-bin correlations in the ATLAS data.

This addendum to Ref.~\cite{TOPQ-2024-12} follows a similar approach,
comparing the distributions measured
in Ref.~\cite{TOPQ-2024-12} with various predictions with or without
quasi-bound-state contributions, normalised according to a theoretical
prediction. In a second step, the \mll, \dphill\ and
\dphillvmll\ distributions are also fitted to predictions where the normalisation
of the quasi-bound-state contribution is left free. The analysis exploits the
precision of the full ATLAS Run~2 dataset, and takes into account bin-to-bin
correlations in the uncertainties in both the experimental measurements and
the predictions. The predictions are described in Section~\ref{sa:mcmod}
and the analysis methodology in Section~\ref{sa:meth}. Results are given
in Section~\ref{sa:res} and conclusions drawn in Section~\ref{sa:conc}.

\subsection{Simulated event samples}\label{sa:mcmod}

The measured distributions were compared with particle-level predictions in the
fiducial region defined by the \ttem\  measurement~\cite{TOPQ-2024-12},
namely an opposite-charge
$e\mu$ pair, with each lepton satisfying $\pt>20$\,\GeV\ and $|\eta|<2.5$, with
no requirements on jets.\footnote{ATLAS uses a right-handed coordinate system
with its origin at
the nominal interaction point in the centre of the detector, and the $z$ axis
along the beam line. Pseudorapidity is defined in terms of the polar angle
$\theta$ as $\eta=-\ln\tan{\theta/2}$, and transverse momentum and energy
are defined relative to the beam line as $\pt=p\sin\theta$ and
$\et=E\sin\theta$. The azimuthal angle around the beam line is denoted by
$\phi$.
The rapidity is defined as $y=\frac{1}{2}\ln\left(\frac{E+p_z}{E-p_z}\right)$,
where $p_z$ is the $z$-component of the momentum and $E$ is the energy
of the relevant object or system.} For the pQCD \ttbar\ predictions covering
the inclusive phase space, the following three models were considered,
all of which are described in more detail in Ref.~\cite{TOPQ-2024-12}:
\begin{description}
\item[{\scr Powheg\,+\,Pythia8}:] The QCD next-to-leading-order (NLO) matrix
element for \ttbar\ production implemented in the {\ttr hvq} process in
{\scr Powheg Box} v2~\cite{powheghvq}, interfaced to
{\scr Pythia8}~\cite{pythia82}.
\item[{\scr Powheg\,+\,Pythia8}] NNLO rew: The {\scr Powheg\,+\,Pythia8} sample
described above, reweighted to bring the top quark kinematics into agreement
with the predictions of the NNLO QCD\,+\,NLO electroweak calculation
of Ref.~\cite{ttnnlocalc} as described in Ref.~\cite{TOPQ-2020-14}.
The reweighting was performed in the three variables $p_\mathrm{T}(t)$, \mtt\ and
$p_\mathrm{T}(\ttbar)$, and is not the same reweighting as used in the
dedicated ATLAS \ttbar\ threshold study~\cite{TOPQ-2025-11}, which uses
a different set of variables.
\item[{\scr Powheg MiNNLO\,+\,Pythia8}:] The prediction obtained from
the \minnlops\ matched NNLO QCD plus parton shower procedure~\cite{powhegminnlo},
implemented in {\scr Powheg}~\cite{powheg,powheg2,powheg3} and interfaced
to {\scr Pythia8}.
\end{description}
The latter two predictions give a better overall description of the lepton
and dilepton kinematics than the first prediction from
{\scr Powheg\,+\,Pythia8}. For comparison with the \embb\ distributions
measured in Ref.~\cite{TOPQ-2024-12}, which include both \ttbar\ and $Wt$
contributions, and also make fiducial requirements
on the presence of two $b$-tagged jets with $\pt>25$\,\GeV\ and $|\eta|<2.5$,
the \powhegbbfl\ sample was also considered. This sample was generated
with the {\scr Powheg} $bb4l$ generator~\cite{jezobb4l}, which
includes the complete set of $\ell\ell'\nu\bar{\nu}\bbbar$
final states, taking into account interference between the \ttbar\
and $Wt$ final states as well as off-shell and non-resonant effects.
All these samples use the NNPDF3.0 parton distribution function (PDF)
set~\cite{nnpdf3} in the matrix element calculation, the {\scr Pythia8}
A14 tune~\cite{ATL-PHYS-PUB-2014-021}, {\scr EvtGen}~\cite{evtgen}
to handle the decay of $b$- and $c$-flavoured hadrons,
and assume $\mtop=172.5$\,\GeV.

The quasi-bound-state effects were modelled using the two approaches
described in Ref.~\cite{TOPQ-2025-11}, in each case adding to the pQCD
\ttbar\ predictions discussed above. The first approach describes the
formation of a quasi-bound colour-singlet \ttbar\ state using the Green's
function of non-relativistic QCD (NRQCD) in the Coulomb gauge~\cite{topofuks}.
The results were used to reweight the production and
decay of a six-body final state generated with
{\scr MadGraph 3.4.2}~\cite{madgraph} and the NNPDF3.0 PDF set, interfaced to
{\scr Pythia8}.
The inclusive production cross-section was set to 6.43\,pb, as obtained from
analytic calculations~\cite{toponiumxsec}. This model is referred to as
Green's function reweighting or
\ttonia. The second approach is that originally used in the
CMS publication~\cite{CMS-TOP-24-007}, i.e.\ a simplified model with a
pseudo-scalar resonance \etatt\ of mass 343\,\GeV\ and width 2.8\,\GeV.
The resonance was simulated
using {\scr MadGraph 3.5.5} to generate the four-body $WbWb$ final state
that was then interfaced to {\scr Pythia8}. The same cross-section as
for the \ttonia\ approach was assumed.

\subsection{Methodology}\label{sa:meth}

The consistency of each prediction with a measured normalised differential
cross-section was quantified using a $\chi^2$ calculated as
\begin{eqnarray}\label{ea:redchi}
\chi^2 = {\mathbf\Delta}_{(n-1)}^T {\mathbf S}^{-1}_{(n-1)} {\mathbf\Delta}_{(n-1)}\,,
\end{eqnarray}
where ${\mathbf\Delta}_{(n-1)}$ is the vector of differences between the measured
and predicted normalised differential cross-section in each of the $n$
bins, excluding the last one, and ${\mathbf S}^{-1}_{(n-1)}$ is the inverse of
the corresponding
covariance matrix, including both the experimental uncertainties in the
measurement and the uncertainties in the predictions, and their
bin-to-bin correlations through off-diagonal terms. The last bin of each
distribution was excluded to account for the normalisation condition.\footnote{
The $\chi^2$ does not depend on which bin is excluded.}
The uncertainties in the pQCD predictions were calculated as discussed in
Ref.~\cite{TOPQ-2024-12}, and include PDF, QCD renormalisation and factorisation
scale, initial/final state radiation (IFSR) and top quark mass variations.
The NNPDF3.0 replica sets were used to calculate the bin-to-bin correlations
for the PDF uncertainty, whilst the QCD scale, radiation and top quark mass
uncertainties were each assumed to be fully correlated between bins.
For predictions
that include a component from quasi-bound-state effects, added with a fixed
normalisation corresponding to
the assumed cross-section of 6.43\,pb, the same relative uncertainties as in
the pQCD-alone predictions were assumed, and applied to the total prediction.

In a second interpretation, the quasi-bound-state prediction was added to the
pQCD prediction with a normalisation given by a signal strength parameter
$\mu$, where $\mu=1$ corresponds to the predicted cross-section of 6.43\,pb.
The $\chi^2$ of Eq.~(\ref{ea:redchi}) was then minimised to determine the
best-fit value of $\mu$ and its uncertainty. These fits were performed
for the \mll, \dphill\ and \dphillvmll distributions only.
Separate uncertainties for PDF, QCD renormalisation and factorisation
scales, IFSR and the use of {\scr Herwig7}~\cite{herwigpp,herwig71}
rather than {\scr Pythia8} for parton shower
and hadronisation were calculated for the quasi-bound-state contribution, and
considered independent of the uncertainties in the pQCD component.

\subsection{Results}\label{sa:res}

The $\chi^2$ of Eq.~(\ref{ea:redchi}) was calculated for each pQCD model
alone, and for each model with the addition of the \ttonia\ or \etatt\
quasi-bound-state model. The $\chi^2$ values and associated probabilities
for each of the ten one-dimensional
distributions measured in Ref.~\cite{TOPQ-2024-12} are shown in
Table~\ref{ta:reschioned}, and for the three two-dimensional distributions
of \etal, \rapll\ and \dphill\ in bins of \mll\ in Table~\ref{ta:reschitwod}.
The comparisons for predictions involving  \powhegbbfl\ are based on
the measured \embb\ distributions, and the others on the \ttem\ distributions.
The values for the pQCD-alone predictions are the same as those in
Ref.~\cite{TOPQ-2024-12}.

\begin{table}[tp]
\caption{\label{ta:reschioned}$\chi^2$ values (top block) and associated
probabilities
(bottom block) for comparison of normalised measured single-differential
fiducial cross-sections with various predictions without and with
quasi-bound-state contributions simulated with the \ttonia\ and \etatt\ models.
Within each block, the last three rows correspond to comparisons
of the \embb\ cross-sections with combined \ttbar+$Wt$ samples, and the
other rows to \ttem\ cross-sections compared with \ttbar\ samples.}
\footnotesize

\hspace{-10mm}
\begin{tabular}{lrrrrrrrrrr}\toprule
Distribution & \ptl & \etal & \ptll & \mll & \rapll & \dphill & \ptsum & \esum & \ptlmax & \ptlmin \\
$N_\mathrm{dof}$ & 12 & 8 & 10 & 14 & 8 & 19 & 9 & 11 & 11 & 10 \\ \midrule
$\chi^2$ values & \\ \cmidrule{1-1}{\scr Powheg\,+\,PY8}&  35.2 &  19.7 &  12.5 &  49.8 &  11.6 &  22.7 &  34.9 &  14.7 &  32.0 &  27.4 \\
{\scr Powheg\,+\,PY8}\,+\,\ttonia&  30.1 &  20.0 &  11.1 &  29.0 &  11.6 &  19.4 &  31.4 &  11.6 &  26.4 &  24.9 \\
{\scr Powheg\,+\,PY8}\,+\,\etatt&  30.6 &  20.1 &  11.1 &  30.0 &  11.6 &  19.4 &  31.6 &  12.0 &  26.9 &  25.1 \\
{\scr Powheg\,+\,PY8} NNLO rew&  16.4 &  14.9 &  11.7 &  33.6 &  12.8 &  16.5 &  24.6 &  11.3 &  14.3 &  14.6 \\
{\scr Powheg\,+\,PY8} NNLO rew\,+\,\ttonia&  16.7 &  15.1 &  15.4 &  23.8 &  13.0 &  18.8 &  24.6 &  10.4 &  14.1 &  14.2 \\
{\scr Powheg\,+\,PY8} NNLO rew\,+\,\etatt&  16.6 &  15.2 &  15.3 &  23.8 &  13.0 &  18.5 &  24.6 &  10.6 &  14.1 &  14.4 \\
{\scr Powheg MiNNLO\,+\,PY8}&  15.4 &  12.3 &   7.2 &  25.9 &  11.7 &  23.6 &  20.9 &   7.5 &  12.1 &  12.5 \\
{\scr Powheg MiNNLO\,+\,PY8}\,+\,\ttonia&  14.4 &  12.4 &   7.3 &  16.5 &  11.9 &  16.7 &  20.7 &   6.6 &  11.0 &  12.0 \\
{\scr Powheg MiNNLO\,+\,PY8}\,+\,\etatt&  14.4 &  12.5 &   7.3 &  16.5 &  11.9 &  17.2 &  20.6 &   6.7 &  10.9 &  12.0 \\
\cmidrule{1-1}{\scr Powheg} $bb4l$\,+\,{\scr PY8}&  15.5 &  11.2 &   3.1 &  20.5 &  14.7 &  21.4 &  20.5 &  18.6 &  19.2 &  21.8 \\
{\scr Powheg} $bb4l$\,+\,{\scr PY8}\,+\,\ttonia&  14.8 &  11.6 &   4.1 &  14.1 &  14.7 &  19.9 &  18.6 &  15.6 &  18.3 &  19.4 \\
{\scr Powheg} $bb4l$\,+\,{\scr PY8}\,+\,\etatt&  14.6 &  11.6 &   4.1 &  14.0 &  14.7 &  19.9 &  18.8 &  15.9 &  18.2 &  19.5 \\
\midrule[\heavyrulewidth]
$\chi^2$ probabilities & \\
\cmidrule{1-1}{\scr Powheg\,+\,PY8}& $4\ 10^{-4}$ & 0.011 & 0.25 & $7\ 10^{-6}$ & 0.17 & 0.25 & $6\ 10^{-5}$ & 0.20 & $8\ 10^{-4}$ & $2\ 10^{-3}$ \\
{\scr Powheg\,+\,PY8}\,+\,\ttonia& $3\ 10^{-3}$ & 0.010 & 0.35 & 0.011 & 0.17 & 0.43 & $3\ 10^{-4}$ & 0.39 & $6\ 10^{-3}$ & $6\ 10^{-3}$ \\
{\scr Powheg\,+\,PY8}\,+\,\etatt& $2\ 10^{-3}$ & 0.010 & 0.35 & $8\ 10^{-3}$ & 0.17 & 0.43 & $2\ 10^{-4}$ & 0.37 & $5\ 10^{-3}$ & $5\ 10^{-3}$ \\
{\scr Powheg\,+\,PY8} NNLO rew& 0.18 & 0.061 & 0.31 & $2\ 10^{-3}$ & 0.12 & 0.62 & $3\ 10^{-3}$ & 0.41 & 0.22 & 0.15 \\
{\scr Powheg\,+\,PY8} NNLO rew\,+\,\ttonia& 0.16 & 0.057 & 0.12 & 0.048 & 0.11 & 0.47 & $4\ 10^{-3}$ & 0.49 & 0.23 & 0.16 \\
{\scr Powheg\,+\,PY8} NNLO rew\,+\,\etatt& 0.17 & 0.056 & 0.12 & 0.049 & 0.11 & 0.49 & $3\ 10^{-3}$ & 0.47 & 0.23 & 0.16 \\
{\scr Powheg MiNNLO\,+\,PY8}& 0.22 & 0.14 & 0.71 & 0.026 & 0.16 & 0.21 & 0.013 & 0.76 & 0.35 & 0.25 \\
{\scr Powheg MiNNLO\,+\,PY8}\,+\,\ttonia& 0.28 & 0.13 & 0.70 & 0.28 & 0.16 & 0.61 & 0.014 & 0.83 & 0.44 & 0.29 \\
{\scr Powheg MiNNLO\,+\,PY8}\,+\,\etatt& 0.28 & 0.13 & 0.70 & 0.28 & 0.16 & 0.58 & 0.014 & 0.83 & 0.45 & 0.28 \\
\cmidrule{1-1}{\scr Powheg} $bb4l$\,+\,{\scr PY8}& 0.22 & 0.19 & 0.98 & 0.11 & 0.064 & 0.32 & 0.015 & 0.069 & 0.057 & 0.016 \\
{\scr Powheg} $bb4l$\,+\,{\scr PY8}\,+\,\ttonia& 0.25 & 0.17 & 0.94 & 0.45 & 0.065 & 0.40 & 0.029 & 0.16 & 0.076 & 0.035 \\
{\scr Powheg} $bb4l$\,+\,{\scr PY8}\,+\,\etatt& 0.27 & 0.17 & 0.94 & 0.45 & 0.065 & 0.40 & 0.027 & 0.14 & 0.078 & 0.034 \\
\bottomrule
\end{tabular}
\end{table}

\begin{table}[tp]
\caption{\label{ta:reschitwod}$\chi^2$ values (top block) and associated
probabilities
(bottom block) for comparison of normalised measured double-differential
fiducial cross-sections with various simulation
samples. Within each block, the last three rows correspond to comparisons
of the \embb\ cross-sections with combined \ttbar+$Wt$ samples, and the
other rows to \ttem\ cross-sections compared with \ttbar\ samples.}
\small\centering

\begin{tabular}{lrrr}\toprule
Distribution & \etalvmll & \rapllvmll & \dphillvmll \\
$N_\mathrm{dof}$ & 35 & 19 & 39 \\ \midrule
$\chi^2$ values & \\ \cmidrule{1-1}{\scr Powheg\,+\,PY8}&  72.7 &  46.0 &  79.9 \\
{\scr Powheg\,+\,PY8}\,+\,\ttonia&  60.6 &  33.7 &  63.2 \\
{\scr Powheg\,+\,PY8}\,+\,\etatt&  61.0 &  34.4 &  64.2 \\
{\scr Powheg\,+\,PY8} NNLO rew&  71.0 &  33.8 &  71.0 \\
{\scr Powheg\,+\,PY8} NNLO rew\,+\,\ttonia&  64.4 &  27.1 &  71.2 \\
{\scr Powheg\,+\,PY8} NNLO rew\,+\,\etatt&  64.6 &  27.4 &  71.3 \\
{\scr Powheg MiNNLO\,+\,PY8}&  44.8 &  26.7 &  50.2 \\
{\scr Powheg MiNNLO\,+\,PY8}\,+\,\ttonia&  38.1 &  21.4 &  44.3 \\
{\scr Powheg MiNNLO\,+\,PY8}\,+\,\etatt&  37.9 &  21.3 &  44.0 \\
\cmidrule{1-1}{\scr Powheg} $bb4l$\,+\,{\scr PY8}&  47.8 &  23.7 &  66.0 \\
{\scr Powheg} $bb4l$\,+\,{\scr PY8}\,+\,\ttonia&  45.6 &  18.0 &  65.9 \\
{\scr Powheg} $bb4l$\,+\,{\scr PY8}\,+\,\etatt&  45.3 &  18.1 &  65.9 \\
\midrule[\heavyrulewidth]
$\chi^2$ probabilities & \\
\cmidrule{1-1}{\scr Powheg\,+\,PY8}& $2\ 10^{-4}$ & $5\ 10^{-4}$ & $1\ 10^{-4}$ \\
{\scr Powheg\,+\,PY8}\,+\,\ttonia& $5\ 10^{-3}$ & 0.020 & $8\ 10^{-3}$ \\
{\scr Powheg\,+\,PY8}\,+\,\etatt& $4\ 10^{-3}$ & 0.017 & $7\ 10^{-3}$ \\
{\scr Powheg\,+\,PY8} NNLO rew& $3\ 10^{-4}$ & 0.020 & $1\ 10^{-3}$ \\
{\scr Powheg\,+\,PY8} NNLO rew\,+\,\ttonia& $2\ 10^{-3}$ & 0.10 & $1\ 10^{-3}$ \\
{\scr Powheg\,+\,PY8} NNLO rew\,+\,\etatt& $2\ 10^{-3}$ & 0.096 & $1\ 10^{-3}$ \\
{\scr Powheg MiNNLO\,+\,PY8}& 0.12 & 0.11 & 0.11 \\
{\scr Powheg MiNNLO\,+\,PY8}\,+\,\ttonia& 0.33 & 0.32 & 0.26 \\
{\scr Powheg MiNNLO\,+\,PY8}\,+\,\etatt& 0.34 & 0.32 & 0.27 \\
\cmidrule{1-1}{\scr Powheg} $bb4l$\,+\,{\scr PY8}& 0.073 & 0.21 & $4\ 10^{-3}$ \\
{\scr Powheg} $bb4l$\,+\,{\scr PY8}\,+\,\ttonia& 0.11 & 0.52 & $4\ 10^{-3}$ \\
{\scr Powheg} $bb4l$\,+\,{\scr PY8}\,+\,\etatt& 0.11 & 0.51 & $5\ 10^{-3}$ \\
\bottomrule
\end{tabular}
\end{table}

The measured absolute \ttem\ and \embb\ differential
cross-sections as functions of \mll\ and \dphill\ are shown in
Figure~\ref{fa:distresa}, together with panels showing the ratio of each
normalised prediction to data, demonstrating the effect of adding the \ttonia\
or \etatt\ prediction to each pQCD \ttbar\ or $\ttbar+Wt$ model.
The two-dimensional \dphillvmll\ distribution and corresponding ratios
are shown in Figure~\ref{fa:distresb}.

\begin{figure}[tp]
\parbox{83mm}{\includegraphics[width=76mm]{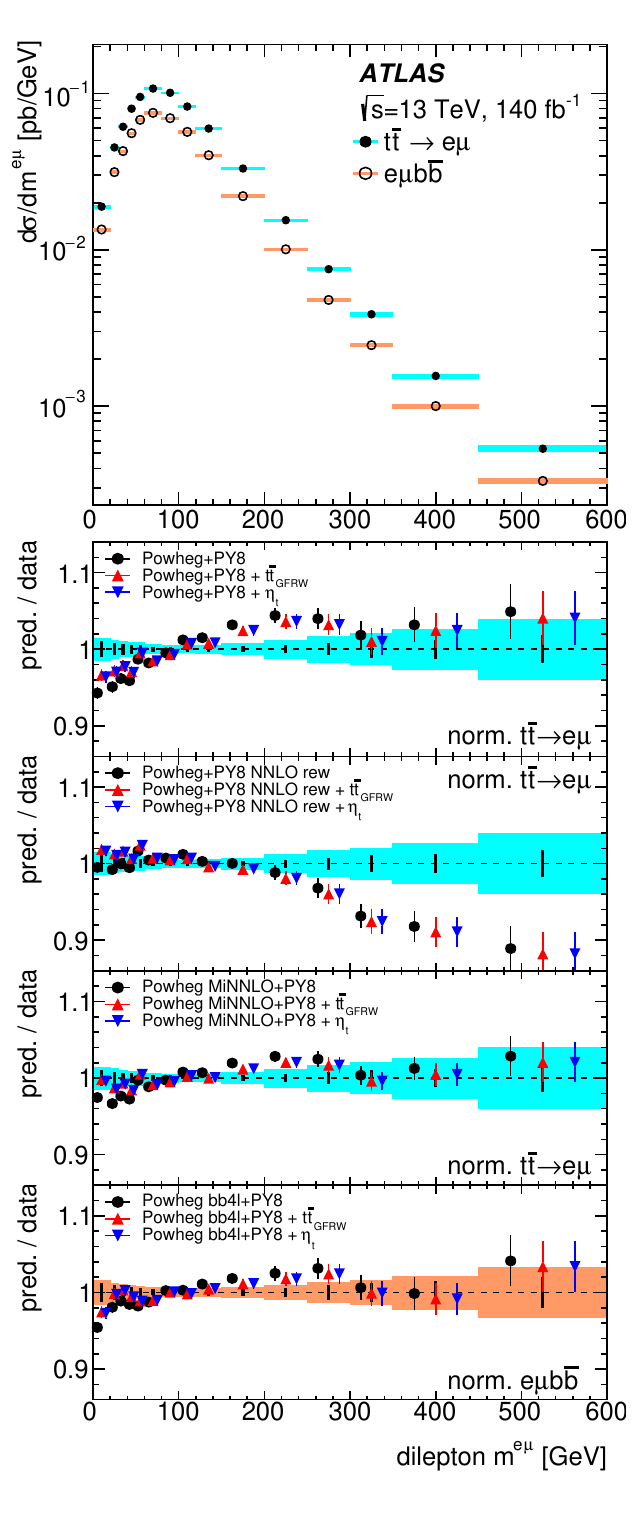}\vspace{-7mm}\center{(a)}}
\parbox{83mm}{\includegraphics[width=76mm]{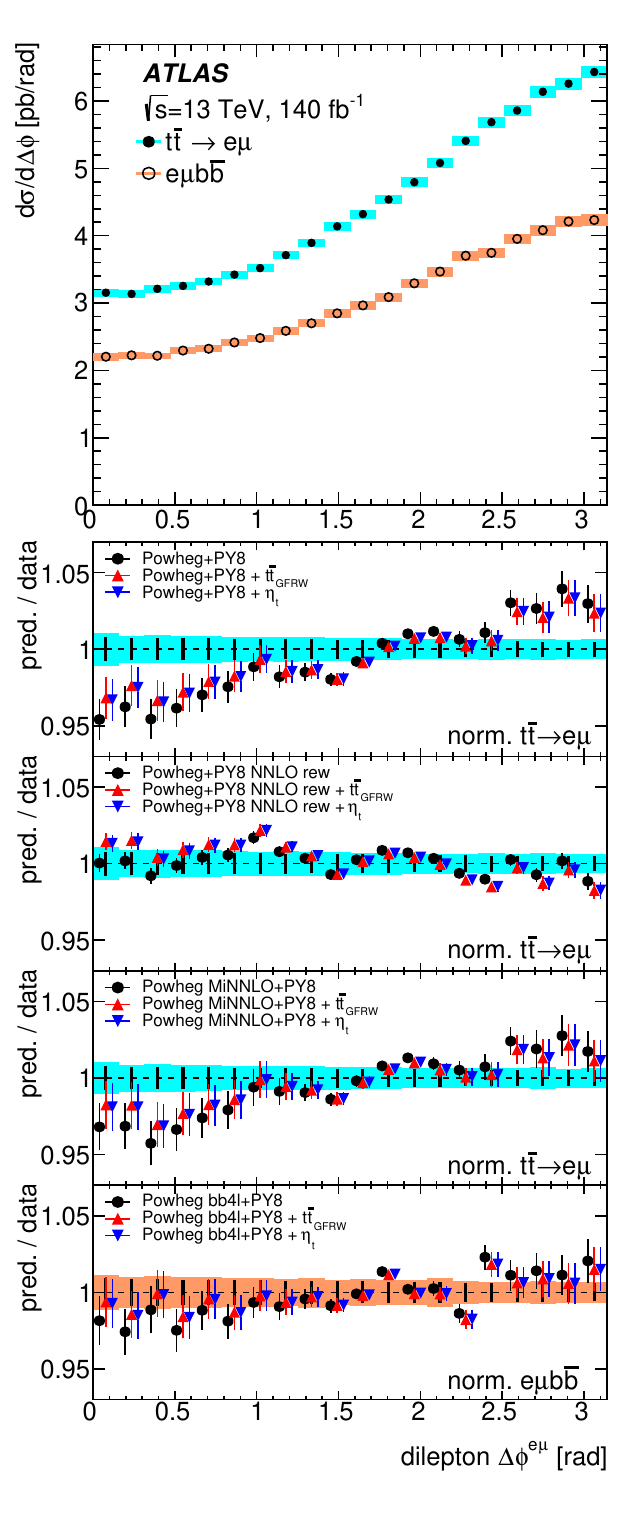}\vspace{-7mm}\center{(b)}}
\caption{\label{fa:distresa}Differential cross-sections as functions of (a) \mll\
and (b) \dphill. The upper panels show the measured absolute \ttem\ (filled points)
and \embb\ (open points) cross-sections, including the contributions from
$W\rightarrow\tau\rightarrow e/\mu$ decays, with the total uncertainties shown
by the shaded bands. The other panels show ratios of various predictions
without and with quasi-bound-state contributions simulated with the \ttonia\
and \etatt\ models to the
measured normalised differential cross-sections, for \ttem\ (middle three panels)
and \embb\ (lower panels). The markers showing the ratios for each prediction
are offset from the bin centres for better visibility, and the total
uncertainty in the prediction is shown by the error bar. The data statistical
uncertainty is shown by the black error bars and the total uncertainty by the
cyan (\ttem) or orange (\embb) band around unity.}
\end{figure}

\begin{figure}[tp]
\centering

\includegraphics[width=160mm]{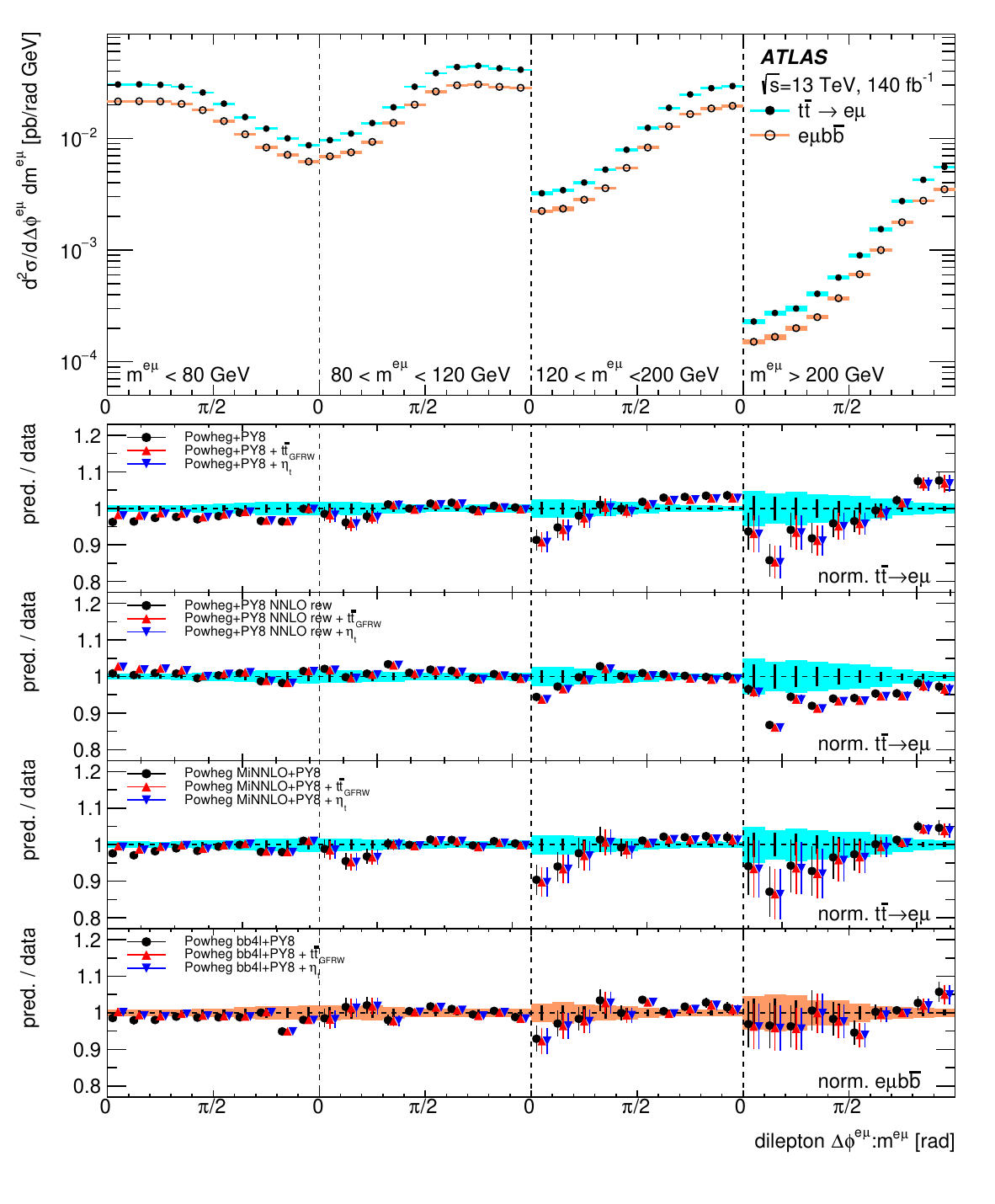}
\caption{\label{fa:distresb}Double-differential cross-section as a function
of \dphill\ and \mll, showing the unrolled cross-sections as a function of
\dphill\ in four bins of \mll. The upper panel shows the measured absolute \ttem\
(filled points)
and \embb\ (open points) cross-sections, including the contributions from
$W\rightarrow\tau\rightarrow e/\mu$ decays, with the total uncertainties shown
by the shaded bands. The other panels show ratios of various predictions
without and with quasi-bound-state contributions simulated with the \ttonia\
and \etatt\ models to the
measured normalised differential cross-sections, for \ttem\ (middle three panels)
and \embb\ (lower panel). The markers showing the ratios for each prediction
are offset from the bin centres for better visibility, and the total
uncertainty in the prediction is shown by the error bar. The data statistical
uncertainty is shown by the black error bars and the total uncertainty by the
cyan (\ttem) or orange (\embb) band around unity.}
\end{figure}

The modelling of the \mll\ distribution is significantly improved by the
addition of the \ttonia\ quasi-bound-state component, reflected both
in the $\chi^2$ values and ratio plots. The \ttonia\ contribution adds
about 3\% to the absolute differential cross-section for $\mll<30$\,\GeV,
reducing to a negligible contribution for $\mll>100$\,\GeV. The
NNLO-reweighted {\scr Powheg\,+\,Pythia8} pQCD prediction shows a significant
deficit with respect to the data at high \mll, which has the effect of reducing
the discrepancy in the normalised distribution at low \mll. However, the
addition of the \ttonia\ contribution still improves the overall agreement
and $\chi^2$ values. The \mll\ distribution of the \etatt\ model is very similar
to that of \ttonia, and the improvements in $\chi^2$ are very similar.

Smaller improvements are seen for the \dphill\ and \dphillvmll\ distributions
when adding the \ttonia\ or \etatt\ contributions to the pQCD \ttbar\
predictions. In the case of the NNLO-reweighted {\scr Powheg\,+\,Pythia8}
prediction, where the modelling of \dphill\ is already good,
adding \ttonia\ slightly increases the $\chi^2$ value. The first \mll\ bin in
the two-dimensional distributions covers $\mll<80$\,\GeV, covering
the entire range where the quasi-bound-state contribution is significant,
and where it varies rapidly as a function of \mll. At low \mll, \dphill\ and
\mll\ are also correlated, due to the requirement that both leptons have
$\pt>20$\,\GeV, which strongly restricts the selected phase space.
The finer bins of the
one-dimensional \mll\ distribution therefore provide greater sensitivity
than the two-dimensional \dphillvmll\ distribution with the present binning.

Tables~\ref{ta:reschioned} and~\ref{ta:reschitwod} also show the $\chi^2$ values
for all the other one- and two-dimensional distributions measured in
Ref.~\cite{TOPQ-2024-12}. For these distributions, the changes in $\chi^2$
from adding the quasi-bound-state contributions are not significant,
demonstrating that these distributions do not provide sensitivity to
\ttbar\ threshold effects.

The results of fitting the quasi-bound-state signal strengths $\mu$ using each
of the \mll, \dphill\ and \dphillvmll\ distributions
are shown numerically in Table~\ref{ta:oniafit} and
graphically in Figure~\ref{fa:oniafit}.
Predictions using the \powhegminnlo\ and NNLO-reweighted
{\scr Powheg\,+\,Pythia8} pQCD \ttbar\ models were fitted together with
the \ttonia\ and
\etatt\ quasi-bound state models using the \ttem\ results, and
\powhegbbfl\ was fitted together with each quasi-bound-state model using
the \embb\ results.

\begin{table}
\caption{\label{ta:oniafit}Summary of the fits for quasi-bound-state
contributions with floating normalisation. The columns show the combination
of pQCD and quasi-bound-state models, the fitted normalised  distribution,
the number of degrees of freedom, the $\chi^2$ at the minimum and the
corresponding fitted value of $\mu$ with its total uncertainty
including statistical, experimental and prediction systematic contributions.
The \ttem\ distributions are fitted for the first four groups, and the
\embb\ distributions for the last four.}
\centering
\begin{tabular}{llrrc}\toprule
Model & Distribution & $N_\mathrm{dof}$ & $\chi^2_\mathrm{min}$ & $\mu$ \\ \midrule{\scr Powheg MiNNLO\,+\,PY8} + \ttonia & \mll & 13 & 16.2 & $1.22^{+0.38}_{-0.38}$\rule[-1mm]{0mm}{6mm} \\
& \dphill & 18 & 14.5 & $2.10^{+0.95}_{-0.78}$\rule[-1mm]{0mm}{6mm} \\
& \dphillvmll & 38 & 43.2 & $0.81^{+0.32}_{-0.31}$\rule[-1mm]{0mm}{6mm} \\
\cmidrule{1-1}{\scr Powheg\,+\,PY8} NNLO rew + \ttonia & \mll & 13 & 23.8 & $1.02^{+0.34}_{-0.33}$\rule[-1mm]{0mm}{6mm} \\
& \dphill & 18 & 16.3 & $0.18^{+0.51}_{-0.48}$\rule[-1mm]{0mm}{6mm} \\
& \dphillvmll & 38 & 67.4 & $0.62^{+0.34}_{-0.33}$\rule[-1mm]{0mm}{6mm} \\
\cmidrule{1-1}{\scr Powheg MiNNLO\,+\,PY8} + \etatt & \mll & 13 & 15.8 & $1.38^{+0.40}_{-0.41}$\rule[-1mm]{0mm}{6mm} \\
& \dphill & 18 & 14.2 & $2.45^{+0.93}_{-0.85}$\rule[-1mm]{0mm}{6mm} \\
& \dphillvmll & 38 & 43.4 & $0.86^{+0.33}_{-0.33}$\rule[-1mm]{0mm}{6mm} \\
\cmidrule{1-1}{\scr Powheg\,+\,PY8} NNLO rew + \etatt & \mll & 13 & 23.6 & $1.14^{+0.37}_{-0.37}$\rule[-1mm]{0mm}{6mm} \\
& \dphill & 18 & 16.4 & $0.19^{+0.54}_{-0.52}$\rule[-1mm]{0mm}{6mm} \\
& \dphillvmll & 38 & 68.4 & $0.54^{+0.35}_{-0.34}$\rule[-1mm]{0mm}{6mm} \\
\midrule
{\scr Powheg} $bb4l$\,+\,{\scr PY8} + \ttonia & \mll & 13 & 14.0 & $1.19^{+0.44}_{-0.45}$\rule[-1mm]{0mm}{6mm} \\
& \dphill & 18 & 19.9 & $1.25^{+1.04}_{-1.02}$\rule[-1mm]{0mm}{6mm} \\
& \dphillvmll & 38 & 63.9 & $0.60^{+0.43}_{-0.42}$\rule[-1mm]{0mm}{6mm} \\
\cmidrule{1-1}{\scr Powheg} $bb4l$\,+\,{\scr PY8} + \etatt & \mll & 13 & 13.7 & $1.34^{+0.47}_{-0.49}$\rule[-1mm]{0mm}{6mm} \\
& \dphill & 18 & 19.8 & $1.44^{+1.10}_{-1.13}$\rule[-1mm]{0mm}{6mm} \\
& \dphillvmll & 38 & 64.3 & $0.58^{+0.44}_{-0.44}$\rule[-1mm]{0mm}{6mm} \\
\bottomrule
\end{tabular}
\end{table}

\begin{figure}[tp]
\parbox{83mm}{\includegraphics[width=83mm]{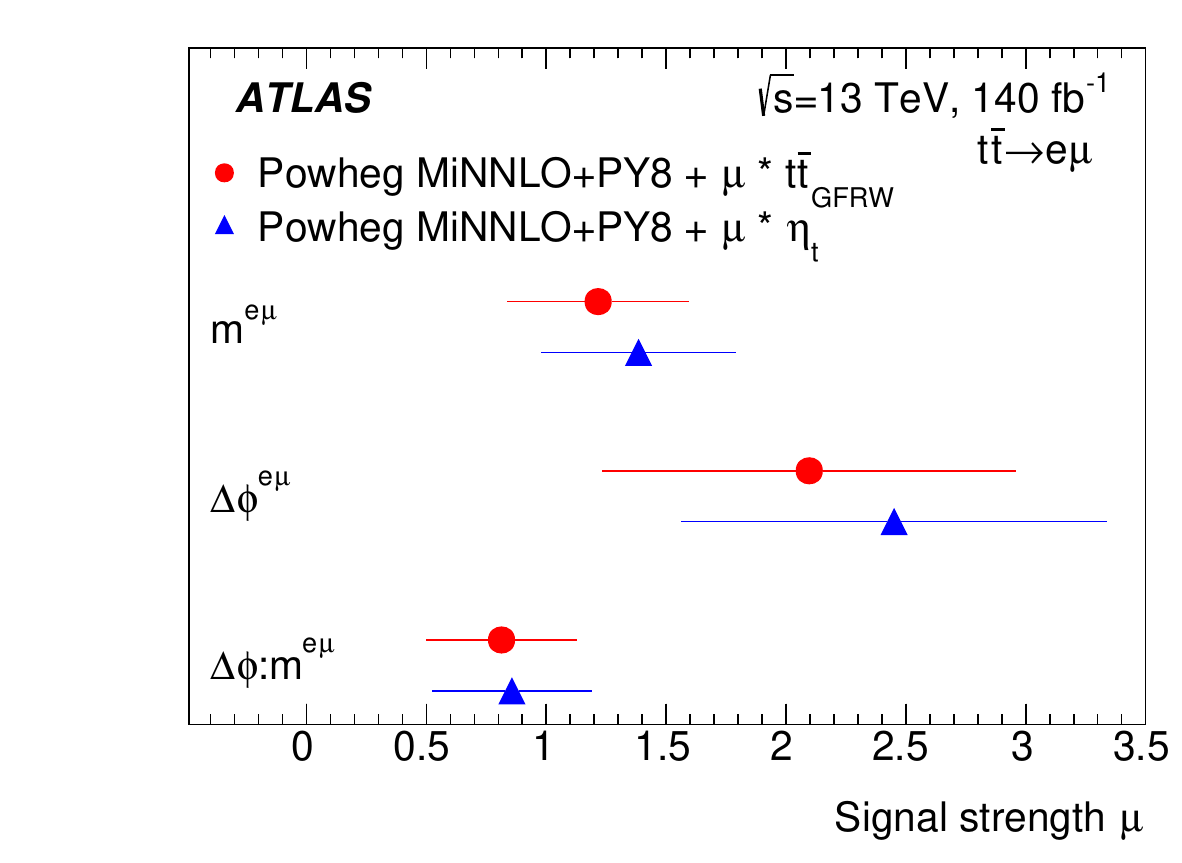}\vspace{-7mm}\center{(a)}}
\parbox{83mm}{\includegraphics[width=83mm]{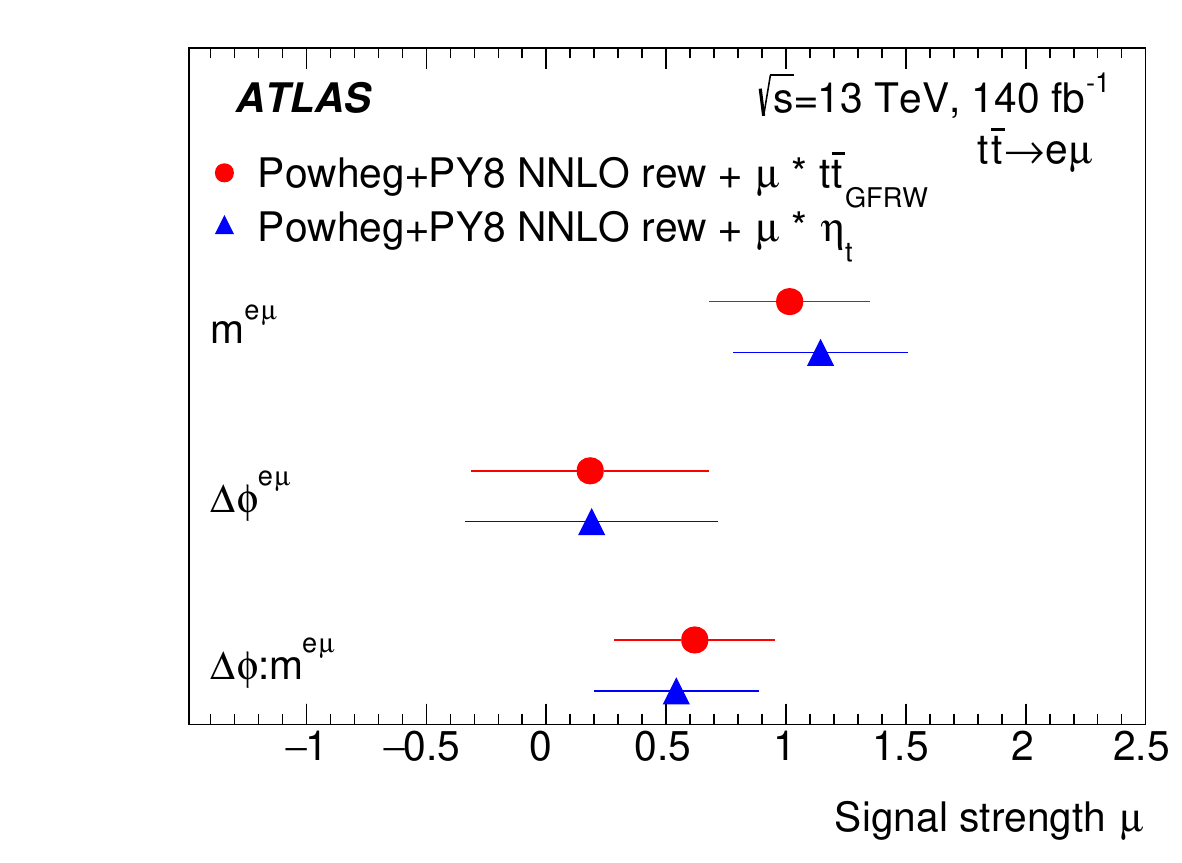}\vspace{-7mm}\center{(b)}}
\parbox{83mm}{\includegraphics[width=83mm]{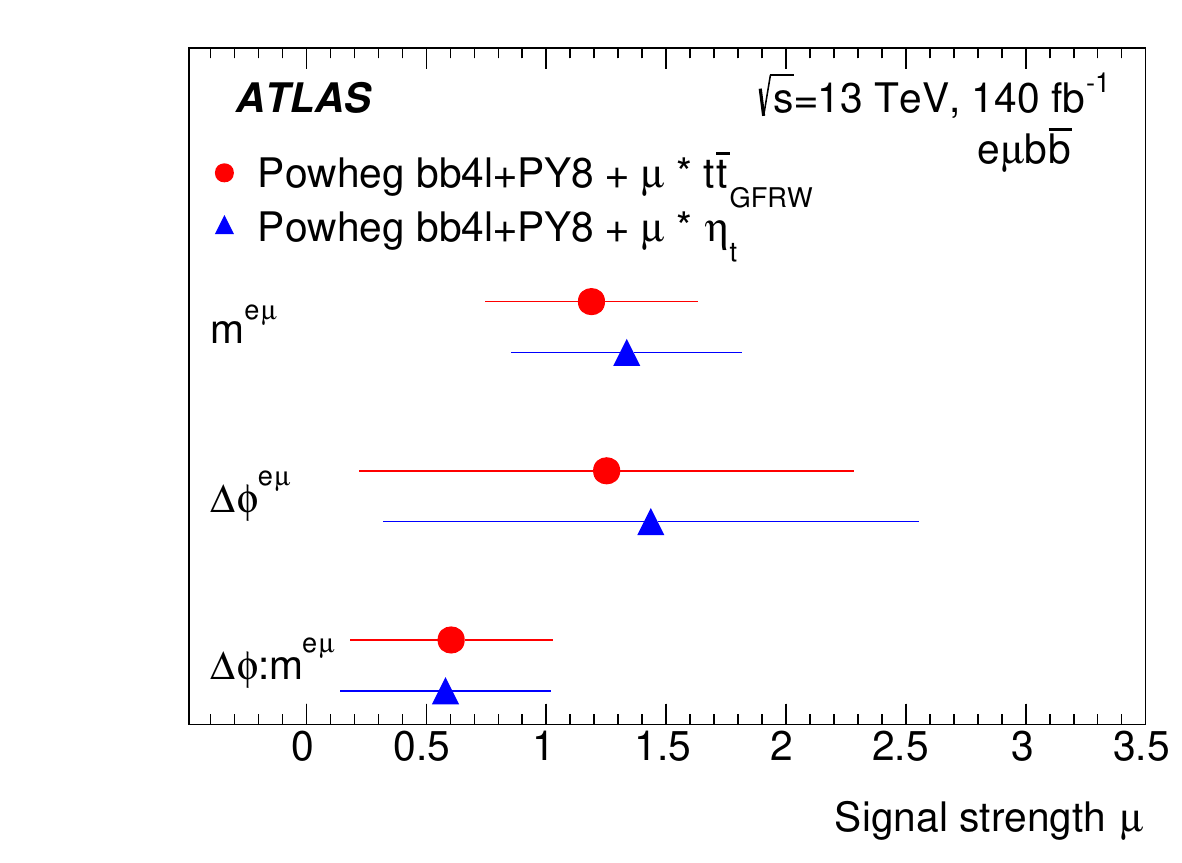}\vspace{-7mm}\center{(c)}}
\caption{\label{fa:oniafit}Fitted $\mu$ values, the signal strength of the
quasi-bound-state contribution,
using \ttonia\ (red circles) and \etatt\ (blue triangles)
models fitted to each of the \mll, \dphill\ and \dphillvmll\ distributions,
with the pQCD contribution modelled using (a)~\powhegminnlo,
(b) NNLO-reweighted {\scr Powheg\,+\,Pythia8} and (c) \powhegbbfl. The
error bars show the full uncertainties from the fit, including statistical,
experimental systematic and prediction-related uncertainties.}
\end{figure}

The fits to the \ttem\ \mll\ distributions all give values of $\mu$ that
are compatible with one, and differ from zero by more than three standard
deviations. The fits to the \embb\ \mll\ distribution with \powhegbbfl\
as the pQCD $\ttbar+Wt$ prediction give $\mu$ values that are compatible with
those from the \ttem\ fits and
differ from zero by more than two standard deviations. The experimental
uncertainties in the \embb\ distributions are larger than those for \ttem,
due to the extra requirements on $b$-tagged jets and the associated
systematic uncertainties.
The fits to the \dphill\ distribution have larger uncertainties and
greater spread than those for \mll,
with the results generally favouring $\mu=1$, but also being compatible
with $\mu=0$ for the NNLO-reweighted {\scr Powheg\,+\,Pythia8} pQCD model
and both quasi-bound-state models. The results from the \dphillvmll\
distribution are compatible with $\mu=1$, but do not strongly exclude
zero, especially in the case of the NNLO-reweighted pQCD model.
In general, the largest contribution to the total uncertainty in $\mu$
comes from the experimental measurement, followed by the uncertainties
in the pQCD predictions. The uncertainties in the quasi-bound-state model
play only a minor role.

The best description of the \mll\ distribution in the \ttem\ selection
is given by the
\powhegminnlo\ pQCD \ttbar\ model combined with either the \ttonia\ or
\etatt\ quasi-bound-state model. Taking \ttonia\ as the more physically
motivated model, the signal strength is
\begin{equation}\nonumber
\mu = 1.22\pm 0.38
\end{equation}
corresponding to an observed cross-section of $7.8\pm 2.4$\,pb, close to
that observed in Ref.~\cite{TOPQ-2025-11} using a fit to angular
observables in fully reconstructed dileptonic \ttbar\ events.

\FloatBarrier

\subsection{Conclusion}\label{sa:conc}

The sensitivity of leptonic differential distributions in
$\ttbar\rightarrow e\mu\nu\bar{\nu}\bbbar$ events to the formation of
quasi-bound states near the \ttbar\ threshold has been explored. Recent
ATLAS measurements of the \mll\ and \dphill\ distributions in $pp$ collisions
at \sxyt\ have been compared with \ttbar\ models incorporating pQCD predictions
with or without the addition of colour-singlet quasi-bound-state formation.
The data are better described by predictions including the latter contribution,
modelled either using a non-relativistic QCD Green's function approach,
or a pseudo-scalar \etatt\ resonance. Fits to the \mll\ distribution for
the \ttem\ selection show
evidence for a quasi-bound-state contribution with a significance exceeding
three standard deviations, and a measured cross-section in agreement with
dedicated studies using the reconstructed \ttbar\ invariant mass distribution
and angular variables.


%
%
\clearpage
 
\begin{flushleft}
\hypersetup{urlcolor=black}
{\Large The ATLAS Collaboration}

\bigskip

\AtlasOrcid[0000-0002-6665-4934]{G.~Aad}$^\textrm{\scriptsize 103}$,
\AtlasOrcid[0000-0001-7616-1554]{E.~Aakvaag}$^\textrm{\scriptsize 17}$,
\AtlasOrcid[0000-0002-5888-2734]{B.~Abbott}$^\textrm{\scriptsize 122}$,
\AtlasOrcid[0000-0002-0287-5869]{S.~Abdelhameed}$^\textrm{\scriptsize 118a}$,
\AtlasOrcid[0000-0002-1002-1652]{K.~Abeling}$^\textrm{\scriptsize 55}$,
\AtlasOrcid[0000-0001-5763-2760]{N.J.~Abicht}$^\textrm{\scriptsize 49}$,
\AtlasOrcid[0000-0002-8496-9294]{S.H.~Abidi}$^\textrm{\scriptsize 30}$,
\AtlasOrcid[0009-0003-6578-220X]{M.~Aboelela}$^\textrm{\scriptsize 45}$,
\AtlasOrcid[0000-0002-9987-2292]{A.~Aboulhorma}$^\textrm{\scriptsize 36e}$,
\AtlasOrcid[0000-0001-5329-6640]{H.~Abramowicz}$^\textrm{\scriptsize 156}$
% ...
\bigskip
\\

% ...
$^{17}$Department for Physics and Technology, University of Bergen, Bergen; Norway.\\
% ...
$^{30}$Physics Department, Brookhaven National Laboratory, Upton NY; United States of America.\\
% ...
$^{36}$$^{(a)}$Facult\'e des Sciences Ain Chock, Universit\'e Hassan II de Casablanca;$^{(b)}$Facult\'{e} des Sciences, Universit\'{e} Ibn-Tofail, K\'{e}nitra;$^{(c)}$Facult\'e des Sciences Semlalia, Universit\'e Cadi Ayyad, LPHEA-Marrakech;$^{(d)}$LPMR, Facult\'e des Sciences, Universit\'e Mohamed Premier, Oujda;$^{(e)}$Facult\'e des sciences, Universit\'e Mohammed V, Rabat;$^{(f)}$Institute of Applied Physics, Mohammed VI Polytechnic University, Ben Guerir; Morocco.\\
% ...
$^{45}$Physics Department, Southern Methodist University, Dallas TX; United States of America.\\
% ...
$^{49}$Fakult\"{a}t Physik, Technische Universit{\"a}t Dortmund, Dortmund; Germany.\\
% ...
$^{55}$II. Physikalisches Institut, Georg-August-Universit\"{a}t G\"ottingen, G\"ottingen; Germany.\\
% ...
$^{103}$CPPM, Aix-Marseille Universit\'e, CNRS/IN2P3, Marseille; France.\\
% ...
$^{118}$$^{(a)}$New York University Abu Dhabi, Abu Dhabi;$^{(b)}$United Arab Emirates University, Al Ain; United Arab Emirates.\\
% ...
$^{122}$Homer L. Dodge Department of Physics and Astronomy, University of Oklahoma, Norman OK; United States of America.\\
% ...
$^{156}$Raymond and Beverly Sackler School of Physics and Astronomy, Tel Aviv University, Tel Aviv; Israel.\\
% ...
$^{a}$ Also at Affiliated with an institute formerly covered by a cooperation agreement with CERN.\\
$^{b}$ Also at An-Najah National University, Nablus; Palestine.\\
$^{c}$ Also at Borough of Manhattan Community College, City University of New York, New York NY; United States of America.\\
$^{d}$ Also at Center for Interdisciplinary Research and Innovation (CIRI-AUTH), Thessaloniki; Greece.\\
$^{e}$ Also at Centre of Physics of the Universities of Minho and Porto (CF-UM-UP); Portugal.\\
$^{f}$ Also at CERN, Geneva; Switzerland.\\
% ...

\end{flushleft}


%

%
%

%

%
%
%
%

\end{document}